%% file: main.tex
\documentclass[10pt]{article}
\usepackage[T1]{fontenc}
\usepackage{amsmath}
\usepackage{amsthm}
\usepackage{amsfonts}
\usepackage{bbm}
\usepackage{mathrsfs}

\usepackage{float}
\usepackage{hyperref}
\usepackage[nameinlink,capitalize]{cleveref}
\usepackage{chngcntr}
\usepackage{cite}
\usepackage[nottoc,numbib]{tocbibind}
\usepackage{subcaption}

\newtheorem{remark}{Remark}

\counterwithin{theorem}{section}
\counterwithin{lemma}{section}
\counterwithin{remark}{section}
\counterwithin{observation}{section}
\counterwithin{definition}{section}

\usepackage{pgf,tikz,pgfplots}
\usetikzlibrary{arrows}
\usetikzlibrary{arrows.meta}
\usetikzlibrary{spy}

\newcommand{\vb}[1]{\mathbf{#1}}
\newcommand{\bm}[1]{\boldsymbol{#1}}

\newcommand{\wrt}{\text{w.r.t.}}

\newcommand{\perm}{\bm{\mathbbm{E}}}
\newcommand{\one}{\bm{\mathbbm{1}}}
\newcommand{\Vol}{V}
\newcommand{\Surf}{A}
\newcommand{\curv}{s}

\newcommand{\pder}[2]{ \dfrac{\partial {#1}}{\partial {#2}}}
\newcommand{\norm}[1]{\| {#1} \|}
\newcommand{\con}[2]{\langle {#1} , \, {#2} \rangle}
\newcommand{\jump}[1]{\ensuremath{[\![#1]\!]} }
\DeclareMathOperator{\at}{\bigg\vert}

\DeclareMathOperator{\ext}{\mathrm{ext}}

\DeclareMathOperator{\polar}{\mathrm{polar}}
\DeclareMathOperator{\sym}{\mathrm{sym}}
\DeclareMathOperator{\dev}{\mathrm{dev}}
\DeclareMathOperator{\sph}{\mathrm{sph}}
\DeclareMathOperator{\skw}{\mathrm{skew}}
\DeclareMathOperator{\Anti}{\mathrm{Anti}}
\DeclareMathOperator{\axl}{\mathrm{axl}}
\DeclareMathOperator{\tr}{\mathrm{tr}}
\DeclareMathOperator{\id}{\mathrm{id}}
\DeclareMathOperator{\cof}{\mathrm{cof}}

\newcommand{\dd}{\mathrm{d}}
\newcommand{\Grad}{\mathrm{D}}
\DeclareMathOperator{\curl}{\mathrm{curl}}
\DeclareMathOperator{\Curl}{\mathrm{Curl}}
\DeclareMathOperator{\di}{\mathrm{div}}
\DeclareMathOperator{\Di}{\mathrm{Div}}

\newcommand{\R}{\mathbb{R}}
\newcommand{\SO}{\mathrm{SO}}
\newcommand{\so}{\mathfrak{so}}

\newcommand{\Sym}{\mathrm{Sym}}
\newcommand{\GL}{\mathrm{GL}}

\newcommand{\C}{\mathit{C}}

\newcommand{\CG}{\mathcal{CG}}
\newcommand{\Ned}{\mathcal{N}}

\newcommand{\defmap}{\bm{\varphi}}
\newcommand{\disp}{\vb{u}}
\newcommand{\rot}{\bm{\omega}}
\newcommand{\Piola}{\bm{S}}
\newcommand{\Double}{\bm{\mathfrak{P}}}
\newcommand{\Dtrac}{\bm{\mathfrak{T}}}
\newcommand{\Dforce}{\bm{\mathfrak{M}}}
\newcommand{\RPiola}{\bm{T}}
\newcommand{\RDouble}{\bm{\mathfrak{S}}}
\newcommand{\defgrad}{\bm{F}}
\newcommand{\Coss}{\overline{\bm{R}}}
\newcommand{\Stretch}{\overline{\bm{U}}}

\newcommand{\elec}{\vb{e}}

\newcommand{\Curva}{\overline{\bm{\mathfrak{A}}}}
\newcommand{\Wry}{\overline{\bm{\mathfrak{K}}}}
\newcommand{\mmag}{\vb{m}}
\newcommand{\hmag}{\vb{h}}
\newcommand{\bmag}{\vb{b}}
\newcommand{\fmag}{\overline{\bm{\mathfrak{g}}}}
\newcommand{\tbmag}{\overline{\bm{\mathfrak{b}}}}
\newcommand{\amag}{\overline{\bm{\mathfrak{a}}}}

\newcommand{\Cm}{\mathbb{C}}
\newcommand{\J}{\mathbb{J}}

\newcommand{\muc}{\mu_{\mathrm{c}}}
\newcommand{\Lc}{L_\mathrm{c}}
\newcommand{\Lm}{\mathbb{L}}
\newcommand{\Fm}{\bm{\mathbbm{F}}}
\newcommand{\Mm}{\bm{\mathbbm{H}}}

\newcommand{\Energy}{\mathcal{I}}
\newcommand{\Work}{\mathcal{W}}

\newcommand{\AS}[1]{{\color{black} #1}}

\makeatletter
\let\@fnsymbol\@arabic
\makeatother

\title{Cosserat micropolar and couple-stress elasticity models of flexomagnetism at finite deformations}

\author{
\normalsize{Adam Sky$^*$}\thanks{$^*$Corresponding author: Adam Sky, Institute of Computational Engineering and Sciences, Department of Engineering, Faculty of Science, Technology and Medicine, University of Luxembourg, 6 Avenue de la Fonte, L-4362 Esch-sur-Alzette, Luxembourg, email: adam.sky@uni.lu}
    , \quad
    \normalsize{David Codony}\thanks{David Codony, Laboratori de Càlcul Numèric (LaCàN), Universitat Politècnica de Catalunya (UPC), Campus Nord UPC-C2, E-08034 Barcelona, Spain, email: david.codony@upc.edu}\,
    \thanks{David Codony, Institut de Matemàtiques de la UPC-BarcelonaTech (IMTech), Barcelona, Spain, email: david.codony@upc.edu}
    , \quad
    \normalsize{Stephan Rudykh}\thanks{Stephan Rudykh, School of Mathematical and Statistical Sciences, College of Science and Engineering, University of Galway, Galway, H91 TK33, Ireland, email: stephan.rudykh@universityofgalway.ie}
    , \quad
    \normalsize{Andreas Zilian}\thanks{Andreas Zilian, Institute of Computational Engineering and Sciences, Department of Engineering, Faculty of Science, Technology and Medicine, University of Luxembourg, 6 Avenue de la Fonte, L-4362 Esch-sur-Alzette, Luxembourg, email: andreas.zilian@uni.lu}
    ,\\
	\normalsize{St\'ephane P. A. Bordas}\thanks{St\'ephane P. A. Bordas, Institute of Computational Engineering and Sciences, Department of Engineering, Faculty of Science, Technology and Medicine, University of Luxembourg, 6 Avenue de la Fonte, L-4362 Esch-sur-Alzette, Luxembourg, email: stephane.bordas@alum.northwestern.edu}\,
    \thanks{Computational Modelling and Data Science, Department of Engineering, University of Exeter, United Kingdom, email: s.bordas@exeter.ac.uk}
    \quad
    \normalsize{and} \quad
	\normalsize{Patrizio Neff}\thanks{Patrizio Neff, Chair for Nonlinear 
		Analysis and Modelling, Faculty of Mathematics, Universit\"{a}t Duisburg-Essen,
		Thea-Leymann Str. 9, 45127 Essen, Germany, email: patrizio.neff@uni-due.de}
}

\begin{document}

\maketitle

\begin{abstract}
We propose geometrically nonlinear (finite) continuum models of flexomagnetism based on the Cosserat micropolar and its descendent couple-stress theory. These models introduce the magneto-mechanical interaction by coupling the micro-dislocation tensor of the micropolar model with the magnetisation vector using a Lifshitz invariant. In contrast to conventional formulations that couple strain-gradients to the magnetisation using fourth-order tensors, our approach relies on third-order tensor couplings by virtue of the micro-dislocation being a second-order tensor.
Consequently, the models permit centrosymmetric materials with a single new flexomagnetic constant, and more generally allow cubic-symmetric materials with two such constants.
We postulate the flexomagnetic action-functionals and derive the corresponding governing equations using both scalar and vectorial magnetic potential formulations, and present numerical results for a nano-beam geometry, confirming the physical plausibility and computational feasibility of the models. 
\\
\vspace*{0.25cm}
\\

\end{abstract}

\section{Introduction}

Flexomagnetism is a contemporarily studied  physical phenomenon where a magnetic field is induced by deformation alone, without the need for external magnetic fields, a time-dependent electric field, or electric currents \cite{TANG2025100878}. Thus, it has clear potential to be utilised in energy generation, electronics, or various magneto-mechanical applications \cite{ZHANG2024112648,BERJAMIN2025104369,ZHANG2025109826,PerezGarcia2025a,PerezGarcia2025b}. Unlike the one-way coupling of magnetostriction \cite{DorfmannMW,Dorfmann2014,LIU2014451}, the effect is a two-way coupling, and in contrast to the two-way coupling of piezomagnetism \cite{Xu20212,XU2020106352,LIU2014451}, an inversion-symmetry breaking crystalline structure is not required. 

The conceptual development of continuum flexomagnetism \cite{Sidhardh2018} has largely mirrored that of its electric analogue, flexoelectricity \cite{tagantsev1986_,catalan2013_,wang2019_,Codony,CodonyFP,CODONY2021104182,Gupta2025_}. 
By analogy, a coupling between strain-gradients and magnetisation has been proposed even in centrosymmetric materials, leading to continuum formulations of flexomagnetism being constructed similarly to flexoelectricity \cite{Qiao,Lee}.
Alternatively, some formulations develop models for simplified geometries such as beams \cite{MALIKAN2021114179,Malikan2022,Li2024,Ray2025,Malikan2022Composite} plates \cite{Xu13022025} and shells \cite{Wang}, directly describing nanoscale shapes within a dimensionally reduced continuum framework.
However, to the best of our knowledge, existing fully three-dimensional flexomagnetic continuum models remain confined to the geometrically linear regime, whereas flexoelectricity has already seen extensions to finite deformations, for instance via finite strain-gradient theories \cite{CODONY2021104182,El-Dhaba2024_} or the full micromorphic model \cite{MCBRIDE2020113320} of Eringen \cite{Eringen1999} and Mindlin \cite{Mindlin1964}.

Unlike flexoelectricity, which is considered to be inherent to all dielectric materials~\cite{RestaPhysRevLett}, flexomagnetism has been reported primarily in materials with long-range magnetic ordering, particularly in antiferromagnetic \cite{Makushko2022Cr2O3}, ferrimagnetic \cite{TANG2025100878,Lee}, and some ferromagnetic systems \cite{Gong}. 
Antiferromagnets such as Cr$_2$O$_3$~\cite{Makushko2022Cr2O3}, commonly known as Chromia, are of particular interest because they feature a relatively high Néel temperature and a centrosymmetric structure. The latter dismisses piezomagnetic effects, leaving flexomagnetism as the dominant magneto-mechanical reaction.
Finally, multiferroic materials, which exhibit simultaneous magnetic and electric ordering, are also possible candidates~\cite{Eliseev}. In contrast, paramagnetic and diamagnetic materials are generally not good candidates for flexomagnetism, as they lack the required magnetic ordering~\cite{Gaeta2021,WHITES2005479}.

Despite ongoing research into flexomagnetism, many challenges remain. These include the determination of the magnitude of flexomagnetic coefficients, the complexity of boundary conditions in higher-order models, and the lack of reliable experimental benchmarks. Moreover, current understanding of the mechanism underlying the flexomagnetic coupling remains incomplete, and no material has yet demonstrated a flexomagnetic response strong enough for real-world device applications.
To address these limitations, current efforts focus on strategic material selection, strain-gradient engineering \cite{Malikan2022Composite,GRECO2024105477,MOCCI2021104643}, and manipulation of internal interactions to enhance the effect~\cite{TANG2025100878}. Multi-scale methods are also being developed, where continuum material parameters are characterised using first-principles calculations \cite{CodonyFP} such as density functional theory (DFT) \cite{Gong,Qiao}.

Generally speaking, the effects of flexomagnetism become pronounced at the nanoscale, where large strain-gradients naturally arise. 
These are neglected in the classical Cauchy continuum, but recovered in higher-order gradient theories or generalised continua.
Since the flexomagnetic effect co-occurs with strain-gradients it is traditionally coupled with them in continuum models.
However, their co-occurrence does not imply that magnetisation truly couples to every component of the strain-gradient. 
Indeed, since magnetic dipoles, unlike electric dipoles, are not composed of spatially separated charges, the intuition underlying the flexoelectric mechanism may not be directly applicable for flexomagnetism.
\AS{Moreover, experimental evidence and first-principles calculations suggest that even flexoelectricity itself is not governed solely by continuum strain gradients, but rather often enhanced or mediated by local structural distortions at defects, interfaces, domain walls, and other nanoscale inhomogeneities \cite{Gao2018,Wu2022}. Recent work on flexomagnetism points in a similar direction, linking magnetic responses to localised edge stresses, bond-angle distortions, and other nanostructure-induced symmetry-breaking effects \cite{Qiu2023}.}
Finally, some conceptual interpretations suggest that the orientation of magnetic dipoles may not be rigidly tied to material curves \cite{Makushko2022Cr2O3}.

In this work, we propose a new phenomenological model for flexomagnetism based on the geometrically exact Cosserat micropolar continuum theory~\cite{Eringen1999,Cosserat,Eremeyev2013,Neff2004,Neff2015Uni,Lankeit2016} in dislocation form \cite{Ghiba}. This theory extends classical elasticity by introducing independent micro-rotations at each material point. The resulting kinematic fields include a micro-rotation tensor $\Coss:\Vol \to \SO(3)$ and a corresponding micro-dislocation tensor $\Curva = \Coss^T \Curl \Coss$, allowing for the modelling of effects not captured by classical theories, such as rotational inertia, size effects, and asymmetric stress distributions \cite{NEFF2006574,Neff2006Couple}. 
\AS{The related couple-stress model may be viewed as a constrained limit case of micropolar elasticity, in which the independent micro-rotation is no longer retained as a fully independent degree of freedom. Both theories have been widely employed to model size-dependent mechanical behaviour in micro- and nano-structured beams, plates, shells, arches, and energy-harvesting systems \cite{Sahmani2025,Sahmani2025b,Shahzad2025,AtifShahzad2023}.}
\AS{Thus, seeing as} the Cosserat model is particularly relevant for materials with a pronounced microstructure or in cases where the scale of deformation is comparable to internal length scales~\cite{Forest2013Cosserat, Eringen1999}, it is a natural candidate for a flexomagnetic model.

In our phenomenological formulation, the magneto-mechanical coupling is introduced through the interaction between the magnetisation field $\vb{m}:\Vol \to \R^3$ and the micro-dislocation tensor 
$\Curva:\Vol \to \R^{3 \times 3}$. Since this coupling arises from the independent micro-rotations $\Coss$, and not from the stretch tensor $\bm{U} = \sqrt{\defgrad^T\defgrad}$, with the deformation tensor $\defgrad = \Grad \defmap$ of the deformation map $\defmap:\Vol \to \Vol_\varphi$, the model is designed to reflect the physical intuition that extension alone may not induce a magnetic response. This aligns with the understanding that magnetic dipoles, unlike electric dipoles, are not composed of spatially separated charges and therefore do not physically deform. Further, because the Cosserat rotation tensor and its associated micro-dislocation are independent variables with their own coupling to deformation, this framework permits magnetic-dipole orientations that are genuinely independent from deformation curves. In fact, the rigidity of this coupling, and thus the magnetisation induced by curvature, is governed by the Cosserat coupling parameters, such as the couple modulus $\muc \ge 0$ \cite{Neff2006Couple}.
Finally, because the coupling occurs between a vector and a second-order tensor, rather than the more involved third-order tensors typical of conventional strain-gradient models, the physically linear regime for cubic-symmetric materials involves at most only two new material parameters. This significantly simplifies the theoretical formulation and reduces the amount of data needed for experimental calibration of the model.

The structure of this paper is as follows. We begin by presenting the geometrically nonlinear micropolar model. Next, we introduce the proposed flexomagnetic coupling. This is followed by a formulation based on the magnetic scalar potential, which is then rewritten in terms of a vector potential using a partial Legendre transform. The resulting model is subsequently examined in detail on a representative nano-beam domain. Lastly, we conclude with a discussion of the main findings and outline directions for future work.

\subsection{Notation}

The following notation is employed throughout this work.
Exceptions are made clear in the precise context.
\begin{itemize}
    \item Vectors are defined as bold lower-case letters $\vb{v}, \, \bm{\xi} \in \R^d$.
    \item Second-order tensors are denoted with bold capital letters $\bm{T}\in \R^{d \times d}$.
    \item Third-order tensors are identified by bold blackboard capital letters $\bm{\mathbbm{M}}\in \R^{d \times d \times d}$.
    \item Fourth-order tensors are designated by the blackboard format $\mathbb{C} \in \R^{d \times d \times d \dots}$.
    \item We denote the Cartesian basis as $\{\elec_1, \, \elec_2, \, \elec_3\}$.
    \item Summation over indices follows the standard rule of repeating indices.
    \item The angle-brackets define scalar products of arbitrary dimensions $\con{\vb{v}}{\vb{u}} = v_i u_i$, $\con{\bm{T}}{\bm{F}} = T_{ij}F_{ij}$.
    \item The matrix product is used to indicate all partial-contractions between a higher-order and a lower-order tensor $\bm{T}\vb{v} = T_{ij} v_j \elec_i$, $\mathbb{C}\bm{T} = C_{ijkl}T_{kl}\elec_i \otimes \elec_j$, $\bm{\mathbbm{M}}\bm{T} = M_{ijk}T_{jk}\elec_i$.
    \item The second-order identity tensor is defined via $\one = \vb{e}_i \otimes \vb{e}_i$, such that $\one \vb{v} = \vb{v}$. 
    \item Volumes, surfaces and curves of the physical domain are identified via $\Vol$, $\Surf$ and $\curv$, respectively.   
    \item The space of invertible matrices with a positive determinant is given by $\GL^+(d) = \{\bm{T} \in \R^{d \times d} \; | \; \det \bm{T} > 0\}$.
    \item Symmetric positive definite matrices are given by  $\Sym^{++}(d) = \{ \bm{T} \in \Sym(d) \; | \; \con{\vb{v}}{\bm{T} \vb{v}} > 0 \; \forall \, \vb{r} \in \R^d \setminus \{0\} \}$.
    \item The space of orthogonal transformations is defined as $\mathrm{O}(d) = \{ \bm{R} \in \R^{d \times d} \;| \; \det \bm{R} = \pm 1 \, , \; \bm{R}^{T}\bm{R} = \one \}$.
    \item Its restriction to rotation matrices is given by $\SO(d) = \{ \bm{R} \in \R^{d \times d} \; | \; \det \bm{R} = 1 \, , \; \bm{R}^{T}\bm{R} = \one \}$.
    \item We define the vector space of skew-symmetric second order tensors as $\so(d) = \{ \bm{T} \in \R^{d \times d} \; | \; \bm{T} = -\bm{T}^{T} \}$.
   \item The space $\so(d)$ is associated with the operators $\skw \bm{T} = (1/2)(\bm{T} - \bm{T}^T) \in \so(d)$, $\Anti \vb{v} = \vb{v} \times \one \in \so(3)$, and its inverse $\axl (\Anti \vb{v}) = \vb{v}$.
   \item The nabla operator is defined as $\nabla = \elec_i \partial_i$.
   \item The left-gradient is given via $\nabla$, such that $\nabla \lambda = \nabla \otimes \lambda$.
    \item The right-gradient is defined for vectors and higher order tensors via $\Grad$, such that $\Grad \vb{v} = \vb{v} \otimes \nabla$.
    \item We define the vectorial divergence as $\di \vb{v} = \con{\nabla}{\vb{v}}$.
    \item The tensor divergence is given by $\Di \bm{T} = \bm{T} \nabla$, implying a single contraction acting row-wise.
    \item The vectorial curl operator reads $\curl \vb{v} = \nabla \times \vb{v}$
    \item For tensors the operator is given by $\Curl \bm{T} = -\bm{T} \times \nabla$, acting row-wise. 
\end{itemize}

\section{The \AS{finite-strain} Cosserat model}

Let the current configuration of a physical object in three dimensional space $\Vol_\varphi \subset \R^3$, composed of material points and their independent orientations, be given by a deformation map that assigns each $\vb{x}$-coordinate a displacement $\vb{u}$ and by an independent rotation field, both as functions of the reference configuration $\Vol \subset \R^3$ 
\begin{align}
    &\vb{x} + \vb{u} = \defmap : \Vol  \to \Vol_\varphi \subset \R^3 \, , && \Coss : \Vol \to \SO(3)  \, ,  
\end{align}
the elastic Cosserat micropolar model can be derived by postulating an equivalence between the deformation gradient and the following multiplicative decomposition 
\begin{align}
    &\defgrad = \Coss \, \Stretch \, , && \defgrad = \Grad \defmap : \Vol \to \GL^+(3) \, , &&  \Stretch : \Vol \to \GL^+(3) \, . 
\end{align}
In this setting, the Biot-type stretch tensor $\Stretch$ can be interpreted as some intermediate configuration before the independent rotation is applied, see \cref{fig:mp}. 
\begin{figure}
    \centering
    \input{figs/mp}
    \caption{Mapping of the reference configuration $\Vol \subset \R^3$ onto the current configuration $\Vol_\varphi \subset \R^3$ by the deformation tensor $\defgrad: \Vol \to \GL^+(3)$. The mapping can be decomposed into the initial application of the Biot-type stretch tensor $\Stretch:\Vol \to \GL^+(3)$, followed by an independent rotation of the material points $\Coss:\Vol \to \SO(3)$. Note that in general, the tangential curves of the various configurations do not agree with the independent orientation of each material point. The correspondence between the two is predominantly governed by the Cosserat couple modulus $\muc$, which can be roughly understood as a rotational spring.}
    \label{fig:mp}
\end{figure}
It is to be noted that $\Stretch$ is not required to be symmetric, such that the decomposition is \textbf{not} the classical polar decomposition.
In the purely Lagrangian quasi-static case, the model is obtained by postulating the action-functional 
\begin{align}
    \int_{\Vol} \Psi_\mathrm{Coss}(\defgrad,\Coss,\Grad \Coss) - \con{\defmap}{\vb{f}} - \con{\Coss}{\Dforce} \, \dd \Vol - \int_{\Surf_{N}^{\defmap}} \con{\defmap}{\vb{t}} \, \dd \Surf - \int_{\Surf_{N}^{\Coss}} \con{\Coss}{\Dtrac} \, \dd \Surf
    \quad \to \quad \min \quad \wrt \quad \{\defmap,\Coss\} \,  \,  , 
\end{align}
where $\vb{f}: \Vol \to \R^3$ represent body (volume) forces, $\Dforce:\Vol \to \R^{3 \times 3}$ are couple-forces, $\vb{t}:\partial\Vol \to \R^3$ are surface tractions, and $\Dtrac:\partial\Vol \to \R^{3 \times 3}$ are couple-tractions. For the internal energy of the system to be frame-invariant, the energy density function must satisfy
\begin{align}
    \Psi_\mathrm{Coss}(\defgrad,\Coss,\Grad \Coss) = \Psi_\mathrm{Coss}(\bm{Q}\defgrad,\bm{Q}\Coss,\Grad [\bm{Q}\Coss]) = \Psi_\mathrm{Coss}(\bm{Q}\defgrad,\bm{Q}\Coss, \bm{Q} \Grad \Coss) \quad \forall \, \bm{Q}\in \SO(3) \, , 
\end{align}
for any superimposed rigid body motion $\defmap \mapsto \bm{Q}\defmap + \vb{c}$ and $\Coss \mapsto \bm{Q} \Coss$ with $\vb{c} \in \R^3$ and $\bm{Q} \in \SO(3)$. Considering the energy density function for a fixed point $\vb{x}_i \in \R^3$ while choosing $\bm{Q} = \Coss^T|_{\vb{x}_i}$ we observe 
\begin{align}
\Psi_\mathrm{Coss}(\bm{Q}\defgrad,\bm{Q}\Coss,\bm{Q}\Grad \Coss) \at_{\vb{x}_i} &= \Psi_\mathrm{Coss}(\Coss^T\defgrad,\Coss^T\Coss,\Coss^T\Grad \Coss) \at_{\vb{x}_i} 
\notag\\
&= \Psi_\mathrm{Coss}(\Coss^T\defgrad,\one,\Coss^T\Grad \Coss) \at_{\vb{x}_i} = \Psi_\mathrm{Coss}(\Stretch,\Coss^T\Grad \Coss) \at_{\vb{x}_i} \, ,
\end{align}
since $\Stretch = \Coss^T \defgrad$ by definition, and the function clearly cannot depend on the constant identity tensor $\one \in \R^{3\times3}$. At this point it is common to propose an additive split in the energy density across the strain and curvature functions
\begin{align}
    \Psi_\mathrm{Coss}(\Stretch,\Coss^T\Grad \Coss) = \Psi_\mathrm{mp}(\Stretch) + \Psi_\mathrm{curv}(\Coss^T\Grad \Coss) \, .
\end{align}
A general structure of a non-negative isotropic function for $\Psi_\mathrm{mp}(\Stretch)$ can be given using a weighted Cartan decomposition by the form
\begin{align}
    \Psi_\mathrm{mp}(\Stretch) &= \dfrac{\lambda_\mathrm{e}}{2} \tr(\Stretch-\one)^2 + \mu_\mathrm{e} \norm{\sym(\Stretch - \one)}^2 + \muc\norm{\skw(\Stretch - \one)}^2 
    \notag \\ 
    & = \dfrac{1}{2} \con{\sym(\Stretch-\one)}{\Cm \sym(\Stretch-\one)} + \muc\norm{\skw\Stretch}^2
    \\
    & = \frac{1}{2}\norm{\sym(\Stretch - \one)}_{\Cm}^2 + \muc\norm{\skw\Stretch}^2 \, , \notag   
\end{align}
such that $\Psi_\mathrm{mp}(\one) = 0$, $\Grad_{\Stretch}\Psi_\mathrm{mp}|_{\one} = 0$, and the material tensor reads (see \cref{ap:fourth} for tensor-forms)
\begin{align}
    \Cm = \lambda_\mathrm{e} \one \otimes \one + 2 \mu_\mathrm{e} \J \in \R^{3 \times 3 \times 3 \times 3} \, .   
\end{align}
The tensor $\Stretch -\one$ is known as the Cosserat strain tensor. The constants $\lambda_\mathrm{e} \geq 0$ and $\mu_\mathrm{e} > 0$ are the classical elastic Lam\'e material constants, and $\muc \geq 0$ is the Cosserat couple modulus. Its coupling effect can be observed via the expansion
\begin{align}
    &2\muc\norm{\skw\Stretch}^2 = \dfrac{\muc}{2} \norm{\Stretch - \Stretch^T}^2 = \dfrac{\muc}{2} \norm{\Coss^T \defgrad - \defgrad^T \Coss}^2 = \dfrac{\muc}{2} \norm{\Coss^T \bm{R} \, \bm{U} - \bm{U}\bm{R}^T \Coss}^2  \, , 
\end{align}
where $\defgrad = \bm{R} \, \bm{U}$ is the true polar decomposition such that
\begin{align}
    \bm{R} =\polar (\bm{F}) \, ,&& \bm{R}:  \Vol \to \SO(3) \, , && \bm{U}:\Vol \to \Sym^{++}(3) \, .
\end{align}
Clearly, any deviation of $\norm{\Coss^T \bm{R} \, \bm{U} - \bm{U}\bm{R}^T \Coss}^2$ from zero depends on the geodesic distance between $\Coss^T\bm{R}$ and the constant identity $\one$, subsequently stretched by $\bm{U}$. This value is then scaled by the material parameter $\muc$. Next, considering the curvature energy we note that gradients of a rotation tensor are fully controlled by its Curl, cf. \cite{SO3} and \cref{ap:curl}. Since the Curl of a second-order tensor yields a second-order tensor, we now adopt this form to redefine the curvature energy density as
\begin{align}
    \Psi_\mathrm{curv} = \Psi_\mathrm{curv}(\Coss^T \Curl \Coss) = \Psi_\mathrm{curv}(\Curva) \, , && \Curva = \Coss^T \Curl \Coss \, , && \Curva : \Vol \to \R^{3\times 3} \, , 
\end{align}
which allows for compact notation and a simpler material tensor. Namely, we can postulate the isotropic curvature energy function to be \cite{Ghiba} 
\begin{align}
    \Psi_\mathrm{curv}(\Curva) &= \dfrac{\mu_\mathrm{e} \Lc^2}{2} (\alpha_1 \norm{\dev \sym \Curva}^2 + \alpha_2 \norm{\skw \Curva}^2 + \alpha_3 \norm{\sph \Curva}^2) = \dfrac{\mu_\mathrm{e} \Lc^2}{2} \con{\Curva}{\Lm \Curva} = \dfrac{\mu_\mathrm{e} \Lc^2}{2} \norm{\Curva}^2_{\Lm} \, ,
\end{align}
where $\Lc \geq 0$ is the characteristic length-scale parameter that governs the energy of higher order effects, and the tensor of dimensionless weights takes the form (see \cref{ap:fourth} for tensor-forms)
\begin{align}
    &\Lm = \alpha_1 \mathbb{D} \mathbb{S} + \alpha_2 \mathbb{A} + \alpha_3 \mathbb{V} \in \R^{3 \times 3\times 3\times 3} \, .
\end{align}
A deduction of the constant identity or a comparable modification of the energy density function is not needed here since for a superimposition $\bm{Q}\Coss$ one already finds $\Curva = (\bm{Q}\Coss)^T\Curl (\bm{Q}\Coss) = \Coss^T\Curl \Coss$.
Putting it all together, the internal mechanical energy of the Cosserat model is given by
\begin{align}
    \boxed{
    \begin{aligned}
    \Energy_\mathrm{Coss}(\defmap,\Coss) = 
    \dfrac{1}{2}\int_{\Vol} \underbrace{\norm{\sym(\Stretch - \one)}^2_{\Cm} + 2\muc \norm{\skw\Stretch}^2}_{2\Psi_\mathrm{mp}}  + \underbrace{\mu_\mathrm{e} \Lc^2 \norm{\Curva}_{\Lm}^2}_{2\Psi_\mathrm{curv}} \, \dd \Vol \, ,
\end{aligned}
    }
\end{align}
and the external work by 
\begin{align}
    \boxed{
    \begin{aligned}
    \Work_\mathrm{Coss}(\defmap,\Coss) = \int_\Vol \con{\defmap}{\vb{f}} + \con{\Coss}{\Dforce} \, \dd \Vol + \int_{\Surf_N^{\defmap}} \con{\defmap}{\vb{t}} \, \dd \Surf + \int_{\Surf_N^{\Coss}} \con{\Coss}{\Dtrac} \, \dd \Surf \, . 
\end{aligned}
    }
\end{align}
Combined, they define the minimisation problem 
\begin{align}
    \boxed{
    \begin{aligned}
    \{\widehat{\defmap}, \widehat{\Coss} \}  &=  \arg \min_{\{\defmap,\Coss\}}  \, [  \Energy_\mathrm{Coss}(\defmap,\Coss) - \Work_\mathrm{Coss}(\defmap,\Coss) ]  \, .  
\end{aligned}
    }
\end{align}

\subsection{Variational formulation}

In order to retrieve a weak form from the action-functional we take variations with respect to the product-variable $\{\defmap,\Coss\}$. The virtual Biot-type stretch reads
\begin{align}
    \delta \Stretch = \delta \Coss^T \defgrad + \Coss^T \delta \defgrad = \Coss^T(\Coss \delta \Coss^T \defgrad + \Grad \delta \defmap) = \Coss^T(\Grad \delta \defmap - \delta \rot \times \defgrad) \, ,
\end{align}
where $\delta \defmap$ and $\delta \rot$ are admissible virtual displacements and rotations
\begin{align}
    &\delta \defmap = \delta (\vb{x} + \disp) = \delta \disp  \, , && \delta \rot  = -\axl(\Coss \delta \Coss^T) = \axl(\delta \Coss \, \Coss^T) \, .
\end{align}
Analogously, the virtual curvature is given by
\begin{align}
    \delta \Curva = \delta \Coss^T \Curl \Coss + \Coss^T \Curl \delta \Coss 
    &= \Coss^T[\Coss \delta \Coss^T \Curl \Coss + \Curl (\delta \Coss \, \Coss^T \Coss)] \notag \\
    &= \Coss^T[ \Curl(\delta \rot \times \Coss) - \delta \rot \times \Curl \Coss ]   \, .
\end{align}
Now, starting with the micropolar energy density we find
\begin{align}
    \delta \Psi_\mathrm{mp}(\Stretch) = \con{\delta \Stretch }{ \Grad_{\Stretch} \Psi_\mathrm{mp} } = \con{\delta \Stretch}{\RPiola} \, , 
\end{align}
where $\RPiola$ is a relative \textbf{non-symmetric} Biot-type stress tensor
\begin{align}
    \RPiola = \Grad_{\Stretch} \Psi_\mathrm{mp}(\Stretch) = \Cm \sym(\Stretch - \one) + 2\muc \skw\Stretch \, . 
\end{align}
The variation of the curvature energy yields
\begin{align}
    \delta \Psi_\mathrm{curv}(\Curva) = \con{\delta \Curva}{\Grad_{\Curva}\Psi_\mathrm{curv}} &= \con{\delta \Curva}{\RDouble} \, ,
\end{align}
where $\RDouble$ is a relative Piola-type double-stress tensor
\begin{align}
    \RDouble = \Grad_{\Curva}\Psi_\mathrm{curv}(\Curva) = \mu_\mathrm{e} \Lc^2 \Lm\Curva \, .
\end{align}
Consequently, the variational form reads
\begin{align}
    \int_\Vol \con{\delta \Stretch}{\RPiola} + \con{\delta \Curva}{\RDouble} \, \dd \Vol = \int_\Vol \con{\delta\defmap}{\vb{f}} + \con{\Anti\delta\rot}{\Dforce\Coss^T} \, \dd \Vol + \int_{\Surf_N^{\defmap}} \con{\delta\defmap}{\vb{t}} \, \dd \Surf + \int_{\Surf_N^{\Coss}} \con{\Anti\delta\rot}{\bm{T}\Coss^T} \, \dd \Surf \, .
\end{align}

\subsection{Boundary value problem}
We retrieve the boundary value problem using integration by parts of the internal virtual work.
Starting with $\delta\Stretch$ we find
\begin{align}
    \int_\Vol \con{\delta \Stretch}{\RPiola} \, \dd\Vol &= \int_\Vol \con{\Coss^T(\Grad \delta \defmap - \delta \rot \times \defgrad)}{\RPiola} \, \dd\Vol = \int_\Vol \con{\Grad \delta \defmap - \delta \rot \times \defgrad}{\Coss \, \RPiola} \, \dd\Vol
    \notag \\
    &= -\int_\Vol  \con{\delta \defmap}{\Di(\Coss \, \RPiola)} + 2\con{\delta \rot  }{\axl(\Coss \, \RPiola\defgrad^T} \, \dd\Vol  + \int_{\Surf_N^{\defmap}} \con{\delta \defmap}{(\Coss \, \RPiola)\vb{n}} \, \dd \Surf \, .
\end{align}
Let $\Piola$ be the non-rotated Piola stress tensor 
\begin{align}
    \RPiola = \Coss^T \Piola \quad \Rightarrow \quad \Piola = \Coss \, \RPiola \, ,  
\end{align}
we substitute to retrieve
\begin{align}
    \int_\Vol \con{\delta \Stretch}{\RPiola} \, \dd\Vol = -\int_\Vol  \con{\delta \defmap}{\Di\Piola} + 2\con{\delta \rot  }{\axl(\Piola\defgrad^T)} \, \dd\Vol  + \int_{\Surf_N^{\defmap}} \con{\delta \defmap}{\Piola\vb{n}} \, \dd \Surf \, .
\end{align}
Analogously, for $\delta \Curva$ we find
\begin{align}
    \int_\Vol \con{\delta \Curva}{\RDouble} \, \dd \Vol &= \int_\Vol \con{\Coss^T[ \Curl(\delta \rot \times \Coss) - \delta \rot \times \Curl \Coss ]}{\RDouble} \, \dd \Vol 
    = \int_\Vol \con{ \Curl(\delta \rot \times \Coss) - \delta \rot \times \Curl \Coss }{\Coss\RDouble} \, \dd \Vol
    \notag \\
    &= \int_\Vol \con{ \delta \rot \times \Coss}{\Curl(\Coss\RDouble)} - \con{ \delta \rot \times \Curl \Coss }{\Coss\RDouble} \, \dd \Vol - \int_{\Surf_N^{\Coss}} \con{\delta \rot \times \Coss}{(\Coss\RDouble)(\Anti \vb{n})} \, \dd \Surf
    \notag \\
    &= 2\int_\Vol \con{ \delta \rot}{\axl[\Curl(\Coss\RDouble)\Coss^T - \Coss\RDouble(\Curl \Coss)^T]} \, \dd \Vol - 2\int_{\Surf_N^{\Coss}} \con{\delta \rot}{\Coss\axl[\RDouble(\Anti \vb{n})]} \, \dd \Surf \, .
\end{align}
Consequently, the boundary value problem in dislocation form reads
\begin{align}
    \begin{aligned}
        -\Di \Piola &= \vb{f} && \text{in} && \Vol \, , 
        \\
        -\axl(\Piola \defgrad^T) + \axl[\Curl(\Coss\RDouble)\Coss^T - \Coss\RDouble(\Curl \Coss)^T]  &= \axl(\Dforce\Coss^T)  && \text{in} && \Vol \, , 
        \\
        \Piola \vb{n} &= \vb{t} && \text{on} && \Surf_N^{\defmap} \, , 
        \\
        -\Coss\axl[\RDouble(\Anti \vb{n})] &= \axl(\Dtrac \, \Coss^T) && \text{on} && \Surf_N^{\Coss} \, ,
        \\
        \defmap &= \widehat{\defmap} && \text{on} && \Surf_D^{\defmap} \, ,
        \\
        \Coss &= \widehat{\Coss} && \text{on} && \Surf_D^{\Coss} \, ,
    \end{aligned}
\end{align}
where it is to noted that the Dirichlet boundary of $\Coss$ is permitted to vanish  $|\Surf_D^{\Coss}| = 0$.
Another formulation can be derived by using the alternative form of $\delta \Curva$ from \cref{ap:variation}
\begin{align}
    \delta \Curva = \tr[(\Grad \delta \rot)^T \Coss] \one - (\Grad \delta \rot)^T \Coss \, .
\end{align}
By definition, the Piola double-stress tensor reads
\begin{align}
    \Double = \tr (\RDouble^T)\Coss - \Coss \RDouble^T  \quad \Rightarrow \quad \RDouble = \dfrac{1}{2}\tr(\Double^T\Coss)\one - \Double^T\Coss \, . 
\end{align}
Thus, the internal virtual work of the curvature reads
\begin{align}
    &\int_\Vol \con{\tr[(\Grad \delta \rot)^T \Coss] \one - (\Grad \delta \rot)^T \Coss}{(1/2)\tr(\Double^T\Coss)\one - \Double^T\Coss} \, \dd \Vol = 
    \notag \\
    &\int_\Vol \dfrac{3}{2} \tr[(\Grad \delta \rot)^T \Coss]\tr(\Double^T\Coss) - \tr[(\Grad \delta \rot)^T \Coss]\tr(\Double^T\Coss) - \dfrac{1}{2}\tr[(\Grad \delta \rot)^T \Coss]\tr(\Double^T\Coss) + \con{(\Grad \delta \rot)^T \Coss}{\Double^T\Coss}  \, \dd \Vol \, , 
\end{align}
which simplifies to
\begin{align}
    \int_\Vol \con{\delta \Curva}{\RDouble} \, \dd \Vol = \int_\Vol \con{(\Grad \delta \rot)^T \Coss}{\Double^T\Coss} \, \dd \Vol = \int_\Vol \con{\Grad \delta \rot }{\Double} \, \dd \Vol = -\int_\Vol \con{\delta \rot}{\Di\Double} \, \dd \Vol + \int_{\Surf_N^{\Coss}} \con{\delta \rot}{\Double \vb{n}} \, \dd \Surf \, .
\end{align}
Consequently, the boundary value problem can also be written as
\begin{align}
    \boxed{
\begin{aligned}
        -\Di \Piola &= \vb{f} && \text{in} && \Vol \, , 
        \\
        -\Di \Double -2\axl(\Piola \defgrad^T)  &= 2\axl(\Dforce\Coss^T)  && \text{in} && \Vol \, , 
        \\
        \Piola \vb{n} &= \vb{t} && \text{on} && \Surf_N^{\defmap} \, , 
        \\
        \Double \vb{n} &= 2\axl(\Dtrac \, \Coss^T) && \text{on} && \Surf_N^{\Coss} \, ,
        \\
        \defmap &= \widehat{\defmap} && \text{on} && \Surf_D^{\defmap} \, ,
        \\
        \Coss &= \widehat{\Coss} && \text{on} && \Surf_D^{\Coss} \, ,
\end{aligned}
}
\end{align}
for which $|\Surf_D^{\Coss}| = 0$ is also permissible.
The latter is the classical Eringen form of the Cosserat micropolar model \cite{Eremeyev2013} that usually adopts the wryness tensor as its curvature measure 
\begin{align}
    \Wry = \dfrac{1}{2}\tr(\Curva^T)\one - \Curva^T \, .
\end{align}
The main difference between the Cosserat micropolar theory and classical Cauchy finite elasticity can be made apparent using the second differential equation, being the balance of angular momentum. Namely, with the transformation of the first Piola stress tensor \cite{Eremeyev2013,Altenbach2018}, it becomes clear that the Cauchy stress can entail skew-symmetric components to equilibrate couple-forces
\begin{align}
    \Piola = (\det \defgrad) \bm{\sigma}\defgrad^{-T} \quad \Rightarrow \quad -\Di \Double -2(\det \defgrad)(\axl\bm{\sigma}) = 2\axl(\Dforce\Coss^T) \, .
\end{align}

\section{The flexomagnetic postulate}

\AS{In the following we introduce derivation of the flexomagnetic model. The derivation proceeds in four stages. First, the flexomagnetic coupling postulate is introduced in terms of the micro-dislocation tensor and magnetisation. Second, magnetostatic energies are incorporated and reduced to a single magnetic variable using partial Legendre transforms. Third, scalar- and vector-potential formulations are derived. Finally, the corresponding variational forms and boundary value problems are obtained.}

With the Cosserat model at hand, we proceed to contrive a novel phenomenological flexomagnetic model. Let the orientation of a local magnetic dipole be given by a director attached to each material point, then its reorientation is governed by the Cosserat rotation tensor $\Coss$.  At the same time, an arbitrary reorientation of neighbouring dipoles under deformation in a specimen may prove negligible, depending on the characteristic length of the region $\Lc$. Further, simply coupling $\Coss$ with the magnetisation of the medium is generally non-objective, as it implies that rigid body motions in a quasi-static context may induce a magnetisation. Conversely, if the curvature energy of the micro-dislocation governed by $\Psi_\mathrm{curv}(\Curva) = \mu_\mathrm{e}(\Lc^2/2)\norm{\Curva}_{\Lm}^2$ is high, then $\Coss$ changes rapidly within a non-negligible region of the specimen, implying that \textbf{many} magnetic dipoles are changing their orientation \textbf{non-uniformly}, as illustrated in \cref{fig:mag}.
In other words, a magnetic polarisation (magnetisation) is induced.
\begin{figure}
    \centering
    \input{figs/mag}
    \caption{A net-magnetisation induced by a non-uniform change in the orientation of magnetic dipoles over a non-negligible region of the domain. The change in orientation is governed by $\Coss:\Vol \to \SO(3)$, while its non-uniformity and impact with respect to the size of the domain is indicated by the curvature energy $\Psi_\mathrm{curv}(\Curva)$.}
    \label{fig:mag}
\end{figure}
Thus, we postulate that the change in micro-dislocation governs the interaction with the magnetisation of the medium, and therefore with the magnetic field.   
The corresponding coupling energy is defined as
\begin{align}
    \Psi_\mathrm{fxm} = \Psi_\mathrm{fxm} (\Curva,\mmag) \, ,  && \mmag:\Vol \to \R^3 \, , 
\end{align}
where $\mmag$ is the magnetisation in the reference configuration. 
Notably, by postulating the coupling directly in the reference configuration it is inherently objective.
Kinematically, the proposed coupling postulate has two significant implications. Firstly, non-uniform extension on its own cannot produce a magnetic reaction. This is in clear contrast to flexoelectricity \cite{CODONY2021104182}, and is motivated by fact that magnetic dipoles, unlike electric dipoles, are not composed of two opposite charges at a distance, and are thus presumed non-deformable. 
Secondly, by the fact that the orientation of each material point given by $\Coss$ is independent of the tangent curves given by $\defgrad$, cf. \cref{fig:mp}, it is truly the individual rotation of each point that contributes to the net magnetic reaction. By construction, the extent to which this rotation manifests as actual deformation is determined by the material parameters of the Cosserat model. Now, to make the coupling term explicit, we require a constitutive relation. Let the coupling be physically linear, then there exists a third-order tensor of flexomagnetic coefficients $\R^{3 \times 3 \times 3}\ni\Mm:\R^{3 \times 3} \to \R^3$ such that the flexomagnetic energy density can be defined as the Lifshitz invariant 
\begin{align}
    &\Psi_\mathrm{fxm}(\Curva,\mmag) = -\mu_0\con{\mmag}{\Mm \Curva}  \, ,
\end{align}
where $\mu_0$ is the permeability of free space. \AS{As a reciprocal term in a monolithic model, it encompasses both direct and converse flexomagnetic effects while avoiding an explicit $\Grad \mmag$ contribution, which would introduce additional magnetisation-gradient boundary conditions associated with surface torques or spin-pinning and require separate assumptions on the surface and interface physics \cite{Rohart}.}
The structure of $\Mm$ can be given explicitly for isotropic centrosymmetric materials via
\begin{align}
    \Mm_\mathrm{iso} = \eta_\mathrm{iso}\perm = \eta_\mathrm{iso} \varepsilon_{ijk} \elec_i \otimes \elec_j \otimes \elec_k \, , && \varepsilon_{ijk} = \left \{ \begin{matrix}
        1 & \text{for even permutations}  \\
        -1 & \text{for odd permutations} \\
        0 & \text{otherwise}
    \end{matrix} \right . \, ,
\end{align}
where $\varepsilon_{ijk}$ is the Levi-Civita permutation symbol, $\perm \in \R^{3\times3\times3}$ is the corresponding permutation tensor, and $\eta_\mathrm{iso} \in \R$ is the isotropic flexomagnetic coefficient. Notably, the permutation tensor is the only centrosymmetric third-order (pseudo-) tensor. 
Since $\Curva$ is a second-order tensor this might seem unphysical at first, as this type of pairing is impossible in piezomagnetism due to inversion symmetry of the crystalline structure of the material. Namely, for piezomagnetism a contradiction arises in the analogous definition with the strain tensor $\con{\mmag}{\perm (\Stretch - \one)} = \con{\mmag}{\perm \Stretch} \mapsto \con{-\one\mmag}{\perm (-\one)\Stretch(-\one)} = \con{-\mmag}{\perm\Stretch} \neq \con{\mmag}{\perm\Stretch}$ since $\Stretch$ does not flip its sign under the inversion $-\one \in \mathrm{O}(3)$, while $\mmag$ does. The only constant centrosymmetric third-order tensor that can accommodate this behaviour is the zero tensor, making centrosymmetric piezomagnetism impossible. Notwithstanding, the more general rule is that \textit{a constitutive tensor is allowed to be non-zero only if its parity matches the parities of the fields it connects to under symmetry operations}. In the case of inversion symmetry this implies that if $\mmag$ is odd (flips sign under inversion) and $\perm$ is even (the Levi-Civita tensor does not flip its sign), then $\Curva$ must be odd to accommodate the sign change. And indeed, the micro-dislocation tensor is odd, as it represents an algebraic tensor-order reduction of the odd third-order tensor $\Coss^T \Grad \Coss \mapsto (-\one)^3\Coss^T \Grad \Coss = - \Coss \Grad \Coss$ without loss of information, cf. \cref{ap:curl}.
Next, for more general non-centrosymmetric yet cubic-symmetric crystals, the third-order tensor can also take the form \cite{Powell2010}
\begin{align}
    \Mm_\mathrm{cub} = \eta_\mathrm{cub} |\varepsilon_{ijk}| \elec_i \otimes \elec_j \otimes \elec_k \, ,
\end{align}
where $\eta_\mathrm{cub} \in \R$ is the corresponding flexomagnetic coefficient. Notably, $|\varepsilon_{ijk}| \elec_i \otimes \elec_j \otimes \elec_k$ is a polar cubic-symmetric tensor that flips its sign under inversion.
Putting it all together, physically linear cubic-symmetric materials can have at most two independent material parameters.
Interestingly, these two third-order tensors split $\Curva$ across skew-symmetric and symmetric contributions
\begin{align}
    & \Mm \Curva =  (\Mm_\mathrm{iso} + \Mm_\mathrm{cub})\Curva = \Mm_\mathrm{iso} \skw\Curva + \Mm_\mathrm{cub}\sym\Curva \, ,
\end{align}
while omitting its diagonal components, as shown in \cref{ap:fourth}. 
As a curiosity, because the diagonal part of the curvature tensor plays no role in the energy of cubic-symmetric materials, one can also express it as
\begin{align}
    \boxed{\Mm \Curva = -\Mm \Wry^T} \quad \Rightarrow \quad -\mu_0\con{\mmag}{\Mm \Curva} = \mu_0\con{\mmag}{\Mm \Wry^T} = \Psi_\mathrm{fxm}(\mmag,\Wry) \, ,
\end{align}
such that the formulation naturally caries over also to the Eringen form of the Cosserat micropolar model.
Now, let $\mmag_\varphi:\Vol_\varphi \to \R^3$ be the magnetisation field in the current configuration, for crystals with cubic symmetry the so called demagnetisation energy density of a solid medium at a constant temperature, also known as the magnetostatic energy density, is given by (recall, densities are 3-forms s.t. $\Psi \, \dd \Vol_\varphi = \Psi (\det \defgrad)\, \dd \Vol$)
\begin{align}
    &\Psi_\mathrm{demag}(\mmag_\varphi) = \dfrac{\mu_0\det\defgrad}{2} \con{\mmag_\varphi}{ \chi_\mathrm{m}^{-1}  \one\mmag_\varphi} = \dfrac{\mu_0\det\defgrad}{2\chi_\mathrm{m}} \norm{\mmag_\varphi}^2 \, , && \chi_\mathrm{m} = \mu_\mathrm{r} - 1 =  \dfrac{\mu}{\mu_0} - 1 \, ,
\end{align}
since the only second-order tensor with cubic-symmetry is the isotropic identity tensor $\one \in \R^{3 \times 3}$ \cite{Powell2010}. 
The additional material constants are known as the relative permeability $\mu_r$, and the magnetic susceptibility $\chi_\mathrm{m}$, otherwise expressed via the total permeability of the material $\mu$.  
In this work, we define the magnetisation field to act on infinitesimal surfaces $\dd \vb{A}_\varphi$ (i.e. it is a 2-form), implying that it transforms by the contravariant Piola transformation
\begin{align}
    \mmag_\varphi = \dfrac{1}{\det \defgrad} \defgrad \mmag \, , 
\end{align}
While this may appear ambiguous due to its additive nature with the magnetic field $\hmag_\varphi$ which transforms differently, this transformation rule is intrinsically needed for coherence with subsequent Legendre transforms, and is justified by the requirements of frame-invariance \cite{bresciani2023existence}, thermodynamic consistency \cite{Roubíček2018}, and the force-interpretation of the magnetostatic Maxwell stress tensor and its associated traction vector \cite{DorfmannMW}.
Now, by inserting the transformation, the demagnetisation energy expressed via the magnetisation of the reference configuration $\mmag$ reads
\begin{align}
    \Psi_\mathrm{demag}(\defgrad,\mmag) = \dfrac{\mu_0\det \defgrad}{2\chi_\mathrm{m}} \norm{ (\det\defgrad)^{-1} \defgrad \mmag}^2 = \dfrac{\mu_0}{2\chi_\mathrm{m}\det\defgrad} \norm{\defgrad \mmag}^2 = \dfrac{\mu_0}{2\chi_\mathrm{m}\det \defgrad} \con{\defgrad^{T}\defgrad}{ \mmag \otimes \mmag} \, ,
\end{align}
where $\defgrad^{T} \defgrad = \bm{C}$ is the Cauchy--Green strain tensor.
Since $\defgrad^T\defgrad = \Stretch^T\Coss^T \Coss \, \Stretch = \Stretch^T \Stretch$ and $\det\defgrad = \det(\Coss \, \Stretch) = (\det\Coss)(\det \Stretch) = \det \Stretch$, we can also express the energy density using the Biot-type stretch tensor of the Cosserat micropolar model
\begin{align}
    \Psi_\mathrm{demag}(\Stretch,\mmag) = \dfrac{\mu_0}{2\chi_\mathrm{m}\det \Stretch} \con{\Stretch^{T}\Stretch}{ \mmag \otimes \mmag} = \dfrac{\mu_0}{2\chi_\mathrm{m}\det \Stretch} \norm{\Stretch\mmag}^2 \, .
\end{align}
The reason to introduce this seemingly unnecessary adjustment is in order to allow for the  subsequent derivation of stress tensors that are consistent with the Cosserat model.
Next, the magnetic enthalpy of free space induced by the magnetic field of the current configuration $\hmag_\varphi$ is
\begin{align}
    \Psi_\mathrm{mag}(\defgrad,\hmag_\varphi) = \dfrac{\mu_0\det\defgrad}{2} \norm{\hmag_\varphi}^2 \, .
\end{align}
By its physical nature, the magnetic field acts on infinitesimal curves $\dd \vb{\curv}_\varphi$ (i.e., it is a 1-form) and thus transforms via the covariant Piola transformation
\begin{align}
    \hmag_\varphi = \bm{F}^{-T} \hmag \, .
\end{align}
Hence, in the reference configuration the magnetic enthalpy is given by
\begin{align}
    \Psi_\mathrm{mag}(\Stretch,\hmag) = \dfrac{\mu_0 \det \Stretch}{2} \norm{\Stretch^{-T}\hmag}^2 = \dfrac{\mu_0 \det \Stretch}{2} \con{\Stretch^{-1}\Stretch^{-T}}{ \hmag \otimes \hmag}  \, ,
\end{align}
where we again exploited the relations of the deformation gradient $\defgrad$ and the Biot-type stretch tensor $\Stretch$.
Now, with the magnetostatic energies and enthalpy at hand, the complete set of energies and enthalpies of the coupled system is given by the sum
\begin{align}
   \Psi_\mathrm{mp}(\Stretch) + \Psi_\mathrm{curv}(\Curva) + \Psi_\mathrm{demag}(\Stretch,\mmag) + \Psi_\mathrm{fxm}(\Curva,\mmag) - \Psi_\mathrm{mag}(\Stretch,\hmag) \, .
\end{align}
Evidently, the latter expression entails two magnetostatic variables. Namely, while the magnetic enthalpy is expressed in terms of the magnetic field $\hmag$, the flexomagnetic coupling and the demagnetisation energy are functions of the magnetisation field $\mmag$. 
In order to achieve a coupled system that is expressed in a single magnetostatic variable, we now employ the partial Legendre transform
\begin{align}
    \inf_{\mmag}[\Psi_\mathrm{mp}(\Stretch) + \Psi_\mathrm{curv}(\Curva) + \Psi_\mathrm{demag}(\Stretch,\mmag) + \Psi_\mathrm{fxm}(\Curva,\mmag) - \Psi_\mathrm{mag}(\Stretch,\hmag) - \mu_0\con{\mmag}{\hmag}] \, . 
\end{align}
This approach is justified as a way to recover the complete magnetic enthalpy, cf. \cref{ap:electro}, and by the physical interpretation of these enthalpies as generating forces that act on the medium \cite{DorfmannMW}.
By the transform, the corresponding magnetic field is given by the stationary point
\begin{align}
    \Grad_\mmag [\Psi_\mathrm{demag}(\Stretch,\mmag) +  \Psi_\mathrm{fxm}(\Curva,\mmag) - \mu_0\con{\mmag}{\hmag} ] = 0 \quad \Rightarrow \quad \hmag = \dfrac{1}{\mu_0}\Grad_\mmag [ \Psi_\mathrm{demag}(\Stretch,\mmag) + \Psi_\mathrm{fxm}(\Curva,\mmag) ] \, . 
\end{align}
The derivative of the demagnetisation energy reads
\begin{align}
    \Grad_\mmag \Psi_\mathrm{demag}(\Stretch,\mmag) = \dfrac{\mu_0}{\chi_\mathrm{m}\det \Stretch} \Stretch^{T} \Stretch \mmag \, , 
\end{align}
and the derivative of the flexomagnetic energy is given by
\begin{align}
    \Grad_\mmag \Psi_\mathrm{fxm}(\Curva,\mmag) = -\mu_0\Mm \Curva \, .
\end{align}
Thus, we find an invertible expression for the magnetic field
\begin{align}
    \hmag = \dfrac{1}{\chi_\mathrm{m}\det \Stretch} \Stretch^{T} \Stretch \mmag -\Mm \Curva \quad \Rightarrow \quad \mmag = \chi_\mathrm{m}(\det \Stretch) \Stretch^{-1} \Stretch^{-T} (\hmag + \Mm \Curva) \, ,
\end{align}
which simply implies that the magnetisation is the additive product of two contributions, being the susceptibility-dependent magnetisation and the curvature-induced magnetisation. 
Now, inserting this identity in $\Psi_\mathrm{demag}$ yields
\begin{align}
    \Psi_\mathrm{demag}(\Stretch,\mmag)\at_{\mmag = \mmag(\Stretch,\Curva,\hmag)} = \dfrac{\chi_\mathrm{m}\mu_0\det \Stretch}{2}\norm{\Stretch^{-T} (\hmag + \Mm \Curva)}^2 \, .
\end{align}
Analogously, $\Psi_\mathrm{fxm}$ transforms to
\begin{align}
    \Psi_\mathrm{fxm}(\Curva,\mmag)\at_{\mmag = \mmag(\Stretch,\Curva,\hmag)} &= -\chi_\mathrm{m}\mu_0(\det \Stretch) \con{\Stretch^{-1} \Stretch^{-T}(\hmag + \Mm \Curva)}{\Mm \Curva} 
    \notag \\
    &= -\chi_\mathrm{m}\mu_0(\det \Stretch) \con{ \Stretch^{-T}(\hmag + \Mm \Curva)}{\Stretch^{-T}(\Mm \Curva)} \, ,  
\end{align}
and finally, $\con{\mmag}{\hmag}$ reads
\begin{align}
    \mu_0\con{\mmag}{\hmag}\at_{\mmag = \mmag(\Stretch,\Curva,\hmag)} = \chi_\mathrm{m}\mu_0(\det \Stretch)  \con{\Stretch^{-1} \Stretch^{-T}(\hmag + \Mm \Curva)}{ \hmag } = \chi_\mathrm{m}\mu_0(\det \Stretch)  \con{\Stretch^{-T}(\hmag + \Mm \Curva)}{ \Stretch^{-T}\hmag } \, .
\end{align}
As such, the full sum is given by 
\begin{align}
    [\Psi_\mathrm{demag}(\Stretch,\mmag) + \Psi_\mathrm{fxm}(\Curva,\mmag) - \mu_0\con{\mmag}{\hmag}]\at_{\mmag = \mmag(\Stretch,\Curva,\hmag)} = -\dfrac{\chi_\mathrm{m}\mu_0\det \Stretch}{2} \norm{\Stretch^{-T}(\hmag + \Mm \Curva)}^2 \, . 
\end{align}
The resulting term is the combined flexomagnetic and demagnetisation enthalpies expressed by the magnetic field
\begin{align}
    \Psi_\mathrm{fmd}(\Stretch,\Curva,\hmag) = \dfrac{\chi_\mathrm{m}\mu_0\det \Stretch}{2} \norm{\Stretch^{-T}(\hmag + \Mm \Curva)}^2 \, .
\end{align}
With the all the enthalpies now governed by the magnetic field, we can finally formulate the internal functional of mechanical energies and magnetic enthalpies 
\begin{align}
    \boxed{
    \begin{aligned}
    \Energy_\mathrm{fxm}(\defmap,\Coss,\hmag) =\dfrac{1}{2}\int_\Vol & \overbrace{\norm{\sym(\Stretch - \one)}^2_{\Cm} + 2\muc \norm{\skw\Stretch}^2}^{2\Psi_\mathrm{mp}}  + \overbrace{\mu_\mathrm{e} \Lc^2 \norm{\Curva}_{\Lm}^2}^{2\Psi_\mathrm{curv}} 
    \\
    & -[\underbrace{\chi_\mathrm{m}\mu_0(\det \Stretch) \norm{\Stretch^{-T}(\hmag + \Mm \Curva)}^2}_{2\Psi_\mathrm{fmd}}  + \underbrace{\mu_0 (\det \Stretch) \norm{\Stretch^{-T}\hmag}^2}_{2\Psi_\mathrm{mag}}]\, \dd \Vol \, .
\end{aligned}
    } \label{eq:flexoh}
\end{align}
Let there be no impressed currents in the medium $\vb{j} = 0$ or on its surface, and the external work be purely mechanical, then the coupled physics yield the saddle-point problem
\begin{align}
    \boxed{
    \{\widehat{\defmap}, \widehat{\Coss} \}, \widehat{\hmag}  =  \arg \min_{\{\defmap,\Coss\}} \max_{\hmag} \, [  \Energy_\mathrm{fxm}(\defmap,\Coss,\hmag) - \Work_\mathrm{Coss}(\defmap,\Coss) ]  \, ,  
    }
\end{align}
where $\{\widehat{\defmap}, \widehat{\Coss} \}, \widehat{\hmag}$ are the minimisers and maximiser, respectively. \AS{
The formulation inherits the variational structure of finite-strain Cosserat elasticity and magnetostatics, supporting the existence of minimisers and maximisers. A rigorous analysis of uniqueness and stability is left for future work.
}

\subsection{Magnetic scalar potential}
For vanishing impressed currents $\vb{j} = 0$, the Amp\'ere--Maxwell law of magnetostatics \cite{Zaglmayr2006} implies 
\begin{align}
    \curl_\varphi \hmag_{\varphi} = \curl_\varphi(\defgrad^{-T}\hmag) = \dfrac{1}{\det \defgrad}\defgrad (\curl \hmag) = 0 \, , 
\end{align}
where we exploited the Piola transformations of the curl operator, cf. \cite{SKY2025322}. Consequently, on a contractible domain, $\hmag$ can be expressed with a scalar magnetic potential
\begin{align}
    \boxed{
    \begin{aligned}
        \hmag = - \nabla \psi \, ,  
    \end{aligned}
    }
    && \psi : \Vol \to \R \, .
\end{align}
Correspondingly, the magnetic enthalpies are given by
\begin{align}
    \Psi_\mathrm{fmd}(\Stretch,\Curva,\psi) = \dfrac{\chi_\mathrm{m}\mu_0\det \Stretch}{2} \norm{\Stretch^{-T}(\Mm \Curva - \nabla\psi )}^2 \, ,
\end{align}
and 
\begin{align}
    \Psi_\mathrm{mag}(\Stretch,\psi) = \dfrac{\mu_0 \det \Stretch}{2} \norm{\Stretch^{-T}\nabla\psi}^2 \, ,
\end{align}
such that the enthalpy functional reads
\begin{align}
    \boxed{
    \begin{aligned}
    \Energy_\mathrm{fxm}(\defmap,\Coss,\psi) = \dfrac{1}{2}\int_\Vol & \overbrace{\norm{\sym(\Stretch - \one)}^2_{\Cm} + 2\muc \norm{\skw\Stretch}^2}^{2\Psi_\mathrm{mp}}  + \overbrace{\mu_\mathrm{e} \Lc^2 \norm{\Curva}_{\Lm}^2}^{2\Psi_\mathrm{curv}}
    \\
    &- [ \underbrace{\chi_\mathrm{m}\mu_0(\det \Stretch)\norm{\Stretch^{-T}(\Mm \Curva - \nabla\psi )}^2}_{2\Psi_\mathrm{fmd}}  + \underbrace{\mu_0 (\det \Stretch) \norm{\Stretch^{-T}\nabla\psi}^2}_{2\Psi_\mathrm{mag}} ]\, \dd \Vol \, .
\end{aligned}
    }
\end{align}
In the case of non-vanishing impressed current $\vb{j} \neq 0$, the formulation becomes more involved. Its treatment is postponed to the next sections, pending further derivations.   
The variation of the magnetic enthalpies reads
\begin{align}
    \delta [\Psi_\mathrm{fmd} (\Stretch,\Curva,\hmag) + \Psi_\mathrm{mag} (\Stretch,\hmag)] = \con{\delta \Stretch}{\Grad_{\Stretch}(\Psi_\mathrm{fmd}+\Psi_\mathrm{mag})} + \con{\delta \Curva}{\Grad_{\Curva}\Psi_\mathrm{fmd}} + \con{\delta \hmag}{\Grad_{\hmag}(\Psi_\mathrm{fmd}+\Psi_\mathrm{mag})} \, ,
\end{align}
where $\delta \hmag$ is an admissible virtual magnetic field.
For the first term we find
\begin{align}
    \Grad_{\Stretch}[\Psi_\mathrm{fmd}(\Stretch,\Curva,\hmag)+\Psi_\mathrm{mag}(\Stretch,\hmag)] =   & \dfrac{\chi_\mathrm{m}\mu_0\det \Stretch}{2} \norm{\Stretch^{-T}(\hmag + \Mm \Curva)}^2 \Stretch^{-T}   \notag \\
    & - \chi_\mathrm{m}\mu_0 (\det \Stretch) [\Stretch^{-T}(\hmag + \Mm \Curva)] \otimes [\Stretch^{-T}(\hmag + \Mm \Curva)] \Stretch^{-T} \notag \\[1ex]
    & + \dfrac{\mu_0 \det \Stretch}{2} \norm{\Stretch^{-T}\hmag}^2 \Stretch^{-T}  \notag \\
    &   - \mu_0 (\det \Stretch) (\Stretch^{-T}\hmag)  \otimes (\Stretch^{-T}\hmag) \Stretch^{-T}   \, ,
\end{align}
for which the relevant mathematical identities in the derivation can be found in \cref{ap:tensorgrads}.
Expanding the norms while substituting $\chi_\mathrm{m}\mu_0 = \mu - \mu_0$, and rearranging the terms yields
\begin{align}
    & \dfrac{\mu \det \Stretch}{2}\norm{\Stretch^{-T}\hmag}^2\Stretch^{-T} - \mu(\det \Stretch) (\Stretch^{-T}\hmag) \otimes (\Stretch^{-T}\hmag) \Stretch^{-T} 
    \notag \\
    &+ \dfrac{(\mu - \mu_0)\det \Stretch}{2} \norm{\Stretch^{-T}(\Mm \Curva)}^2\Stretch^{-T} - (\mu - \mu_0)(\det \Stretch) [\Stretch^{-T}(\Mm \Curva)] \otimes [\Stretch^{-T}(\Mm \Curva)]\Stretch^{-T} 
     \\
    &+(\mu - \mu_0)(\det \Stretch) \con{\Stretch^{-T}\hmag}{\Stretch^{-T}(\Mm \Curva)} \Stretch^{-T} - 2(\mu - \mu_0)(\det \Stretch)\sym[(\Stretch^{-T}\hmag) \otimes \Stretch^{-T}(\Mm \Curva)]\Stretch^{-T}   \, , \notag
\end{align}
where we recognise the rotated Piola--Maxwell stress tensor of magnetostriction \cite{LIU2014451,DorfmannMW,Dorfmann2014}
\begin{align}
    \boxed{
    \begin{aligned}
        \RPiola_\mathrm{Mxl} = (\det \Stretch) \bar{\bm{\sigma}}_\mathrm{Mxl} \Stretch^{-T} =(\det \Stretch) \bigg [ \underbrace{\mu (\Stretch^{-T}\hmag) \otimes (\Stretch^{-T}\hmag)  - \dfrac{\mu}{2}\norm{\Stretch^{-T}\hmag}^2\one}_{\mu (\Coss^T\hmag_\varphi) \otimes (\Coss^T\hmag_\varphi)  - \dfrac{\mu}{2}\norm{\hmag_\varphi}^2\one} \bigg] \Stretch^{-T}  \, ,
    \end{aligned}
    }
\end{align}
and define the remaining parts as the rotated Piola--Maxwell flexomagnetic curvature stress tensor
\begin{align}
    \boxed{
    \begin{aligned}
        \RPiola_\mathrm{fmc} = (\det \Stretch)  \bar{\bm{\sigma}}_\mathrm{fmc} \Stretch^{-T} =& (\det \Stretch)  [\mu-\mu_0]\bigg [  [\Stretch^{-T}(\Mm \Curva)] \otimes [\Stretch^{-T}(\Mm \Curva)]  - \dfrac{\mu-\mu_0}{2}\norm{\Stretch^{-T}(\Mm \Curva)}^2\one \bigg] \Stretch^{-T} \, ,
    \end{aligned}
    }
\end{align}
and the rotated flexomagnetic curvature coupling stress tensor
\begin{align}
    \boxed{
    \begin{aligned}
        \RPiola_\mathrm{fmcc} = (\det \Stretch)  \bar{\bm{\sigma}}_\mathrm{fmcc} \Stretch^{-T} = (\det \Stretch) \bigg [ 2\chi_\mathrm{m}\mu_0\sym[(\Stretch^{-T}\hmag) \otimes \Stretch^{-T}(\Mm \Curva)] - \chi_\mathrm{m}\mu_0 \con{\Stretch^{-T}\hmag}{\Stretch^{-T}(\Mm \Curva)}  \bigg ]\Stretch^{-T} \, ,
    \end{aligned}
    }
\end{align}
such that
\begin{align}
    \Grad_{\Stretch}[\Psi_\mathrm{fmd}(\Stretch,\Curva,\hmag)+\Psi_\mathrm{mag}(\Stretch,\hmag)] =  -  \RPiola_\mathrm{Mxl} - \RPiola_\mathrm{fmc} - \RPiola_\mathrm{fmcc} \, .
\end{align}
Note that this equation justifies the negative sign of the magnetic enthalpies as it ensures that magnetically induced stresses enter the equation additively to the purely mechanical stresses.
The second term in the variation of the magnetic enthalpies reads
\begin{align}
    \boxed{
    \Grad_{\Curva} \Psi_\mathrm{fmd}(\Stretch,\Curva,\hmag) = -\RDouble_\mathrm{fxm} = (\mu - \mu_0) (\det \Stretch) [\Stretch^{-1}\Stretch^{-T}(\hmag + \Mm \Curva)] \Mm \, ,
    }
\end{align}
representing a rotated flexomagnetic Piola-type double-stress tensor.
Finally, the last term governed by the magnetic field $\hmag$ reads
\begin{align}
    \Grad_{\hmag} [\Psi_\mathrm{fmd}(\Stretch,\Curva,\hmag)+\Psi_\mathrm{mag}(\Stretch,\hmag)] &= (\mu - \mu_0) (\det \Stretch) \Stretch^{-1} \Stretch^{-T} (\hmag + \Mm \Curva) + \mu_0 (\det \Stretch) \Stretch^{-1} \Stretch^{-T} \hmag  \, , \notag \\
    & = (\mu - \mu_0) (\det \Stretch) \Stretch^{-1} \Stretch^{-T} (\Mm \Curva) + \mu (\det \Stretch) \Stretch^{-1} \Stretch^{-T} \hmag
\end{align}
where we recognise the \textbf{classical} magnetic induction \cite{Zaglmayr2006,LIU2014451,Dorfmann2014} in the reference configuration
\begin{align}
    \boxed{
    \begin{aligned}
        \vb{b} = \mu (\det \Stretch) \Stretch^{-1} \Stretch^{-T} \hmag = \mu (\det \defgrad) \defgrad^{-1} \defgrad^{-T} \hmag = \mu (\det \defgrad) \defgrad^{-1}  \hmag_{\varphi} = (\det \defgrad) \defgrad^{-1} \vb{b}_\varphi \, ,
    \end{aligned}
    }
    \label{eq:btoh}
\end{align}
and analogously define the (reference) magnetisation induced by flexure as
\begin{align}
    \boxed{
    \begin{aligned}
        \fmag = \chi_\mathrm{m} (\det \Stretch) \Stretch^{-1} \Stretch^{-T} (\Mm \Curva) = \chi_\mathrm{m} (\det \defgrad) \defgrad^{-1} \defgrad^{-T} (\Mm \Curva) \, .
    \end{aligned}
    }
\end{align}
Consequently, $\vb{b} + \mu_0\fmag$ can be understood as the total magnetic induction in the reference configuration
\begin{align}
\boxed{
\begin{aligned}
    \tbmag = \vb{b} + \mu_0 \fmag =  (\det \defgrad) \defgrad^{-1} \defgrad^{-T} (\mu\hmag + \chi_\mathrm{m}\mu_0  \Mm \Curva) = (\det \defgrad) \defgrad^{-1}  \underbrace{\mu_0 \defgrad^{-T} (\hmag + \chi_\mathrm{m}\hmag + \chi_\mathrm{m} \Mm \Curva)}_{\tbmag_\varphi} \, .  
\end{aligned}
}
\end{align}
The additional virtual work performed by $\delta \hmag = -\nabla\delta\psi$ now reads
\begin{align}
    -\int_\Vol \con{\delta \hmag}{\underbrace{\bmag + \mu_0 \fmag}_{\tbmag}} \, \dd \Vol = \int_\Vol \con{\nabla \delta \psi}{\tbmag} \, \dd \Vol = -\int_\Vol \con{\delta \psi}{\di \tbmag} \, \dd \Vol + \int_{\Surf_N^{\psi}} \delta \psi \underbrace{\con{\tbmag}{\vb{n}}}_{\mathfrak{b}_\mathrm{n}} \, \dd \Surf \, ,
\end{align}
where we integrated by parts.
The latter immediately implies the existence of a prescribed normal magnetic induction on the Neumann boundary, which can be accommodated with additional magnetostatic external work
\begin{align}
    \boxed{
    \Work_\mathrm{ms}(\psi) = \int_{\Surf_N^{\psi}} \con{\psi}{\mathfrak{b}_\mathrm{n}} \, \dd \Surf \, ,
    }
\end{align}
such that the full saddle-point problem reads
\begin{align}
    \boxed{
    \begin{aligned}
    \{\widehat{\defmap}, \widehat{\Coss} \}, \widehat{\psi}  &=  \arg \min_{\{\defmap,\Coss\}} \max_{\psi} \, [  \Energy_\mathrm{fxm}(\defmap,\Coss,\psi) - \Work_\mathrm{Coss}(\defmap,\Coss) + \Work_\mathrm{ms}(\psi) ]  \, ,
\end{aligned}
    }
\end{align}
and the previous variation changes into
\begin{align}
     \int_\Vol \con{\nabla \delta \psi}{\tbmag} \, \dd \Vol - \int_{\Surf_N^{\psi}} \delta \psi \, \mathfrak{b}_\mathrm{n} \, \dd \Surf  = -\int_\Vol \con{\delta \psi}{\di \tbmag} \, \dd \Vol + \int_{\Surf_N^{\psi}} \delta \psi (\con{\tbmag}{\vb{n}} - \mathfrak{b}_\mathrm{n}) \, \dd \Surf \, . 
\end{align}
Consequently, the complete variational functional reads
\begin{align}
    &\int_\Vol \con{\delta \Stretch}{\RPiola + \RPiola_\mathrm{Mxl} + \RPiola_\mathrm{fmc} + \RPiola_\mathrm{fmcc}} + \con{\delta \Curva}{\RDouble + \RDouble_\mathrm{fxm}} + \con{\nabla\delta \psi}{\vb{b} + \mu_0\fmag} \, \dd \Vol 
    \\
    &\qquad \qquad = \int_\Vol \con{\delta \defmap}{\vb{f}} + \con{\Anti\delta \bm{\omega}}{\Dforce \Coss^T} \, \dd \Vol + \int_{\Surf_N^{\defmap}} \con{\delta \defmap}{\vb{t}} \, \dd \Surf + \int_{\Surf_N^{\Coss}} \con{ \Anti \delta \bm{\omega}}{\bm{T}\Coss^T} \, \dd \Surf + \int_{\Surf_N^{\psi}} \delta \psi \,   \mathfrak{b}_\mathrm{n} \, \dd \Surf \, , \notag 
\end{align}
where the right-hand side is the external virtual work. From the variational formulation it becomes clear that the coupling introduces additional stress and double-stress tensors acting on the kinematical variables. To accommodate them in the boundary value problem, it suffices to introduce the rotated total stress tensor
\begin{align}
    \RPiola_\mathrm{tot} = \RPiola + \RPiola_\mathrm{Mxl} + \RPiola_\mathrm{fmc} + \RPiola_\mathrm{fmcc} \, , 
\end{align}
and the rotated total double-stress tensor
\begin{align}
    \RDouble_\mathrm{tot} = \RDouble + \RDouble_\mathrm{fxm} \, ,
\end{align}
as well as their non-rotated counterparts
\begin{align}
    \Piola_\mathrm{tot} = \Coss \, \RPiola_\mathrm{tot} \, , && \Double_\mathrm{tot} = \tr(\RDouble_\mathrm{tot}^T)\Coss - \Coss \RDouble_\mathrm{tot}^T \, . 
\end{align}
Accordingly, the coupled boundary value problem reads
\begin{align}
    \boxed{
    \begin{aligned}
        -\Di \Piola_\mathrm{tot} &= \vb{f} && \text{in} && \Vol \, , 
        \\
        -\Di \Double_\mathrm{tot} -2\axl(\Piola_\mathrm{tot} \defgrad^T)  &= 2\axl(\Dforce\Coss^T)  && \text{in} && \Vol \, , 
        \\
        \di \tbmag  &= 0 && \text{in} && \Vol \, , 
        \\
        \Piola_\mathrm{tot} \vb{n} &= \vb{t} && \text{on} && \Surf_N^{\defmap} \, , 
        \\
        \Double_\mathrm{tot} \vb{n} &= 2\axl(\Dtrac \, \Coss^T) && \text{on} && \Surf_N^{\Coss} \, ,
        \\
         \con{\tbmag}{\vb{n}} &= \mathfrak{b}_\mathrm{n} && \text{on} && \Surf_N^{\psi} \, ,
        \\
        \defmap &= \widehat{\defmap} && \text{on} && \Surf_D^{\defmap} \, ,
        \\
        \Coss &= \widehat{\Coss} && \text{on} && \Surf_D^{\Coss} \, ,
        \\
        \psi &= \widehat{\psi} && \text{on} && \Surf_D^{\psi} \, .
\end{aligned}
    }
\end{align}
In general, prescribing a physically consistent normal component of the magnetic induction $\mathfrak{b}_\mathrm{n}$ on the boundary is not trivial without prior knowledge of the solution. Prescribing it to zero is akin to magnetic insulation and can be done under the assumption that magnetic field lines are tangent to the boundary, while still respecting the fundamental concept of no magnetic monopoles. \AS{Thus,} the more common alternative is to embed the object of interest within a surrounding region of free space \AS{$\Vol \subset \Vol_0 \approx \R^3$} and impose zero boundary conditions \AS{$\mathfrak{b}_\mathrm{n} = 0$} at the outer edge of this region \AS{$\partial \Vol_0$}, which must be sufficiently far from the physical object \AS{$\Vol$}. In such a scenario, the complete domain is composed of multiple intersecting volumes, making the treatment of interfaces critical. For example, let the domain be composed of two intersecting volumes $\Vol = \Vol_1 \cup \Vol_2$ with the common interface $\Xi = \Vol_1 \cap \Vol_2$ and boundary $\partial\Vol = \{\partial \Vol_1 \cup \partial \Vol_2\} \setminus \Xi$, then by the fact that $\di \tbmag = 0$ holds everywhere we get
\begin{align}
    \int_{\Vol} \di \tbmag \, \dd \Vol &=  \int_{\partial\Vol} \con{\tbmag}{\vb{n}} \, \dd \Surf = 0 \, , \notag \\
    \int_{\Vol_1} \di \tbmag_1 \, \dd \Vol &=  \int_{\partial\Vol_1} \con{\tbmag_1}{\vb{n}_1} \, \dd \Surf =  \int_{\partial\Vol_1 \setminus \Xi} \con{\tbmag_1}{\vb{n}_1} \, \dd \Surf + \int_{\Xi} \con{\tbmag_1}{\vb{n}_1} \, \dd \Surf = 0 \, ,  \\
    \int_{\Vol_2} \di \tbmag_2 \, \dd \Vol &=  \int_{\partial\Vol_2} \con{\tbmag_2}{\vb{n}_2} \, \dd \Surf =  \int_{\partial\Vol_2 \setminus \Xi} \con{\tbmag_2}{\vb{n}_2} \, \dd \Surf + \int_{\Xi} \con{\tbmag_2}{\vb{n}_2} \, \dd \Surf = 0 \, , \notag
\end{align}
where $\tbmag_1 = \tbmag|_{\Vol_1}$ and $\tbmag_2 = \tbmag|_{\Vol_2}$.
Rearranging and summing the last two equations therefore leads to
\begin{align}
    \int_{\Xi} \con{\tbmag_1}{\vb{n}_1} \, \dd \Surf + 
    \overbrace{\int_{\partial\Vol_1 \setminus \Xi} \con{\tbmag_1}{\vb{n}_1} \, \dd \Surf 
    + \int_{\partial\Vol_2 \setminus \Xi} \con{\tbmag_2}{\vb{n}_2} \, \dd \Surf}^{\int_{\partial\Vol} \con{\tbmag}{\vb{n}} \, \dd \Surf = 0} + \int_{\Xi} \con{\tbmag_2}{\vb{n}_2} \, \dd \Surf &= \int_{\Xi} \con{\tbmag_1 - \tbmag_2}{\vb{n}_1} \, \dd \Surf 
    \notag \\
    &= \jump{\con{\tbmag}{\vb{n}}}\at_{\Xi} = 0 \,, 
\end{align}
which affirms that the total magnetic induction must be continuous in its normal direction on interfaces. The possibility of a tangential jump is addressed in the next section, where impressed currents are considered.

\subsection{Magnetic vector potential and impressed currents}
The formulation of the action-functional in terms of the magnetic field $\hmag$ has the disadvantage of yielding a saddle-point problem, which is usually more difficult to solve. Further, it complicates the introduction of non-vanishing impressed currents $\vb{j} \neq 0$. 
To rectify both, we reformulate the problem in terms of the total magnetic induction $\tbmag$. To do so, we again employ a partial Legendre transform
\begin{align}
    \sup_{\hmag}[\Psi_\mathrm{mp}(\Stretch) + \Psi_\mathrm{curv}(\Curva) -\Psi_\mathrm{fmd}(\Stretch,\Curva,\hmag) - \Psi_\mathrm{mag}(\Stretch,\hmag) + \con{\hmag}{\tbmag}] \, .
\end{align} 
Since the transform is given by the stationary point, we again recover
\begin{align}
    \tbmag = \Grad_\hmag[\Psi_\mathrm{fmd}(\Stretch,\Curva,\hmag) + \Psi_\mathrm{mag}(\Stretch,\hmag)] &= \underbrace{(\det \Stretch) \Stretch^{-1} \Stretch^{-T}}_{(\det \defgrad) \defgrad^{-1} \defgrad^{-T}} (\mu\hmag + \chi_\mathrm{m}\mu_0  \Mm \Curva) 
     \, .
\end{align}
and can express the inverse relation between $\tbmag$ and $\hmag$ as
\begin{align}
    \hmag = \dfrac{1}{\mu} \bigg [ \dfrac{1}{\det \Stretch} \Stretch^{T} \Stretch \tbmag - \chi_\mathrm{m}\mu_0  (\Mm \Curva) \bigg ]  =  \dfrac{1}{\mu}  \bigg [ \dfrac{1}{\det \defgrad} \defgrad^T \defgrad\tbmag - \chi_\mathrm{m}\mu_0  (\Mm \Curva) \bigg ] \, . 
\end{align}
Inserting the latter in the combined flexomagnetic and demagnetisation enthalpy $\Psi_\mathrm{fmd}$ yields
\begin{align}
    \Psi_\mathrm{fmd}(\Stretch,\Curva,\hmag)\at_{\hmag = \hmag(\Stretch,\Curva,\amag)} 
    &= \dfrac{\chi_\mathrm{m}\mu_0 \det \Stretch}{2}  \norm{(\mu \det \Stretch)^{-1} \Stretch \tbmag - \mu^{-1}\chi_\mathrm{m}\mu_0 \Stretch^{-T} (\Mm \Curva) + \Stretch^{-T}(\Mm \Curva)}^2 
    \notag \\
    &= \dfrac{\chi_\mathrm{m} \mu_0 \det \Stretch}{2\mu^2}  \norm{(\det \Stretch)^{-1} \Stretch \tbmag + \mu_0 \Stretch^{-T} (\Mm \Curva) }^2 \, .
\end{align}
For the magnetisation enthalpy $\Psi_\mathrm{mag}$ one finds
\begin{align}
    \Psi_\mathrm{mag}(\Stretch,\hmag)\at_{\hmag = \hmag(\Stretch,\Curva,\amag)} 
    &= \dfrac{\mu_0 \det \Stretch}{2\mu^2} \norm{(\det \Stretch)^{-1} \Stretch \tbmag - \chi_\mathrm{m}\mu_0 \Stretch^{-T} (\Mm \Curva)}^2 \, .
\end{align}
Finally, the scalar product reads
\begin{align}
    \con{\hmag}{\tbmag}\at_{\hmag = \hmag(\Stretch,\Curva,\amag)} &= \dfrac{1}{\mu}\con{\Stretch^{T} [ (\det \Stretch)^{-1}  \Stretch \tbmag - \chi_\mathrm{m}\mu_0 \Stretch^{-T} (\Mm \Curva)  ]}{\tbmag} 
    \notag \\
    &= \dfrac{\mu\det\Stretch}{\mu^2}\con{(\det \Stretch)^{-1}  \Stretch \tbmag - \chi_\mathrm{m}\mu_0 \Stretch^{-T} (\Mm \Curva)}{(\det \Stretch)^{-1}\Stretch\tbmag} \, .
\end{align}
Consequently, after some manipulation, the sum of the terms is given by
\begin{align}
    [-\Psi_\mathrm{fxm}(\Stretch,\Curva,\hmag) -&\Psi_\mathrm{mag}(\Stretch,\hmag) + \con{\hmag}{\tbmag}]\at_{\hmag = \hmag(\Stretch,\Curva,\amag)}  
    \notag \\
    &= \dfrac{\det \Stretch}{2\mu} \norm{(\det \Stretch)^{-1}  \Stretch\tbmag - \chi_\mathrm{m}\mu_0\Stretch^{-T} (\Mm \Curva)}^2 - \dfrac{\chi_\mathrm{m} \mu_0 \det \Stretch}{2}\norm{\Stretch^{-T}(\Mm \Curva)}^2 \, ,
\end{align}
from which we extract the energy of the \textbf{classical} magnetic induction
\begin{align}
    \Psi_\mathrm{b}(\Stretch,\Curva,\tbmag) = \dfrac{\det \Stretch}{2\mu} \norm{\underbrace{(\det \Stretch)^{-1}  \Stretch\tbmag - \chi_\mathrm{m}\mu_0\Stretch^{-T} (\Mm \Curva)}_{\Coss^T\bmag_\varphi}}^2 \, ,
\end{align}
and the flexomagnetic energy induced by curvature 
\begin{align}
    \Psi_\mathrm{fmc}(\Stretch,\Curva) = \dfrac{\chi_\mathrm{m} \mu_0 \det \Stretch}{2}\norm{\Stretch^{-T}(\Mm \Curva)}^2 \, .
\end{align}
Now, by Gauss's law for magnetism there exist no magnetic charges, such that
\begin{align}
    \di_\varphi\tbmag_\varphi =  \di_\varphi \bigg ( \dfrac{1}{\det \defgrad} \defgrad  \tbmag \bigg ) = \dfrac{1}{\det \defgrad} \di \tbmag = 0 \, ,
\end{align}
where we exploited the Piola transformations of the divergence operator, cf. \cite{SKY2025322}.
Therefore, on a contractible domain there exists the vectorial potential 
\begin{align}
    \boxed{
    \tbmag = \curl \amag 
     \, , } && \amag : \Vol \to \R^3 \, ,
\end{align}
such that the energy of the magnetic induction can be written in terms of $\amag$
\begin{align}
    \Psi_\mathrm{b}(\Stretch,\Curva,\amag) = \dfrac{\det \Stretch}{2\mu} \norm{(\det \Stretch)^{-1}  \Stretch\curl \amag - \chi_\mathrm{m}\mu_0\Stretch^{-T} (\Mm \Curva)}^2 \, .
\end{align}
Thus, the full energy functional reads
\begin{align}
    \boxed{
    \begin{aligned}
    \Energy_\mathrm{fxm}(\defmap,\Coss,\amag) = \dfrac{1}{2}\int_\Vol & \overbrace{\norm{\sym(\Stretch - \one)}^2_{\Cm} + 2\muc \norm{\skw\Stretch}^2}^{2\Psi_\mathrm{mp}}  + \overbrace{\mu_\mathrm{e} \Lc^2 \norm{\Curva}_{\Lm}^2}^{2\Psi_\mathrm{curv}} 
    \\
    & + \underbrace{\dfrac{\det \Stretch}{\mu} \norm{(\det \Stretch)^{-1}  \Stretch \curl \amag - \chi_\mathrm{m}\mu_0\Stretch^{-T} (\Mm \Curva) }^2}_{2\Psi_\mathrm{b}} - \underbrace{\chi_\mathrm{m} \mu_0 (\det \Stretch)\norm{\Stretch^{-T}(\Mm \Curva)}^2}_{2\Psi_\mathrm{fmc}} \, \dd \Vol \, .
\end{aligned}
    }
\end{align}
\begin{remark}
    For the assessment of flexomagnetic parameters it can be useful to introduce a redefinition of the flexomagnetic material tensor as
    \begin{align}
        \boxed{
        \bm{\mathbbm{G}} = \chi_\mathrm{m}\mu_0\Mm \, ,
        } 
    \end{align}
    and directly measure its $\gamma$-coefficients
    \begin{align}
        \bm{\mathbbm{G}} = \gamma_\mathrm{iso} \perm + \gamma_\mathrm{cub}|\varepsilon_{ijk}|\vb{e}_i \otimes \vb{e}_j \otimes \vb{e}_k \, .
    \end{align}
    Accordingly, the energy functional turns to
    \begin{align}
    \boxed{
    \begin{aligned}
    \Energy_\mathrm{fxm}(\defmap,\Coss,\amag) = \dfrac{1}{2}\int_\Vol & \overbrace{\norm{\sym(\Stretch - \one)}^2_{\Cm} + 2\muc \norm{\skw\Stretch}^2}^{2\Psi_\mathrm{mp}}  + \overbrace{\mu_\mathrm{e} \Lc^2 \norm{\Curva}_{\Lm}^2}^{2\Psi_\mathrm{curv}} 
    \\
    & + \underbrace{\dfrac{\det \Stretch}{\mu} \norm{(\det \Stretch)^{-1}  \Stretch \curl \amag - \Stretch^{-T}(\bm{\mathbbm{G}}\Curva) }^2}_{2\Psi_\mathrm{b}} - \underbrace{\chi_\mathrm{m} \mu_0 (\det \Stretch)\norm{\Stretch^{-T}(\Mm \Curva)}^2}_{2\Psi_\mathrm{fmc}} \, \dd \Vol \, .
\end{aligned}
    }
\end{align}
    This is especially useful for  materials with a permeability close to that of free space $\mu \approx \mu_0$ that in fact can also exhibit the flexomagnetic effect \cite{Makushko2022Cr2O3}. The corresponding energy is then 
    \begin{align}
    \boxed{
    \begin{aligned}
    \Energy_\mathrm{fxm}(\defmap,\Coss,\amag) = \dfrac{1}{2}\int_\Vol & \overbrace{\norm{\sym(\Stretch - \one)}^2_{\Cm} + 2\muc \norm{\skw\Stretch}^2}^{2\Psi_\mathrm{mp}}  + \overbrace{\mu_\mathrm{e} \Lc^2 \norm{\Curva}_{\Lm}^2}^{2\Psi_\mathrm{curv}} 
    \\
    & + \underbrace{\dfrac{\det \Stretch}{\mu} \norm{(\det \Stretch)^{-1}  \Stretch \curl \amag - \Stretch^{-T}(\bm{\mathbbm{G}}\Curva) }^2}_{2\Psi_\mathrm{b}} \, \dd \Vol \, ,
\end{aligned}
    }
    \label{eq:fxmg}
\end{align}
since $\chi_\mathrm{m}\mu_0 \approx 0$. 
\end{remark}
The variation of the internal energy with respect to $\amag$ yields
\begin{align}
    \int_\Vol \dfrac{\det \Stretch}{\mu}\con{(\det \Stretch)^{-1} \Stretch \curl \delta \amag}{(\det \Stretch)^{-1}  \Stretch\curl \amag - \chi_\mathrm{m}\mu_0\Stretch^{-T} (\Mm \Curva)} \, \dd \Vol \, . 
    \label{eq:varb}
\end{align}
In the case that the medium is not free of impressed currents, the Amp\'ere--Maxwell law reads \cite{Zaglmayr2006}
\begin{align}
    \curl \hmag = \vb{j} \, , && \vb{j} = (\det \defgrad) \defgrad^{-1} \vb{j}_\varphi \, , 
\end{align}
where $\vb{j}:\Vol \to \R^3$ are impressed currents in the reference configuration, and $\vb{j}_\varphi:\Vol_\varphi \to \R^3$ are their counterpart in the current configuration. Thus, inserting the vector potential of $\tbmag$ and the identity of $\hmag$ into this equation gives
\begin{align}
    \dfrac{1}{\mu}\curl [(\det \Stretch)^{-1} \Stretch^{T} \Stretch \curl \amag - \chi_\mathrm{m}\mu_0  (\Mm \Curva)] = \vb{j} \, .
\end{align}
Testing with an admissible virtual potential $\delta \amag$ 
\begin{align}
    \int_\Vol \con{\delta \amag}{\mu^{-1}\curl [(\det \Stretch)^{-1} \Stretch^{T} \Stretch \curl \amag - \chi_\mathrm{m}\mu_0  (\Mm \Curva)]} \, \dd \Vol = \int_\Vol\con{\delta \amag}{\vb{j}} \, \dd \Vol 
\end{align}
and integrating by parts yields
\begin{align}
    \int_\Vol &(\det\Stretch) \con{(\det\Stretch)^{-1}\Stretch\curl \delta \amag}{\mu^{-1}[(\det\Stretch)^{-1}\Stretch\curl \amag - \chi_\mathrm{m}\mu_0 \Stretch^{-T}(\Mm \Curva)]} \, \dd \Vol 
     \\
    &= \int_\Vol\con{\delta \amag}{\vb{j}} \, \dd \Vol +  \int_{\Surf_N^{\amag}} \con{\delta \amag}{\underbrace{\mu^{-1}[(\det \Stretch)^{-1} \Stretch^T  \Stretch \curl \amag - \chi_\mathrm{m}\mu_0  (\Mm \Curva)]}_{\hmag}\times \vb{n} } \, \dd \Surf
    \, . \notag
\end{align}
On the Neumann boundary of $\amag$, the tangential part of $\hmag$ is given by impressed surface currents
\begin{align}
    \vb{k} = \hmag \times \vb{n}  \, .
\end{align}
Thus, the variational form reads
\begin{align}
    \int_\Vol &\dfrac{\det\Stretch}{\mu} \con{(\det\Stretch)^{-1}\Stretch\curl \delta \amag}{(\det\Stretch)^{-1}\Stretch\curl \amag - \chi_\mathrm{m}\mu_0 \Stretch^{-T}(\Mm \Curva)} \, \dd \Vol  
    = \int_\Vol\con{\delta \amag}{\vb{j}} \, \dd \Vol + \int_{\Surf_N^{\amag}} \con{\delta \amag}{\vb{k}} \, \dd \Surf \, .
\end{align}
Evidently, the left-hand side of the latter equation and the left-hand side of the variation of $\Psi_\mathrm{b}$ with respect to $\amag$ in \cref{eq:varb} are the same. Consequently, the right-hand of the previous equation represents additional virtual work by $\delta \amag$ that must be equilibrated. The corresponding external magnetostatic work is accordingly given by
\begin{align}
    \boxed{
    \Work_\mathrm{ms}(\amag) = \int_\Vol\con{\delta \amag}{\vb{j}} \, \dd \Vol + \int_{\Surf_N^{\amag}} \con{\delta \amag}{\vb{k}}  \, \dd \Surf  \, .
    }
\end{align}
Together, the energy functional and full work functional give rise to the minimisation problem
\begin{align}
    \boxed{
    \begin{aligned}
    \{\widehat{\defmap}, \widehat{\Coss} \}, \widehat{\amag}  &=  \arg \min_{\{\defmap,\Coss,\amag\}}  \, [  \Energy_\mathrm{fxm}(\defmap,\Coss,\amag) - \Work_\mathrm{Coss}(\defmap,\Coss) -\Work_\mathrm{ms}(\amag)]  \, .
\end{aligned}
    }
\end{align}
While the Legendre transform preserves the underlying structure of the problem, the natural expressions for the stress tensors in terms of $\tbmag$ are different. The variation of the magnetic energy reads
\begin{align}
    \delta [\Psi_\mathrm{b} (\Stretch,\Curva,\tbmag) - \Psi_\mathrm{fmc} (\Stretch,\Curva)] = \con{\delta \Stretch}{\Grad_{\Stretch}(\Psi_\mathrm{b}-\Psi_\mathrm{fmc})} + \con{\delta \Curva}{\Grad_{\Curva}(\Psi_\mathrm{b} - \Psi_\mathrm{fmc})} + \con{\delta \tbmag}{\Grad_{\tbmag}\Psi_\mathrm{b}} \, .
\end{align}
By expanding the energy of the magnetic induction
\begin{align}
    \Psi_\mathrm{b}(\Stretch,\Curva,\tbmag) = \dfrac{1}{2\mu \det\Stretch} \norm{\Stretch\tbmag}^2 - \dfrac{\chi_\mathrm{m}\mu_0}{\mu} \con{\tbmag}{\Mm \Curva} + \dfrac{\det \Stretch}{2\mu} \norm{\Stretch^{-T}\chi_\mathrm{m}\mu_0(\Mm \Curva)}^2 \, , 
\end{align}
we recover the tensors
\begin{align}
    \Grad_{\Stretch}[ \Psi_\mathrm{b}(\Stretch,\Curva,\tbmag)  - \Psi_\mathrm{fmc}(\Stretch,\Curva) ] =& (\det \Stretch) \bigg [ \dfrac{1}{\mu} ([\det\Stretch]^{-1}\Stretch \tbmag) \otimes ([\det\Stretch]^{-1}\Stretch \tbmag) - \dfrac{1}{2\mu} \norm{(\det \Stretch)^{-1} \Stretch \tbmag}^2 \one  \bigg] \Stretch^{-T} 
    \notag \\
    &-\dfrac{\det\Stretch}{\mu} \bigg [[\chi_\mathrm{m}\mu_0 \Stretch^{-T}(\Mm \Curva)] \otimes [\chi_\mathrm{m}\mu_0 \Stretch^{-T}(\Mm \Curva)] -  \dfrac{1}{2} \norm{ \chi_\mathrm{m}\mu_0 \Stretch^{-T}(\Mm \Curva)}^2 \one \bigg ] \Stretch^{-T}
    \notag \\
    &+ (\det \Stretch) \bigg [ \chi_\mathrm{m}\mu_0 [\Stretch^{-T}(\Mm \Curva)] \otimes [\Stretch^{-T}(\Mm \Curva)]  - \dfrac{\chi_\mathrm{m}\mu_0}{2}\norm{\Stretch^{-T}(\Mm \Curva)}^2\one \bigg] \Stretch^{-T} 
    \, .  
\end{align}
The first tensor can be defined as a rotated full magnetostrictive and flexomagnetic Piola--Maxwell stress tensor
\begin{align}
    \boxed{
    \RPiola_\mathrm{FMxl} = (\det \Stretch) \bigg [ \dfrac{1}{\mu} ([\det\Stretch]^{-1}\Stretch \tbmag) \otimes ([\det\Stretch]^{-1}\Stretch \tbmag) - \dfrac{1}{2\mu} \norm{(\det \Stretch)^{-1} \Stretch \tbmag}^2 \one  \bigg] \Stretch^{-T} \, ,
    }
\end{align}
whereas the two remaining tensors can be combined into weighted rotated Piola--Maxwell flexomagnetic curvature tensor
\begin{align}
    \boxed{
    \RPiola_\mathrm{wfmc} = (\det\Stretch) \bigg [\dfrac{\chi_\mathrm{m}\mu_0^2}{\mu} [\Stretch^{-T}(\Mm \Curva)] \otimes [ \Stretch^{-T}(\Mm \Curva)] -  \dfrac{\chi_\mathrm{m}\mu_0^2}{2\mu} \norm{ \Stretch^{-T}(\Mm \Curva)}^2 \one \bigg ] \Stretch^{-T} 
    = \dfrac{\mu_0}{\mu} \RPiola_\mathrm{fmc} \, .
    } 
\end{align}
Using the identity $\tbmag = \bmag + \mu_0\fmag$, it can be shown that $\RPiola_\mathrm{FMxl} + \RPiola_\mathrm{wfmc} = \RPiola_\mathrm{Mxl} + \RPiola_\mathrm{fmcc} + \RPiola_\mathrm{fmc}$, confirming the consistency of the Legendre transform.
Analogously, we expand
\begin{align}
    \Grad_{\Curva}[\Psi_\mathrm{b}(\Stretch,\Curva,\tbmag) - \Psi_\mathrm{fmc}(\Stretch,\Curva)] =& -\dfrac{\chi_\mathrm{m}\mu_0}{\mu}\tbmag\Mm + \dfrac{\chi_\mathrm{m}^2\mu_0^2}{\mu} (\det \Stretch)\Stretch^{-1} \Stretch^{-T}(\Mm \Curva) \Mm 
    \notag \\
    &- \chi_\mathrm{m}\mu_0 (\det \Stretch)\Stretch^{-1} \Stretch^{-T}(\Mm \Curva) \Mm \, ,
\end{align}
which can be reformulated into
\begin{align}
   \boxed{
   \RDouble_\mathrm{fxm} = -\chi_\mathrm{m}\mu_0 \bigg [ \dfrac{1}{\mu}\tbmag + \dfrac{\mu_0}{\mu} (\det \Stretch)\Stretch^{-1} \Stretch^{-T}(\Mm \Curva) \bigg ] \Mm \, .
   }
\end{align}
The latter agrees with the definition of $\RDouble_\mathrm{fxm}$ given as a function of the magnetic field $\hmag$.
Finally, the gradient of $\Psi_\mathrm{b}$ with respect to $\tbmag$ reads
\begin{align}
    \Grad_{\tbmag}[\Psi_\mathrm{b}(\Stretch,\Curva,\tbmag)] = \dfrac{1}{\mu \det \Stretch}\Stretch^T \Stretch \tbmag - \dfrac{\chi_\mathrm{m}\mu_0}{\mu}\Mm \Curva  = \hmag \, . 
\end{align}
Thus, the full variational functional reads
\begin{align}
    &\int_\Vol \con{\delta \Stretch}{\RPiola + \RPiola_\mathrm{FMxl} + \RPiola_\mathrm{fmc}} + \con{\delta \Curva}{\RDouble + \RDouble_\mathrm{fxm}} + \con{\curl \delta \amag}{\hmag} \, \dd \Vol 
    \\
    &\quad  = \int_\Vol \con{\delta \defmap}{\vb{f}} + \con{\Anti \delta \bm{\omega}}{\Dforce\Coss^T} + \con{\delta \amag}{\vb{j}} \, \dd \Vol + \int_{\Surf_N^{\defmap}} \con{\delta \defmap}{\vb{t}} \, \dd \Surf + \int_{\Surf_N^{\Coss}} \con{\Anti \delta \bm{\omega}}{\bm{T}\Coss^T} \, \dd \Surf + \int_{\Surf_N^{\amag}} \con{\delta \amag}{\vb{k}} \, \dd \Surf \, ,
    \notag
\end{align}
and the boundary value problem is given by
\begin{align}
    \boxed{
    \begin{aligned}
        -\Di \Piola_\mathrm{tot} &= \vb{f} && \text{in} && \Vol \, , 
        \\
        -\Di \Double_\mathrm{tot} -2\axl(\Piola_\mathrm{tot} \defgrad^T)  &= 2\axl(\Dforce\Coss^T)  && \text{in} && \Vol \, , 
        \\
        \curl \hmag &= \vb{j} && \text{in} && \Vol \, , 
        \\
        \Piola_\mathrm{tot} \vb{n} &= \vb{t} && \text{on} && \Surf_N^{\defmap} \, , 
        \\
        \Double_\mathrm{tot} \vb{n} &= 2\axl(\Dtrac \, \Coss^T) && \text{on} && \Surf_N^{\Coss} \, ,
        \\
        \hmag \times \vb{n} &= \vb{k} && \text{on} && \Surf_N^{\amag} \, ,
        \\
        \defmap &= \widehat{\defmap} && \text{on} && \Surf_D^{\defmap} \, ,
        \\
        \Coss &= \widehat{\Coss} && \text{on} && \Surf_D^{\Coss} \, ,
        \\
        \amag \times \vb{n} &= \widehat{\amag}_t && \text{on} && \Surf_D^{\amag} \, .
\end{aligned}
    }
\end{align}
The boundary Dirichlet condition of the vector potential is simply its tangential component. On a \textit{quasi-perfect electric conductor} this component approaches zero $\widehat{\amag}_t = 0$ as a consequence of Ohm's law \cite{Zaglmayr2006}. 
Considering again the scenario of multiple intersecting domains as in the previous section, then e.g. let the domain be split into two $\Vol = \Vol_1 \cup \Vol_2$ with the interface $\Xi = \Vol_1 \cap \Vol_2$ and boundary $\partial\Vol = \{\partial \Vol_1 \cup \partial \Vol_2\} \setminus \Xi$, the differential equation $\curl \hmag = \vb{j}$ implies
\begin{align}
    -\int_\Vol \curl \vb{h} \, \dd \Vol &= \int_{\partial \Vol} \vb{h} \times \vb{n} \, \dd \Surf = -\int_\Vol \vb{j} \, \dd \Vol \, , \notag \\
    -\int_{\Vol_1} \curl \vb{h}_1 \, \dd \Vol &= \int_{\partial \Vol_1} \vb{h}_1 \times \vb{n}_1 \, \dd \Surf = \int_{\partial \Vol_1\setminus \Xi} \vb{h}_1 \times \vb{n}_1 \, \dd \Surf + \int_{\Xi} \vb{h}_1 \times \vb{n}_1 \, \dd \Surf = -\int_{\Vol_1} \vb{j}_1 \, \dd \Vol \, , 
     \\
    -\int_{\Vol_2} \curl \vb{h}_2 \, \dd \Vol &= \int_{\partial \Vol_2} \vb{h}_2 \times \vb{n}_2 \, \dd \Surf = \int_{\partial \Vol_2\setminus \Xi} \vb{h}_2 \times \vb{n}_2 \, \dd \Surf + \int_{\Xi} \vb{h}_2 \times \vb{n}_2 \, \dd \Surf
    = -\int_{\Vol_2} \vb{j}_2 \, \dd \Vol \, , \notag 
\end{align}
where $\hmag_1 = \hmag|_{\Vol_1}$, $\hmag_2 = \hmag|_{\Vol_2}$, $\vb{j}_1 = \vb{j}|_{\Vol_1}$ and $\vb{j}_2 = \vb{j}|_{\Vol_2}$ but $\vb{j}_1|_{\Xi} \neq \vb{j}_2|_{\Xi}$.
By rearranging and summing the last two equations the latter leads to
\begin{align}
    \int_{\Xi} \vb{h}_1 \times \vb{n}_1 \, \dd \Surf +
    \underbrace{\int_{\partial \Vol_1\setminus \Xi} \vb{h}_1 \times \vb{n}_1 \, \dd \Surf 
    + \int_{\partial \Vol_2\setminus \Xi} \vb{h}_2 \times \vb{n}_2 \, \dd \Surf}_{\int_{\partial \Vol} \vb{h} \times \vb{n} \, \dd \Surf = - \int_\Vol \vb{j} \, \dd \Vol} + \int_{\Xi} \vb{h}_2 \times \vb{n}_2 \, \dd \Surf = \underbrace{-\int_{\Vol_1} \vb{j}_1 \, \dd \Vol - \int_{\Vol_2} \vb{j}_2 \, \dd \Vol}_{-\int_{\Vol} \vb{j} \, \dd \Vol + \int_{\Xi} \vb{k} \, \dd \Surf}
\end{align}
where $\vb{k} = \vb{j}_2|_{\Xi} - \vb{j}_1|_{\Xi}$ stems from the jump of the impressed currents at the interface.
Consequently, there holds
\begin{align}
    \int_{\Xi} \vb{h}_1 \times \vb{n}_1 \, \dd \Surf + \int_{\Xi} \vb{h}_2 \times \vb{n}_2 \, \dd \Surf = \int_{\Xi} (\vb{h}_1 - \vb{h}_2) \times \vb{n}_1 \, \dd \Surf = \int_{\Xi} \vb{k} \, \dd \Surf \quad \Rightarrow \quad \jump{\vb{h} \times \vb{n}}\at_{\Xi} = \vb{k} \, . 
\end{align}
In other words, the tangential jump of the magnetic field
on interfacing boundaries is exactly captured by the impressed surface currents.
Using the transformation from \cref{eq:btoh}, this implies that $\vb{b}$ may exhibit a tangential jump at interfaces, and since $\tbmag$ contains $\vb{b}$, it may tangentially jump as well.

\section{The couple-stress model}

If the Cosserat couple modulus tends to infinity $\muc \to +\infty$, then the associated norm must go to zero $\norm{\skw \Stretch} \to 0$ to ensure finite energies \cite{Neff2004,Neff2006b,Neff2006Couple,Lankeit2016}. This is satisfied according to the polar decomposition \cite{Neff2014Grioli} via
\begin{align}
    &\Coss \to \bm{R} = \polar (\defgrad) = \defgrad \bm{U}^{-1}  && \bm{R}: \Vol \to \SO(3) \, , &&  \bm{U}:\Vol \to \Sym^{++}(3) \, , 
\end{align}
and implies that the Biot-type stretch tensor and the curvature tensors change to
\begin{align}
    \Stretch \to \bm{U} \, , && \Curva \to \bm{\mathfrak{A}} =  \bm{R}^T (\Curl \bm{R}) \, .  
\end{align}
Evidently, the Biot-type stretch tensor degenerates into the classical Biot stretch tensor. 
Owing to the elegance of the bar-notation, one can now recover the flexomagnetic and flexoelectric couple-stress models simply by replacing $\Coss$ with $\bm{R}$ and $\Stretch$ with $\bm{U}$ throughout.
Accordingly, the internal energy density of the micropolar model changes to
\begin{align}
    \Psi_\mathrm{cs}(\defmap) = \dfrac{1}{2} \norm{\sym(\bm{U} - \one)}^2_{\Cm} + \dfrac{\mu_\mathrm{e} \Lc^2}{2} \norm{\bm{\mathfrak{A}}}_{\Lm}^2 \, ,
\end{align}
now representing the couple-stress model, and the corresponding external work reads
\begin{align}
    \boxed{
    \begin{aligned}
    \Work_\mathrm{cs}(\defmap) = \int_\Vol \con{\defmap}{\vb{f}} + \con{\bm{R}}{\Dforce} \, \dd \Vol + \int_{\Surf_N^{\defmap}} \con{\defmap}{\vb{t}} \, \dd \Surf + \int_{\Surf_N^{\bm{R}}} \con{\bm{R}}{\Dtrac} \, \dd \Surf \, . 
\end{aligned}
    }
\end{align}
The action-functional of flexomagnetism with a scalar magentic potential is thus dependent on the internal functional
\begin{align}
    \boxed{
    \begin{aligned}
    \Energy_\mathrm{fxm}(\defmap,\psi) = \dfrac{1}{2}\int_\Vol & \overbrace{\norm{\sym(\bm{U} - \one)}^2_{\Cm} + \mu_\mathrm{e} \Lc^2 \norm{\bm{\mathfrak{A}}}_{\Lm}^2}^{2\Psi_\mathrm{cs}}
    \\
    & - [ \underbrace{\chi_\mathrm{m}\mu_0(\det \bm{U})\norm{\bm{U}^{-T}(\Mm \bm{\mathfrak{A}} - \nabla\psi )}^2}_{2\Psi_\mathrm{fmd}}  + \underbrace{\mu_0 (\det \bm{U}) \norm{\bm{U}^{-T}\nabla\psi}^2}_{2\Psi_\mathrm{mag}} ]\, \dd \Vol \, .
\end{aligned}
    }
\end{align}
reading
\begin{align}
    \boxed{
    \begin{aligned}
    \{\widehat{\defmap},\widehat{\psi} \}  &=  \arg \min_{\defmap} \max_{\psi} \, [  \Energy_\mathrm{fxm}(\defmap,\psi) - \Work_\mathrm{cs}(\defmap) + \Work_\mathrm{ms}(\psi) ]  \, .
\end{aligned}
    }
\end{align}
The formulation in vector potential depends on the energy-functional
\begin{align}
    \boxed{
    \begin{aligned}
    \Energy_\mathrm{fxm}(\defmap,\amag) = \dfrac{1}{2}\int_\Vol & \overbrace{\norm{\sym(\bm{U} - \one)}^2_{\Cm} + \mu_\mathrm{e} \Lc^2 \norm{\bm{\mathfrak{A}}}_{\Lm}^2}^{2\Psi_\mathrm{cs}}
    \\
    & + \underbrace{\dfrac{\det \bm{U}}{\mu} \norm{(\det \bm{U})^{-1}  \bm{U} \curl \amag- \overbrace{\chi_\mathrm{m}\mu_0\bm{U}^{-T} (\Mm \bm{\mathfrak{A}})}^{\bm{U}^{-T} (\bm{\mathbbm{G}} \bm{\mathfrak{A}})}}^2}_{2\Psi_\mathrm{b}} - \underbrace{\chi_\mathrm{m} \mu_0 (\det \bm{U})\norm{\bm{U}^{-T}(\Mm \bm{\mathfrak{A}})}^2}_{2\Psi_\mathrm{fmc}} \, \dd \Vol \, .
\end{aligned}
    }
\end{align}
and is associated with the action-functional
\begin{align}
    \boxed{
    \begin{aligned}
    \{\widehat{\defmap}, \widehat{\amag}\}  &=  \arg \min_{\{\defmap,\amag\}}  \, [  \Energy_\mathrm{fxm}(\defmap,\amag) - \Work_\mathrm{cs}(\defmap) -\Work_\mathrm{ms}(\amag)]  \, .
\end{aligned}
    }
\end{align}

The corresponding boundary value problems change accordingly, now requiring strong enforcement of derivatives of $\defmap$ over $\polar (\Grad \defmap)$.
\begin{remark}
    Notably, by the fact that
\begin{align}
    \Curva \to \bm{\mathfrak{A}} = \bm{R}^T (\Curl \bm{R}) =  (\polar \defgrad)^T \Curl (\polar \defgrad) = (\polar \Grad \defmap)^T \Curl (\polar \Grad \defmap) \, , 
\end{align}
the couple-stress theory represents a second-derivative model, requiring increased smoothness in $\defmap$ for conforming discretisations. Further, its connection to finite strain-gradient theories can be established via the formula
\begin{align}
    \bm{R}^T \Grad \bm{R} = \dfrac{1}{\det \bm{U}} \Anti\left [(\bm{U}[\Curl\bm{U}]^T - \dfrac{1}{2}\tr[\bm{U}(\Curl\bm{U})^T]\one)\bm{U} \right ] \, ,
\end{align}
from \cite{Lankeit2016} using that $\bm{U} = \sqrt{\defgrad^T \defgrad}$.

\end{remark}

\section{Numerical investigation}

\subsection{Discretisation}

The variational formulations presented in the previous sections are nonlinear. Consequently, the solution follows via the Newton--Raphson method, for which the directional derivatives of the functional with respect to iteration-increments are computed using automatic differentiation. For the discretisation of the displacement field we employ vector-valued $\C^0(\Vol)$-continuous cubic Lagrange elements $\disp \in \CG^3(\Vol) \otimes \R^3$. 
The Cosserat rotation tensor is computed using Euler-matrices $\Coss(\bm{\theta}) =  \Coss_z(\theta_z) \Coss_y (\theta_y) \Coss_x(\theta_x)$ with 
\begin{align}
    \Coss_x(\theta_x) = \begin{bmatrix}
1 & 0 & 0 \\
0 & \cos\theta_x & -\sin\theta_x \\
0 & \sin\theta_x & \cos\theta_x
\end{bmatrix} \, , &&
\Coss_y(\theta_y) =
\begin{bmatrix}
\cos\theta_y & 0 & \sin\theta_y \\
0 & 1 & 0 \\
-\sin\theta_y & 0 & \cos\theta_y
\end{bmatrix} \, , &&
\Coss_z(\theta_z) = 
\begin{bmatrix}
\cos\theta_z & -\sin\theta_z & 0 \\
\sin\theta_z & \cos\theta_z & 0 \\
0 & 0 & 1
\end{bmatrix} \, ,
\end{align}
The corresponding rotation vector is discretised using quartic Lagrange elements $\bm{\theta} \in \CG^{4}(\Vol) \otimes \R^3$ to mitigate possible couple-locking effects \cite{CRISFIELD1990131,Harsch}. This does not solve the susceptibility of Euler-matrices to gimbal-locking \cite{Hemingway2018}. \AS{However, although this issue may arise under multiaxial rotations, it does not occur in our uniaxial benchmark tests}.
Now, to retrieve the Curl, we first observe the general structure of an Euler matrix around some Cartesian axis $\xi$ is given by
\begin{align}
    \Coss_\xi (\theta_\xi) = \vb{e}_\xi \otimes \vb{e}_\xi + (\cos\theta_\xi)(\one - \vb{e}_\xi \otimes \vb{e}_\xi) + (\sin \theta_\xi)\Anti \vb{e}_\xi \,,
\end{align}
with \textbf{no summation} over repeating $\xi$-indices. Consequently, its derivative with respect to some arbitrary axis is given by
\begin{align}
    [\Coss_\xi (\theta_\xi)]_{,i} = 
    \theta_{\xi,i}\underbrace{[(\cos \theta_\xi)\Anti \vb{e}_\xi-(\sin\theta_\xi)(\one - \vb{e}_\xi \otimes \vb{e}_\xi)]}_{\widetilde{\bm{R}}_{\xi}} \,.
\end{align}
Thus, the Curl reads
\begin{align}
    \Curl \Coss = -(\Coss_z \Coss_y  \Coss_x)_{,i}(\Anti \vb{e}_i) = - (\widetilde{\bm{R}}_{z} \Coss_{y}  \Coss_{x}) (\Anti \nabla \theta_z) - (\Coss_{z} \widetilde{\bm{R}}_{y}  \Coss_{x}) (\Anti \nabla \theta_y) - (\Coss_{z} \Coss_{y}  \widetilde{\bm{R}}_{x}) (\Anti \nabla \theta_x) \, . 
\end{align}
The correct space for the vectorial potential is the N\'ed\'elec \cite{SKY2024104155,Zaglmayr2006,SKY2024115568} space $\amag \in \Ned_I^p(\Vol)$. However, in general, a vector-potential formulation of magnetostatics is not automatically well-posed. This is because all vectorial potentials of the form $\amag + \nabla \psi$ satisfy  $\tbmag = \curl(\amag + \nabla \psi) = \curl \amag$. To re-enforce uniqueness of the solution it is thus common to introduce the Coulomb gauge \cite{Zaglmayr2006} 
\begin{align}
    -\di \amag = 0 \quad \Rightarrow \quad -\int_\Vol\con{\psi}{\di \amag} \, \dd \Vol = \int_\Vol \con{\nabla \psi}{\amag} \, \dd \Vol \quad \forall\, \psi \in \C^\infty_0(\Vol) \, , 
\end{align}
which effectively implies $\amag \perp \nabla\psi$ for any $\psi \in \CG^{p+1}(\Vol)$ with a vanishing Dirichlet boundary in the discretisation. 
Energetically, the Coulomb gauge can be additively incorporated into the functional 
\begin{align}
    \boxed{
    \Energy_\mathrm{fxm} + \int_\Vol \con{\nabla \psi}{\amag} \, \dd \Vol \, ,
    }
\end{align}
leading to a saddle-point problem. The latter can be avoided if one constructs the vectorial potential as $\amag \in \Ned^p_I(\Vol) \setminus \nabla \CG^{p+1}(\Vol)$ \'a priori. This is possible down to the lowest order N\'ed\'elec elements in the $hp$-finite element method using space-splitting methods \cite{SKY2025322,Zaglmayr2006}, or even fully with non-local approaches \cite{boon2024solversmixedfiniteelement}. Another method is to introduce a small regularisation term  
\begin{align}
    \boxed{
    \Energy_\mathrm{fxm} + \int_\Vol \dfrac{\beta}{2} \norm{\amag}^2 \, \dd \Vol \, ,
    } && 0 < \beta \ll |\mu| \, ,
\end{align}
such that no additional field is needed. For $\beta \to 0$ it can be shown that the solution convergences towards the one that is orthogonal to gradient fields \cite{Zaglmayr2006}.
However, this approach requires solvers that are numerically robust for the very small parameter $\beta \to 0$. In this work we rely on the space-splitting method and the regularisation for the lowest-order basis-functions, and use direct solvers, which is possible due to the relatively small size of the system we consider. \textbf{We do not compute solutions using the scalar magnetic potential in this study, as it typically necessitates the modelling of the surrounding free-space.}

\subsection{Ideal units}

In electromagnetism, the \textit{purest units} for expressing all physical quantities are the four SI base units, being meter (m), kilogram (kg), second (s), and ampere (A). All other units such as, volt (V), coulomb (C), tesla (T), farad (F), and henry (H), can be reduced to combinations of these base units. This is critical, since the underlying choice of units in any simulation must be consistent throughout to be valid. 
Further, in the quasi-static flexomagnetic model time does not play a role, such that newtons can be used. 
Lastly, the units are specifically selected within the metric scale to avoid excessively large or small values, thereby improving the matrix condition number.
Thus, the following units are employed throughout
\begin{center}
    \begin{tabular}{|c|c|}
    \hline  
         nanometer & m = $10^{9}$ nm \\ 
         nano-newton & N = $10^9$ nN \\
         milliampere & A = $10^3$ mA 
         \\\hline  
\end{tabular}
\end{center} 
such that the following conversion of combined units arises 
\begin{center}
    \begin{tabular}{|c|c|}
    \hline  
         nano-newton per nanometer square & $1$ N/mm$^2$ (MPa) = $10^{-3}$ nN/nm$^2$ (GPa) 
         \\
         nano-newton per milliampere nanometer & $1$ N/Am (T) = $10^{-3}$ nN/mAm (kT)
         \\\hline  
\end{tabular}
\end{center}
With these units, the permeability of free space reads 
\begin{align}
    \mu_0 = 4\pi \cdot 10^{-4} \approx 0.00125663706  \, ,
\end{align}
in kilo-tesla nanometer per milliampere kTnm/mA. 
Accordingly, the magnetic induction is measured in kilo-tesla kT, the $\eta$-flexomagnetic coefficients of $\Mm$ in milliampere mA, and the corresponding $\gamma$-flexomagnetic coefficients of $\bm{\mathbbm{G}}$ in kilo-tesla nanometer kTnm, or equivalently in nano-newton per milliampere nN/mA.     
\subsection{A flexomagnetic nano-beam}

\begin{figure}
    \centering
    \begin{subfigure}{0.64\linewidth}
    \centering
        \input{figs/nanobeam}
        \caption{}
    \end{subfigure}
    \begin{subfigure}{0.32\linewidth}
    \centering    
        \includegraphics[width=1\linewidth]{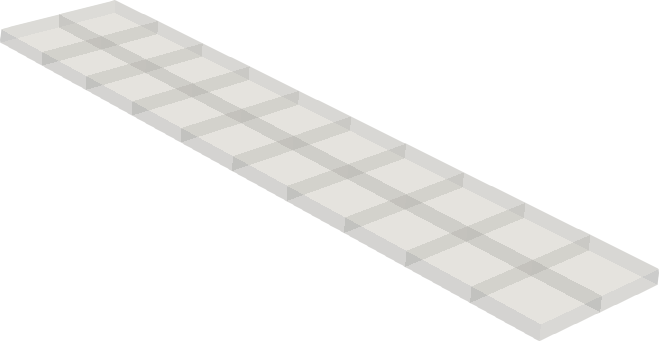}

        \vspace{0.5cm}
        
        \caption{}
    \end{subfigure}
    \caption{(a) Cantilever nano-beam with Dirichlet and Neumann boundary surfaces. (b) Mesh of $20$ hexahedral finite elements with polynomial orders $p \geq 3$, as per field.}
    \label{fig:nanobeam}
\end{figure}

\begin{figure}
    \centering
    \begin{subfigure}{0.32\linewidth}
    \centering
        \input{figs/lctwist}
        \caption{}
    \end{subfigure}
    \begin{subfigure}{0.32\linewidth}
    \centering
        \includegraphics[width=0.85\linewidth]{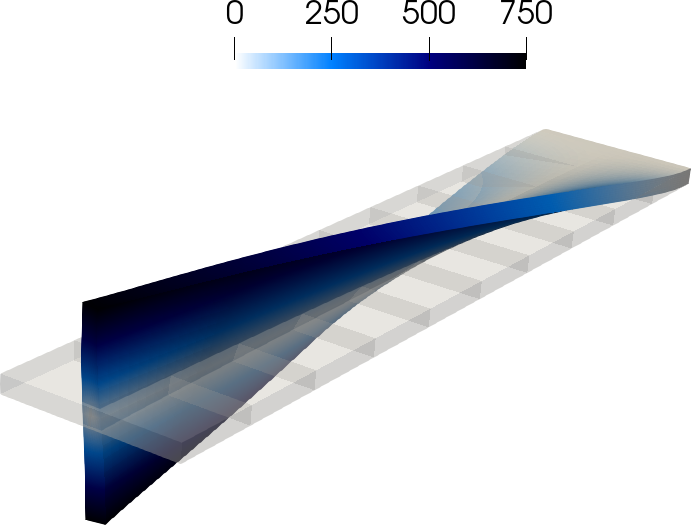}
        \caption{}
    \end{subfigure}
    \begin{subfigure}{0.32\linewidth}
    \centering
        \includegraphics[width=0.85\linewidth]{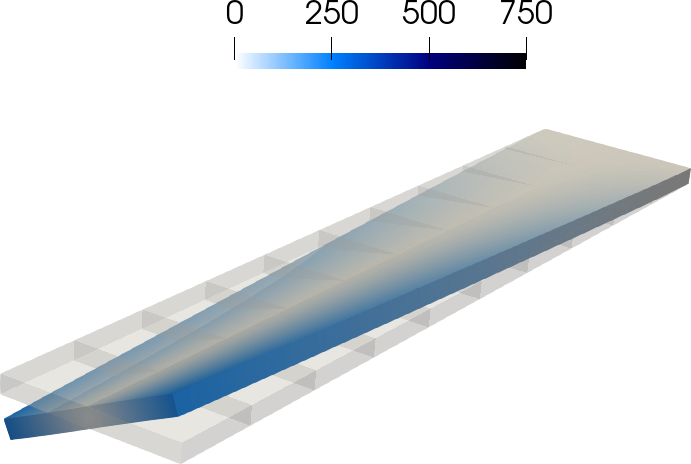}
        \caption{}
    \end{subfigure}
    \caption{(a) Maximal torsion deformation of the beam depending on $\muc$ and $\Lc$ . Beam deformation with $\muc/\mu_\mathrm{e} = 1$ for $\Lc = 1$ (b) and $\Lc = 10^2$ (c).}
    \label{fig:twi}
\end{figure}

In the following we consider a simple cantilever nano-beam given by the reference domain of 
\begin{align}
    \vb{x} = x\vb{e}_1 + y\vb{e}_2 + z\vb{e}_3 \in \bar{\Vol} = [0,5000] \times [-500,500] \times [0,100] \, ,
\end{align}
in nanometers, see \cref{fig:nanobeam}. 
The relevant characteristic length scale for the related strain-gradient theory is around 
$1 \leq \Lc \leq 100$ nm \cite{Codony}.
Further, due to the lack of data on Cosserat material coefficients, we replace the weighted material norm of the mechanical curvature with the classical norm such that the corresponding curvature energy reads $\mu_\mathrm{e}\Lc^2\norm{\Curva}^2$.
The Dirichlet boundary for the displacement field is the back surface of the beam $\Surf_D^{\defmap} = \{\vb{x} \in  \partial V \; | \; x = 0\}$, implying that $\disp|_{x = 0} = 0$.
On the surface at the front of the beam $\Surf_{N_f}^{\defmap} = \{\vb{x} \in  \partial V \; | \; x = 5000\}$ we impose non-vanishing Neumann boundary conditions. 
The boundary for the Cosserat rotation field $\Coss$ is completely Neumann with no imposed values $\Surf_N^{\Coss} = \partial V$.
The beam is defined to be made of Chromium(III) oxide otherwise known as Chromia. Chromia is an antiferromagnetic \AS{purely} centrosymmetric material known to exhibit flexomagnetic effects~\cite{Makushko2022Cr2O3}. Its elastic Lam\'e parameters in gigapascal (GPa) are roughly \cite{Saeki01122011}
\begin{align}
    \lambda_\mathrm{e} = 98 \, , && \mu_\mathrm{e} = 116 \, .
\end{align}
With no availability of the Cosserat coupling modulus or the mechanical characteristic length-scale parameter, we start by characterising the purely mechanical model by varying $\muc$ and $\Lc$. On the Neumann surface $\Surf_{N_f}^{\defmap}$ we impose the traction vector $\vb{t} = 2\cdot10^{-2}(\vb{e}_3 - \vb{e}_1)$, implying $2\cdot10^{-2}$ nN/nm$^2$ in upper and backwards directions. 
\begin{figure}
    \centering
    \begin{subfigure}{0.32\linewidth}
    \centering
        \input{figs/lcbend}
        \caption{}
    \end{subfigure}
    \begin{subfigure}{0.32\linewidth}
    \centering
        \includegraphics[width=0.85\linewidth]{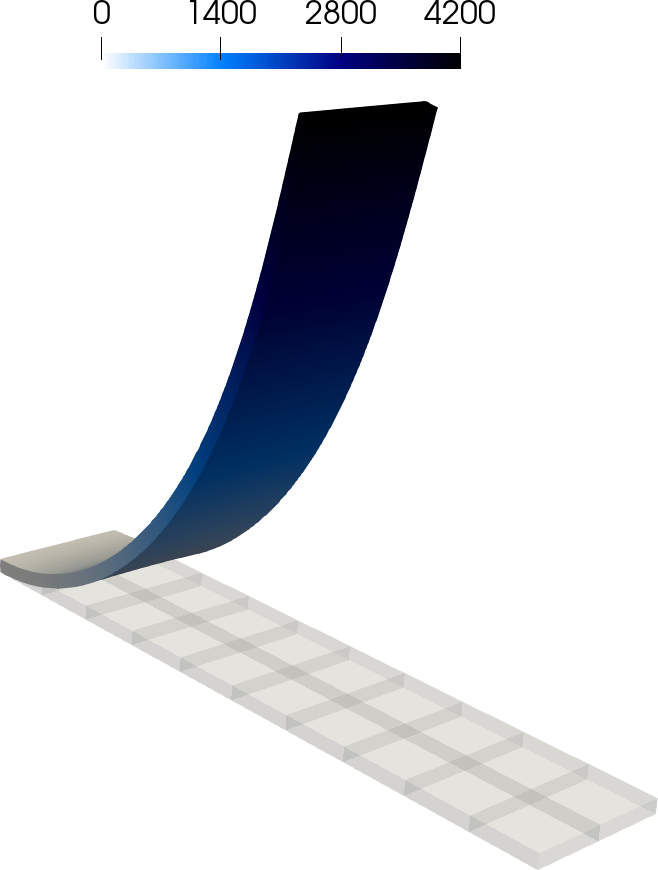}
        \caption{}
    \end{subfigure}
    \begin{subfigure}{0.32\linewidth}
    \centering
        \includegraphics[width=0.85\linewidth]{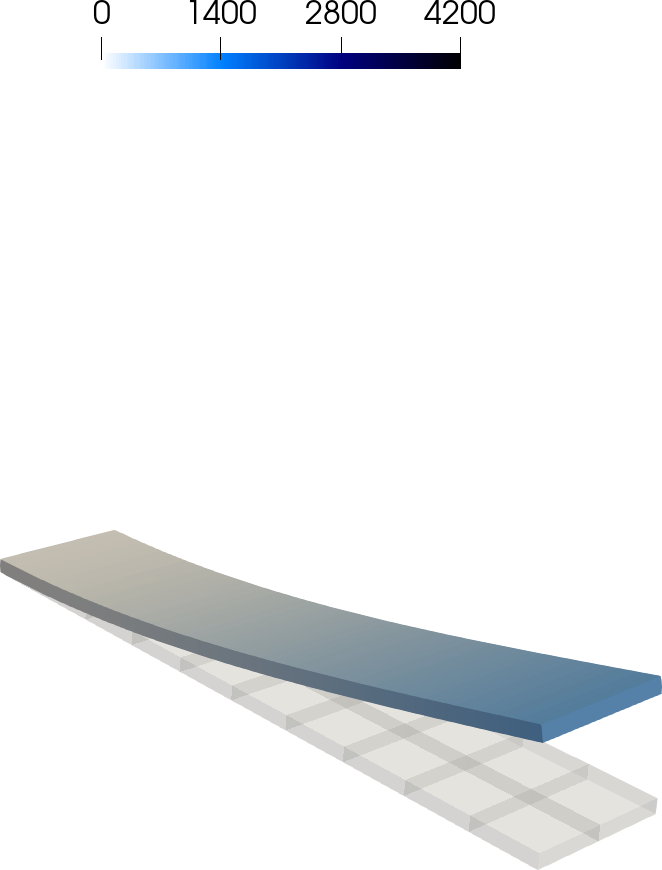}
        \caption{}
    \end{subfigure}
    \caption{(a) Maximal bending deformation of the beam depending on $\muc$ and $\Lc$ . Beam deformation with $\muc/\mu_\mathrm{e} = 1$ for $\Lc = 1$ (b) and $\Lc = 10^2$ (c).}
    \label{fig:muc}
\end{figure}
The results, measuring the corresponding \AS{large} maximal displacement $\max \norm{\disp}$, are listed and displayed in \cref{fig:muc}. Observably, the effect of the Cosserat couple modulus $\muc > 0$ is highly dependent on the characteristic length-scale parameter and in fact, for $\Lc \leq 10$, it becomes negligible. When $\Lc$ approaches the thickness of the domain $\Lc \to 100$, the coupling becomes more pronounced, and the curvature energy leads to a hefty reduction in the observed deformation.
However, increasing $\muc/\mu_\mathrm{e} = 1$ to $\muc/\mu_\mathrm{e} = 10$ makes no significant difference in deformation behaviour.  
Similar observations follow from a torsion test, where the traction vector is redefined as $\vb{t} = 2\cdot10^{-3} \, y(\vb{e}_3 - \vb{e}_2)$. The corresponding \AS{large} twist-deformations are listed in \cref{fig:twi}.
Thus, it is clear that without the orthogonal Cartan-decomposition of the curvature tensor $\Curva$ with weighting material parameters via $\Lm$, the observed behaviour is not dependent on the deformation mode itself. The upper limits of both deformation modes are governed by finite Cauchy elasticity. Finally, we mention that although it is possible to compute solutions with $\muc = 0$ in principle, the procedure is highly unstable and therefore impractical without a dedicated stabilisation technique.  

\begin{figure}
    \centering
    \begin{subfigure}{0.32\linewidth}
    \centering
        \input{figs/gphi}
        \caption{}
    \end{subfigure}
    \begin{subfigure}{0.32\linewidth}
    \centering
        \input{figs/gb}
        \caption{}
    \end{subfigure}
    \begin{subfigure}{0.32\linewidth}
    \centering
        \includegraphics[width=0.85\linewidth]{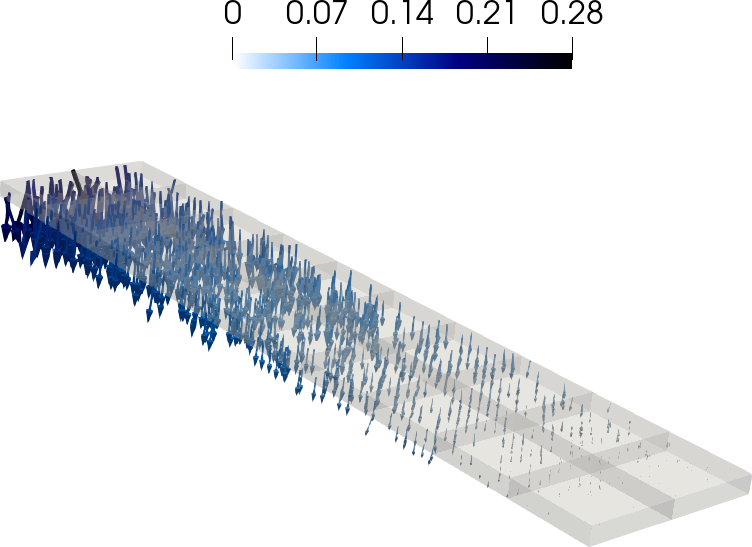}
        \caption{}
    \end{subfigure}
    \begin{subfigure}{0.32\linewidth}
    \centering
        \includegraphics[width=0.85\linewidth]{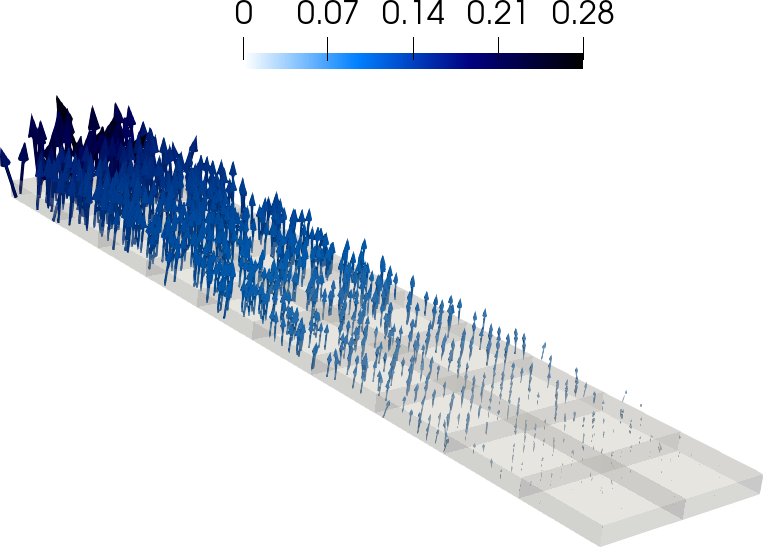}
        \caption{}
    \end{subfigure}
    \begin{subfigure}{0.32\linewidth}
    \centering
        \includegraphics[width=0.85\linewidth]{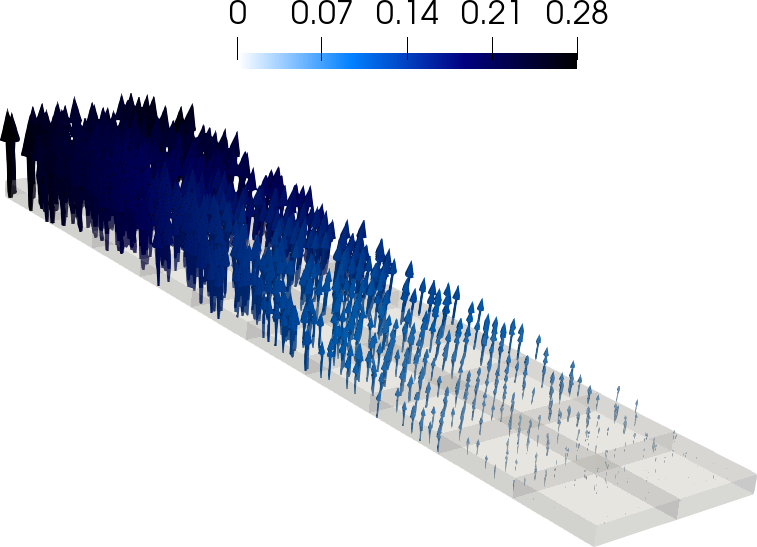}
        \caption{}
    \end{subfigure}
    \begin{subfigure}{0.32\linewidth}
    \centering
        \includegraphics[width=0.85\linewidth]{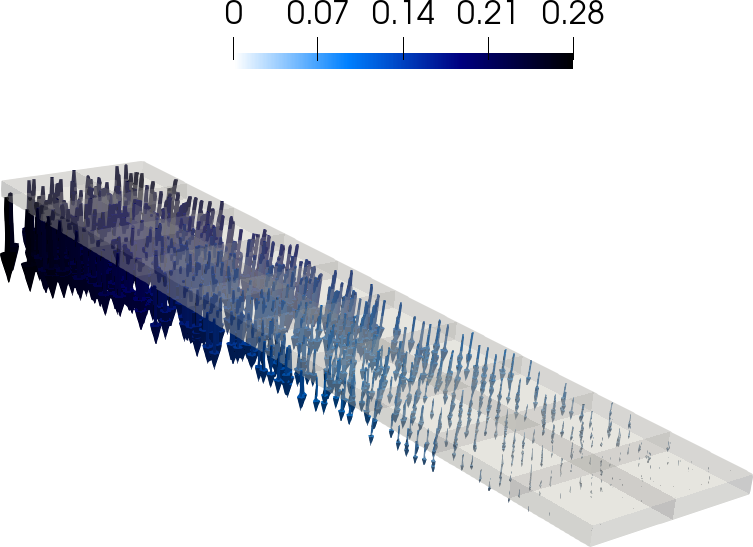}
        \caption{}
    \end{subfigure}
    \begin{subfigure}{0.32\linewidth}
    \centering
        \includegraphics[width=0.85\linewidth]{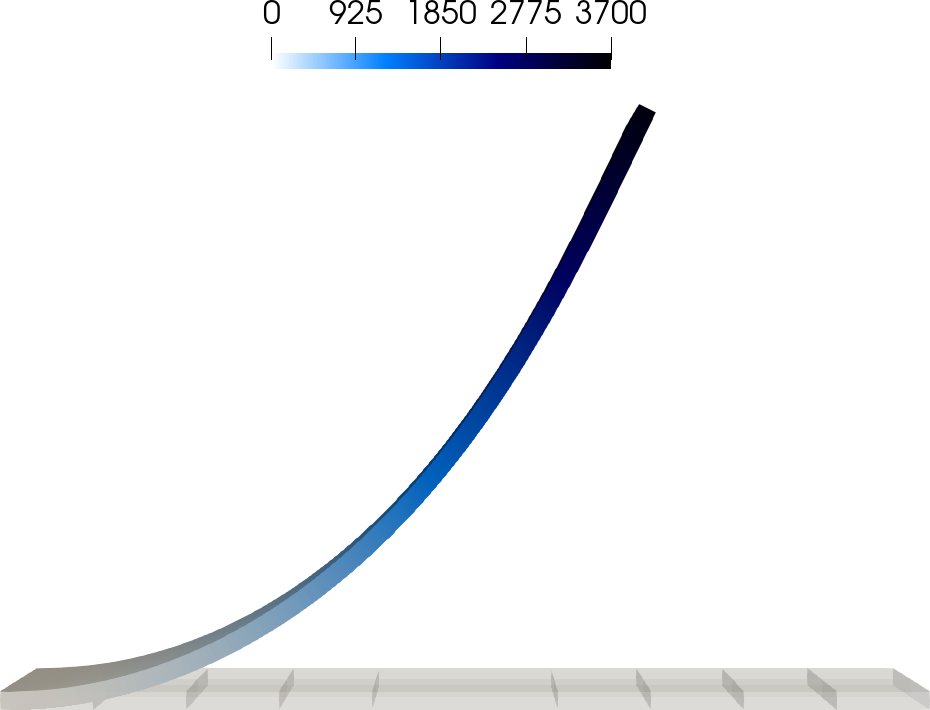}
        \caption{}
    \end{subfigure}
    \begin{subfigure}{0.32\linewidth}
    \centering
        \includegraphics[width=0.85\linewidth]{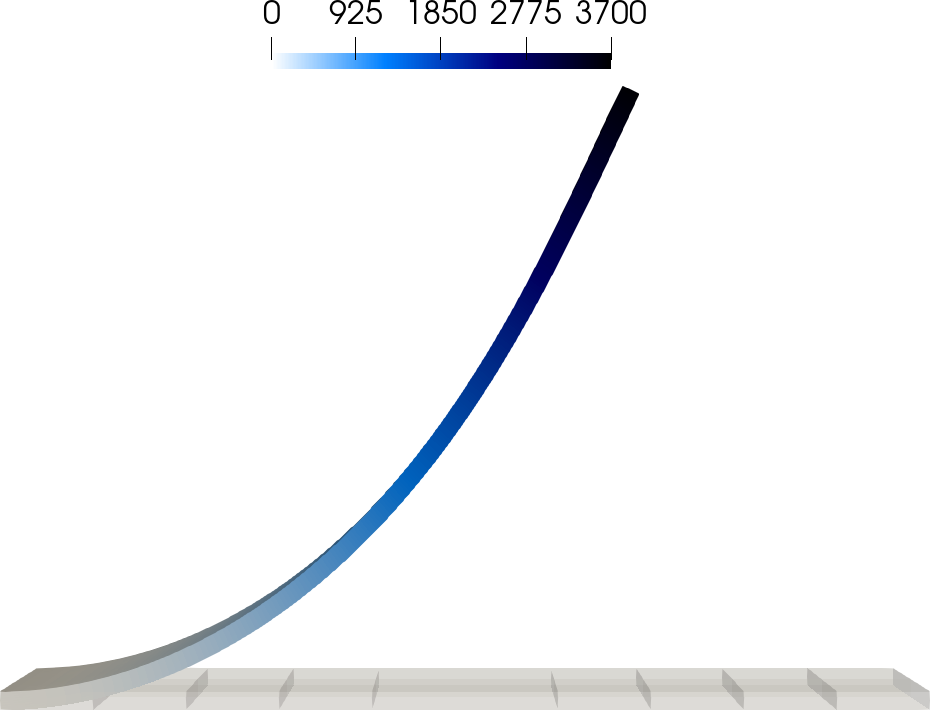}
        \caption{}
    \end{subfigure}
    \caption{Maximal bending deformation for varying $\gamma$ values (a), and the corresponding extremal magnetic induction along the $z$-axis (b). Total magnetic induction $\tbmag$ on the reference domain for $\gamma_\mathrm{iso} = 1000$ (c), $\gamma_\mathrm{iso} = -1000$ (d), $\gamma_\mathrm{cub} = 1000$ (e), and $\gamma_\mathrm{cub} = -1000$ (f). Deformation for $\gamma_\mathrm{iso} = \pm 1000$ (g) and $\gamma_\mathrm{cub} = \pm 1000$ (h).}
    \label{fig:gbend}
\end{figure}

Based on the latter results, we now set $\muc = \mu_\mathrm{e}$ and $\Lc = 10$ for a moderate influence of the mechanical curvature energy on the overall deformation, and move forward to measure the impact of the flexomagnetic coefficients $\gamma_\mathrm{iso}$ and $\gamma_\mathrm{cub}$ on the flexomagnetic model. Even though Chromia is likely to exhibit $\gamma_\mathrm{cub} = 0$ \AS{due to its centrosymmetry}, it is still of interest to explore the disparity between a pure centrosymmetric reaction and a polar \AS{cubic-symmetric} one \AS{without incorporating piezomagnetic contributions associated with cubic symmetry, in order to isolate the purely flexomagnetic response}. 
The permeability of Chromia is approximatively the same as of vacuum 
\begin{align}
    \mu \approx \mu_0 \, , && \mu^{-1} \approx 795.77 \, .
\end{align}
Consequently, one finds $\chi_\mathrm{m}\mu_0 \approx 0$, such that the flexomagnetic curvature energy becomes negligible $\Psi_\mathrm{fmc} \approx 0$. \textbf{Hence, we omit the energy term $\Psi_\mathrm{fmc}$ from the action-functional \'a priori and use the formulation \cref{eq:fxmg} for all benchmarks}. 
\AS{For the vector potential $\amag$ associated with the total magnetic induction field $\tbmag$, natural Neumann boundary conditions are imposed on the entire boundary $\partial \Vol$, implying vanishing surface currents.}
We now repeat the bending test with $\gamma_\mathrm{iso} \in [-1000,1000]$ and $\gamma_\mathrm{cub} = 0$, followed by $\gamma_\mathrm{cub} \in [-1000,1000]$ and $\gamma_\mathrm{iso} = 0$.
The measured deformations and total magnetic induction are listed and visualised in \cref{fig:gbend}.
As expected, whenever $\gamma_\mathrm{iso} = 0$ and $\gamma_\mathrm{cub} = 0$, we retrieve the purely mechanical solution with no magnetic induction. 
While $\gamma_\mathrm{iso} \neq 0$ or $\gamma_\mathrm{cub} \neq 0$ lead to a reduction in the deformation, it appears $\gamma_\mathrm{iso}$ is more pronounced. In terms of the energy distributions, whenever $\gamma_\mathrm{iso} \neq 0$ more energy is allocated to mechanical deformations, whereas $\gamma_\mathrm{cub}$ seams to yield a higher allocation of energy in the magnetic induction field. We also observe the dependence on the sign of $\gamma_\mathrm{iso}$ and $\gamma_\mathrm{cub}$ is expressed in the orientation of the magnetic induction. Lastly, the orientation of the magnetic induction field produced under $\gamma_\mathrm{cub}$ is more aligned with the $z$-axis, than the one induced by $\gamma_\mathrm{iso}$. Regardless, since cubic-symmetry is not too far off from isotropy, the latter test does not yield the a strong difference, which may be required in order to calibrate $\gamma_\mathrm{iso}$ and $\gamma_\mathrm{cub}$ for a specific material.  

\begin{figure}
    \centering
    \begin{subfigure}{0.32\linewidth}
    \centering
        \input{figs/twigphi}
        \caption{}
    \end{subfigure}
    \begin{subfigure}{0.32\linewidth}
    \centering
        \input{figs/twigb}
        \caption{}
    \end{subfigure}
    \begin{subfigure}{0.32\linewidth}
    \centering
        \includegraphics[width=0.85\linewidth]{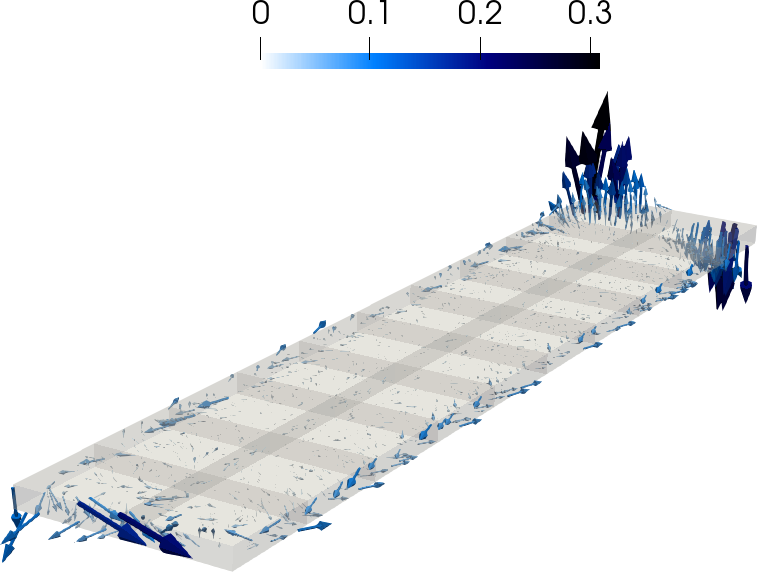}
        \caption{}
    \end{subfigure}
    \begin{subfigure}{0.32\linewidth}
    \centering
        \includegraphics[width=0.85\linewidth]{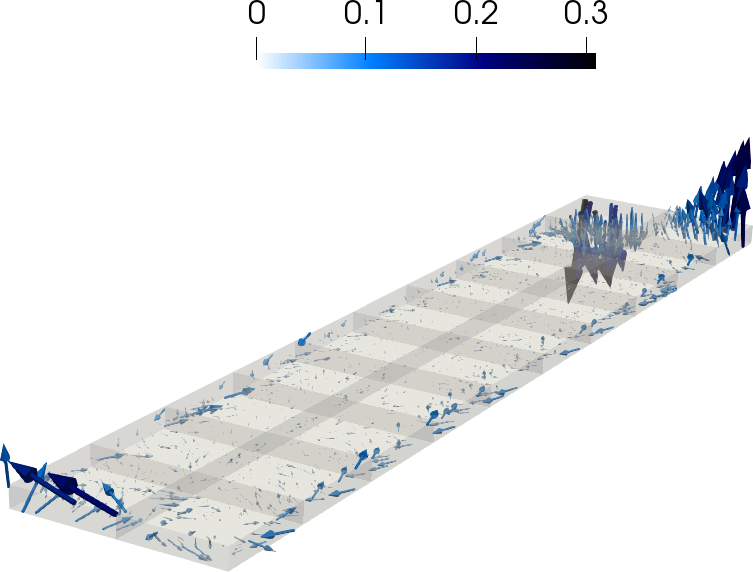}
        \caption{}
    \end{subfigure}
    \begin{subfigure}{0.32\linewidth}
    \centering
        \includegraphics[width=0.85\linewidth]{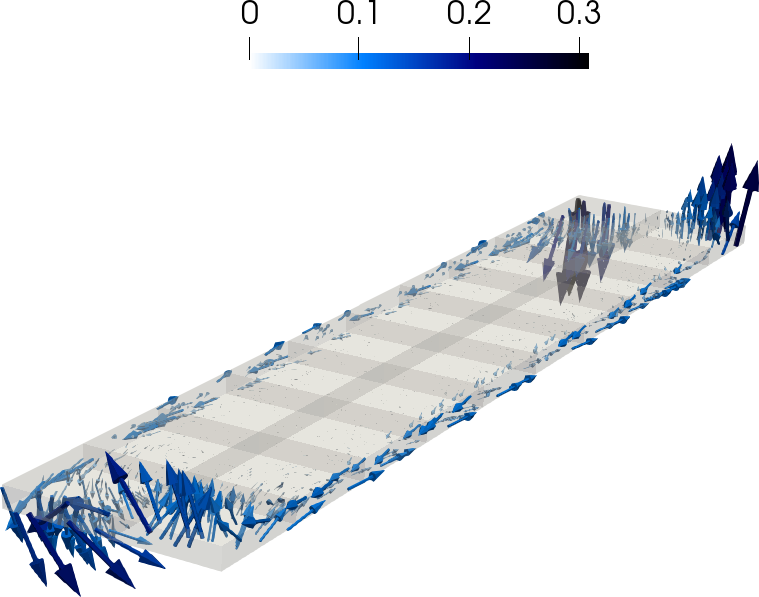}
        \caption{}
    \end{subfigure}
    \begin{subfigure}{0.32\linewidth}
    \centering
        \includegraphics[width=0.85\linewidth]{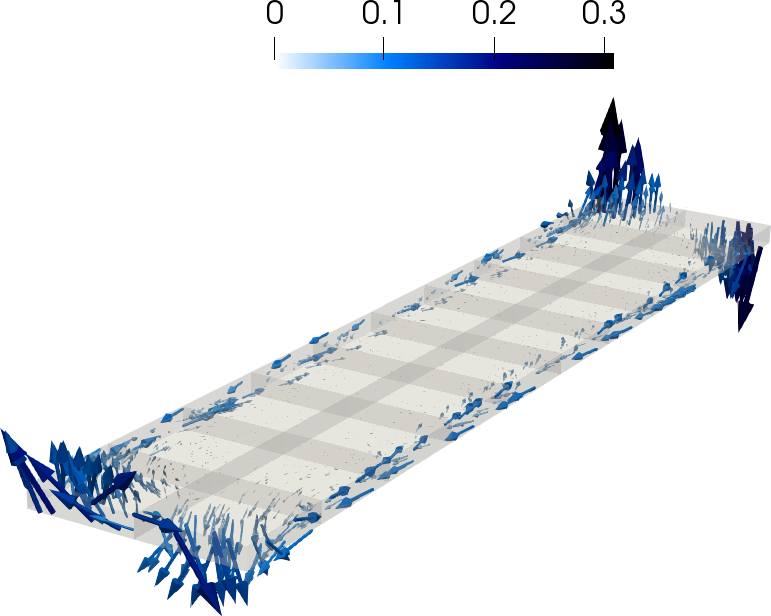}
        \caption{}
    \end{subfigure}
    \caption{Maximal twist deformation for varying $\gamma$ values (a), and the corresponding maximal magnetic induction (b). Total magnetic induction $\tbmag$ on the reference domain for $\gamma_\mathrm{iso} = 1000$ (c), $\gamma_\mathrm{iso} = -1000$ (d), $\gamma_\mathrm{cub} = 1000$ (e), and $\gamma_\mathrm{cub} = -1000$ (f).}
    \label{fig:gtwist}
\end{figure}

In contrast, the twist test does show a clear difference in behaviour, even if actual changes in deformation are very small, as listed and visualised in \cref{fig:gtwist}. In all cases, the magnetic induction concentrates around areas with significant curvature, and its maximal values are practically the same for both $\gamma_\mathrm{iso}$ and $\gamma_\mathrm{cub}$. However, it is more pronounced for $\gamma_\mathrm{cub}$, while yielding virtually no reduction in deformation, as opposed to the influence of $\gamma_\mathrm{iso}$. This trend continues even for $\gamma_\mathrm{iso} = \pm 2000$ or $\gamma_\mathrm{cub} = \pm 2000$, which go beyond the graph. Thus, the twist test represents a viable candidate to distinguish between the two parameters in a calibration procedure. 

\begin{figure}
    \centering
    \begin{subfigure}{0.32\linewidth}
    \centering
        \begin{tikzpicture}[spy using outlines={magnification = 3, circle, size=1.65cm, black, dotted, connect spies}]
        \node [] { \includegraphics[width=0.85\linewidth]{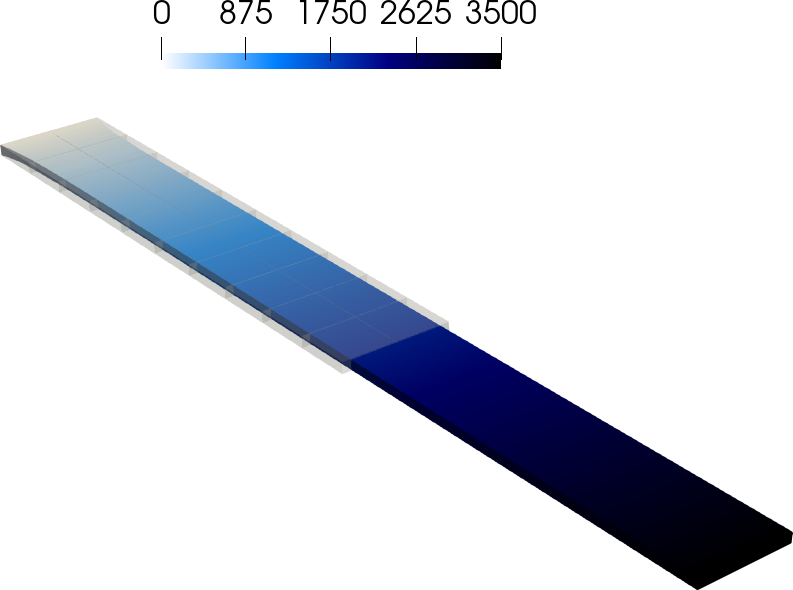} };
        \spy on (-2.1,0.8) in node at (-1.25,-1.05);   
        \end{tikzpicture}
        \caption{}
    \end{subfigure}
    \begin{subfigure}{0.32\linewidth}
    \centering
        \includegraphics[width=0.85\linewidth]{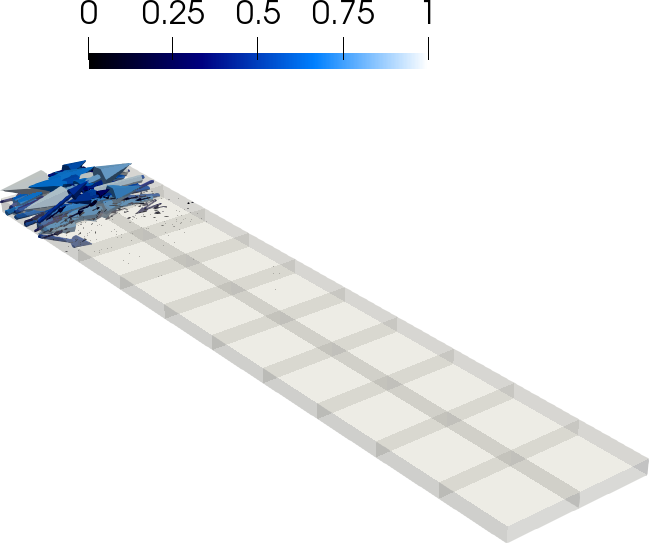}
        \caption{}
    \end{subfigure}
    \begin{subfigure}{0.32\linewidth}
    \centering
        \includegraphics[width=0.85\linewidth]{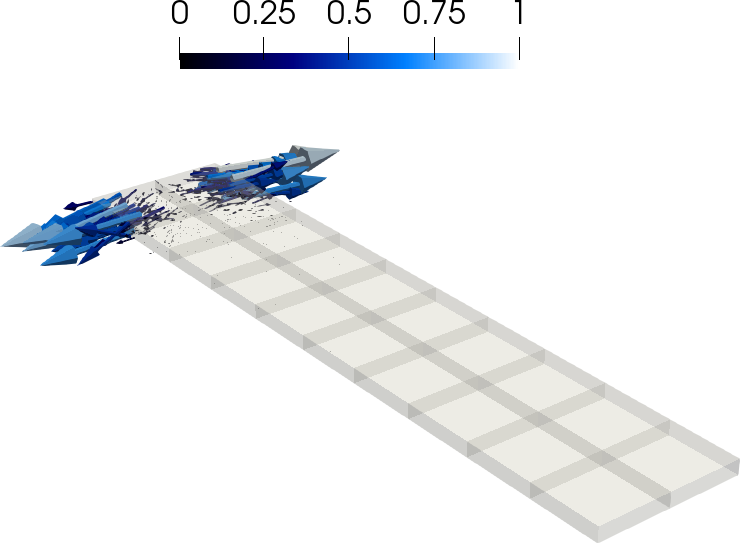}
        \caption{}
    \end{subfigure}
    \begin{subfigure}{0.32\linewidth}
    \centering
        \includegraphics[width=0.85\linewidth]{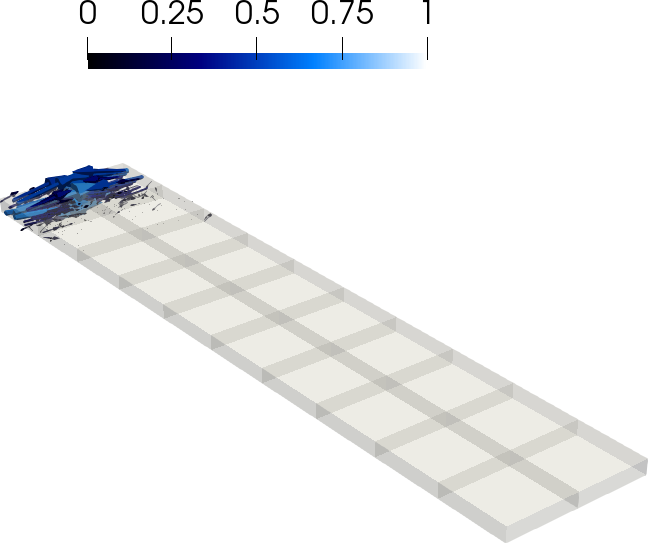}
        \caption{}
    \end{subfigure}
    \begin{subfigure}{0.32\linewidth}
    \centering
        \includegraphics[width=0.85\linewidth]{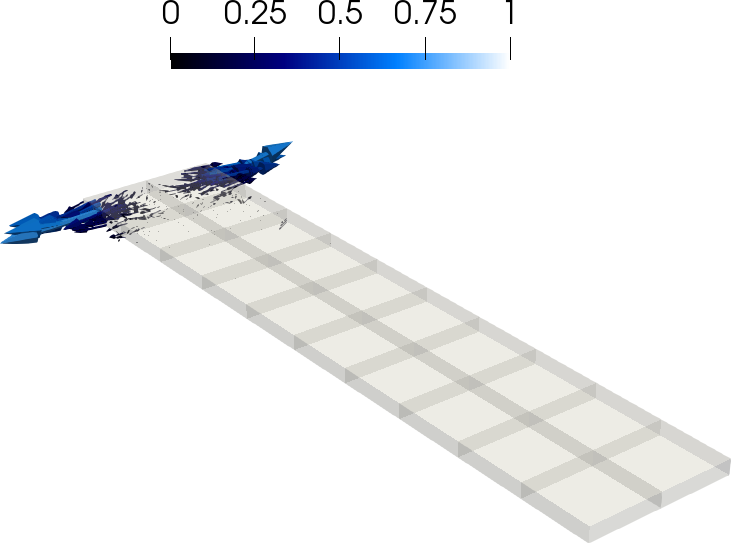}
        \caption{}
    \end{subfigure}
    \caption{Extension deformation for $\gamma_\mathrm{iso} = \pm 1000$ and $\gamma_\mathrm{cub} = \pm 1000$ (a). Total magnetic induction $\tbmag$ on the reference domain for $\gamma_\mathrm{iso} = 1000$ (b), $\gamma_\mathrm{iso} = -1000$ (c), $\gamma_\mathrm{cub} = 1000$ (d), and $\gamma_\mathrm{cub} = -1000$ (e).}
    \label{fig:gext}
\end{figure}

\AS{Unlike prevalent models of the analogous flexoelectric phenomenon \cite{Codony}, our model assumes that a magnetic dipole cannot be treated as a deformable pair of independently displaced charges. Rather, it is represented as a point-wise, indivisible axial dipole whose orientation is not altered by purely axial strain gradients. Consequently, non-uniform uniaxial deformation alone is insufficient to reorient magnetic dipoles and cannot, by itself, induce a flexomagnetic response}. This is verified in \cref{fig:gext}, where the traction vector is set to $\vb{t} = 200\vb{e}_1$. Since the cross‐section at the Dirichlet boundary remains unchanged while the cross‐section under the applied traction contracts, the resulting difference in surface size and shape induces both a stress and strain gradient.
Nevertheless, and despite the substantial resulting extension of $\approx 70 \%$, any magnetic induction is produced solely in the proximity of the support, primarily near the outer-surface of the beam where curvature arises due to transverse contraction.

\begin{figure}
    \centering
    \begin{subfigure}{0.32\linewidth}
    \centering
        \includegraphics[width=0.85\linewidth]{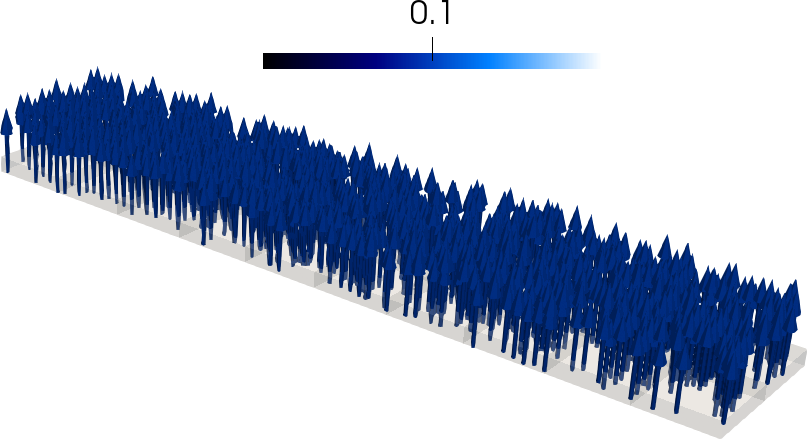}
        \caption{}
    \end{subfigure}
    \begin{subfigure}{0.32\linewidth}
    \centering
        \includegraphics[width=0.85\linewidth]{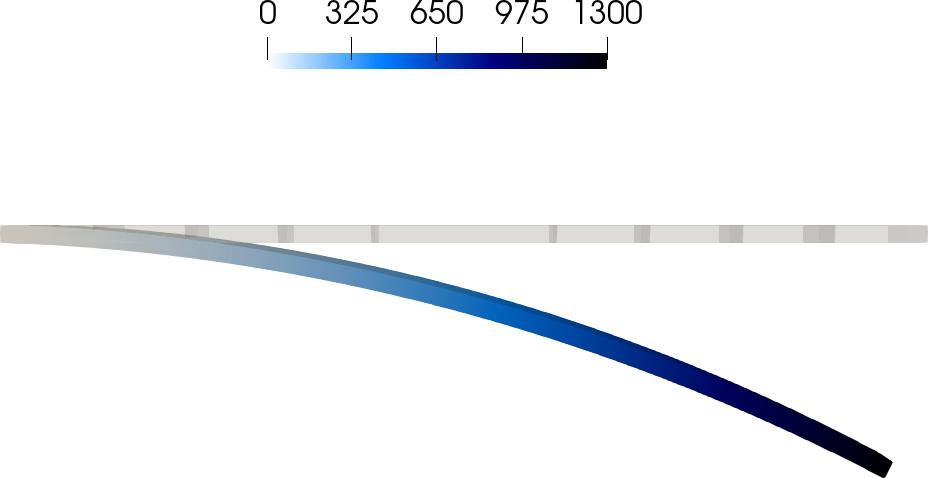}
        \caption{}
    \end{subfigure}
    \begin{subfigure}{0.32\linewidth}
    \centering
        \includegraphics[width=0.85\linewidth]{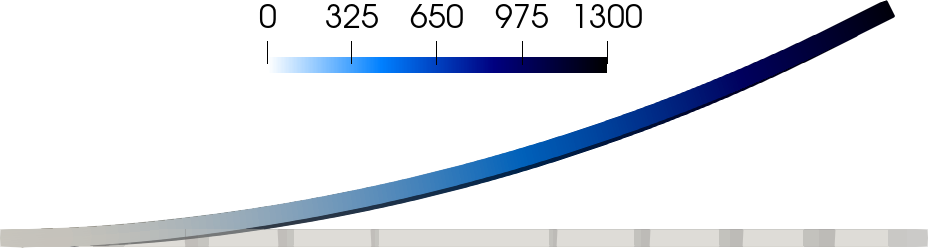}
        \caption{}
    \end{subfigure}
    \begin{subfigure}{0.32\linewidth}
    \centering
        \includegraphics[width=0.85\linewidth]{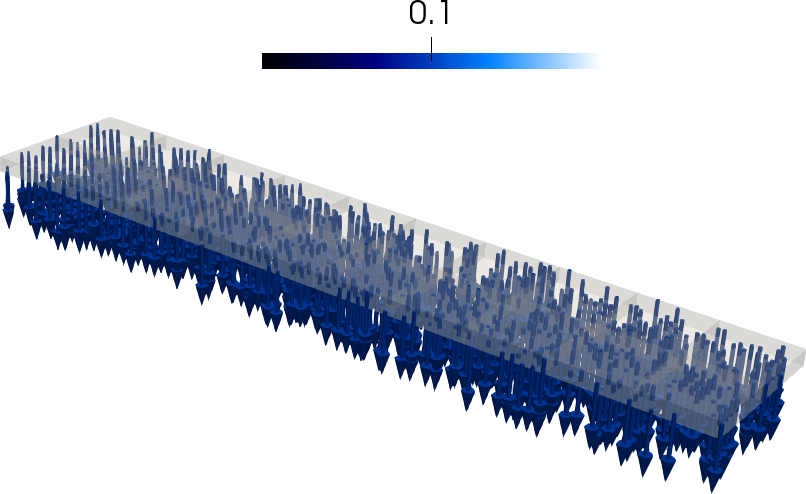}
        \caption{}
    \end{subfigure}
    \begin{subfigure}{0.32\linewidth}
    \centering
        \includegraphics[width=0.85\linewidth]{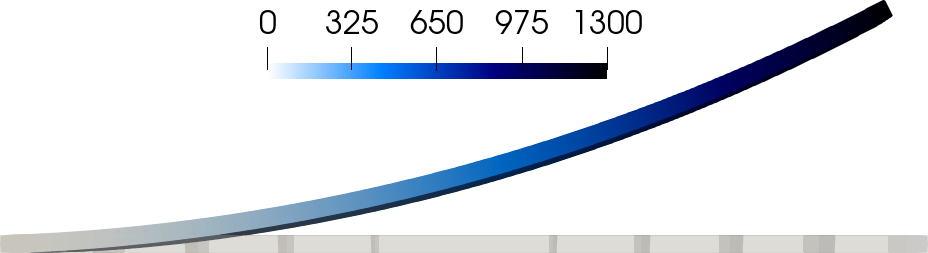}
        \caption{}
    \end{subfigure}
    \begin{subfigure}{0.32\linewidth}
    \centering
        \includegraphics[width=0.85\linewidth]{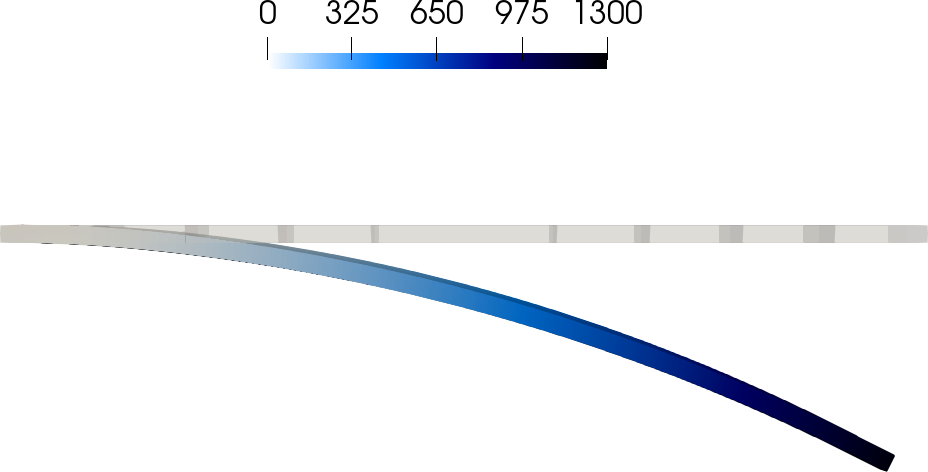}
        \caption{}
    \end{subfigure}
    \caption{Prescribed magnetic induction $\tbmag = 0.1\vb{e}_3$ (a) and corresponding deformation for $\gamma_\mathrm{iso} = 1000$ (b) and $\gamma_\mathrm{cub} = 1000$ (c).
    Prescribed magnetic induction $\tbmag = -0.1\vb{e}_3$ (d) and corresponding deformation for $\gamma_\mathrm{iso} = 1000$ (e) and $\gamma_\mathrm{cub} = 1000$ (f).}
    \label{fig:preb}
\end{figure}

Flexomagnetism is a direction-dependent effect. As such, changes in the direction of the magnetic induction field are expected to flip the direction of the deformation and vice versa. The ability of our model to capture this behaviour is exemplified in \cref{fig:preb}, where we drop the traction vector, and prescribe the magnetic induction field to $\tbmag = \pm 0.1\vb{e}_3$ throughout the domain. Observably, changing the sign of $\tbmag$ without changing the sign of $\gamma_\mathrm{iso}$ or $\gamma_\mathrm{cub}$ changes the direction of the deformation.

\begin{figure}
    \centering
    \begin{subfigure}{0.64\linewidth}
    \centering
        \input{figs/nanobeamtop}
        \caption{}
    \end{subfigure}
    \begin{subfigure}{0.32\linewidth}
    \centering
        \includegraphics[width=0.85\linewidth]{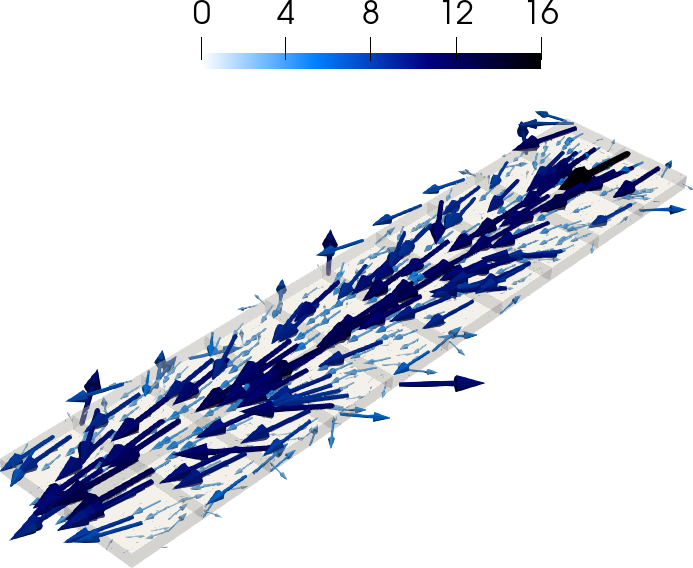}
        \caption{}
    \end{subfigure}
    \begin{subfigure}{0.32\linewidth}
    \centering
        \includegraphics[width=0.85\linewidth]{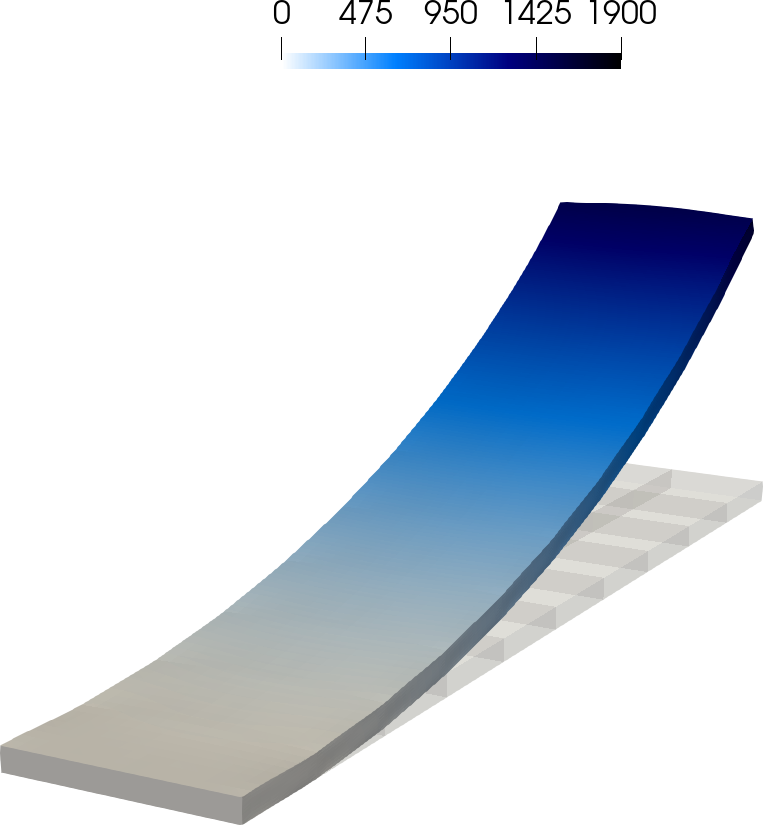}
        \caption{}
    \end{subfigure}
    \begin{subfigure}{0.32\linewidth}
    \centering
        \includegraphics[width=0.85\linewidth]{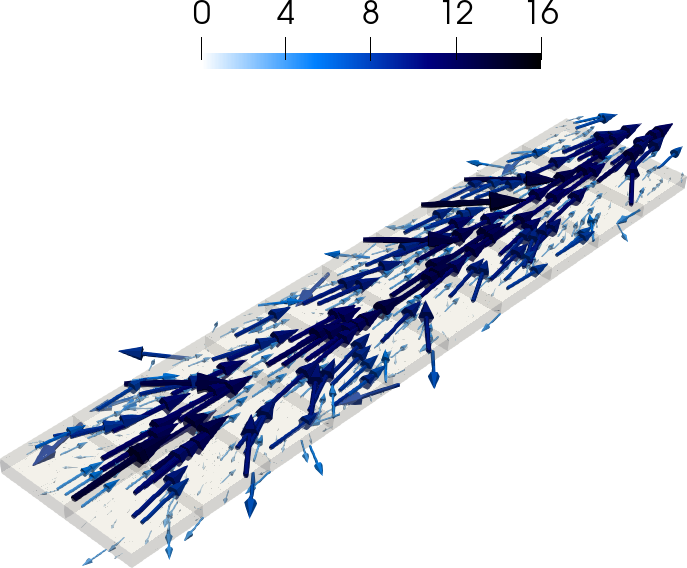}
        \caption{}
    \end{subfigure}
    \begin{subfigure}{0.32\linewidth}
    \centering
        \includegraphics[width=0.85\linewidth]{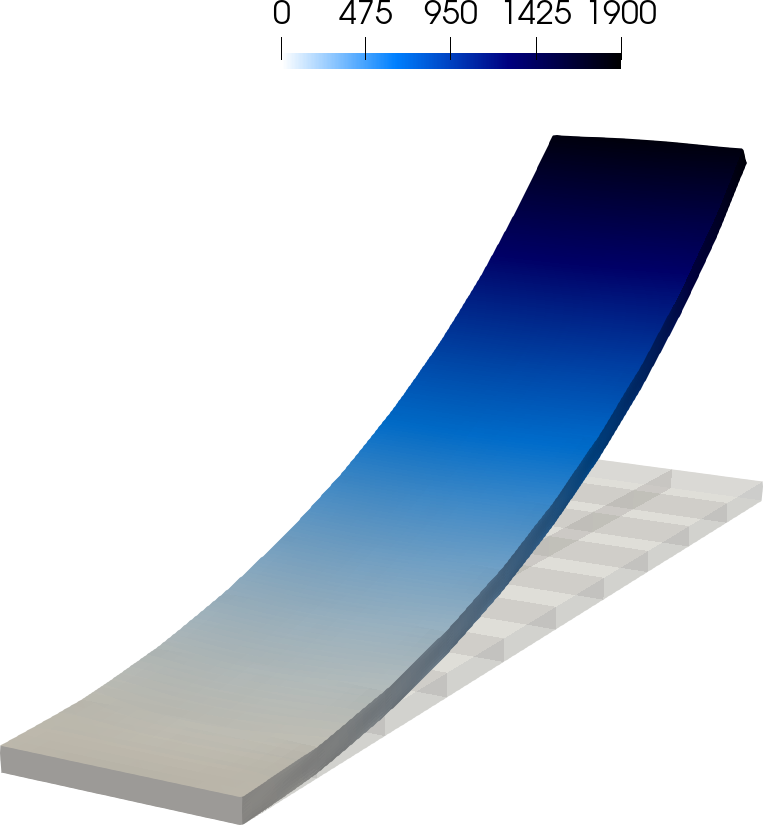}
        \caption{}
    \end{subfigure}
    \caption{The domain $\Vol \subset \R^3$ where impressed current are prescribed on the top outer surface $\Surf_{N_k}^{\tbmag}$ (a). Resulting total induction field (b) for and corresponding deformation (c) for $\con{\vb{k}}{\vb{e}_2} = 1$. Total induction field (d) and deformation (e) for $\con{\vb{k}}{\vb{e}_2} = -1$.}
    \label{fig:jcur}
\end{figure}

Finally, the magnetic induction itself can also be produced by an external stimulus. Thus, on the top surface of the beam we prescribe the impressed surface current $\vb{k} = \pm 1 \vb{e}_2$ in milliampere in the direction of the $y$-axis. The resulting magnetic induction governed by $\gamma_\mathrm{cub} = 1000$ is visualised in \cref{fig:jcur}, and is mostly parallel to the $x$-axis. Irrespective of the sign of $\vb{k}$, the deformation is upwards for an impressed current on the top surface. However, they differ in magnitude. Presumably, as indicated by the previous benchmarks, the deformation is largely due to any projection of the magnetic induction in $z$-direction $\con{\tbmag}{\vb{e}_3}$, which appears to depend more on the physical location of the current in this example.

\subsection{\AS{Comparison with infinitesimal strain-gradient flexomagnetism}}
\AS{
To the best of our knowledge, the present work provides the first fully three-dimensional finite-strain continuum formulation of flexomagnetism. Existing models are either geometrically linear or restricted to reduced beam and plate theories, leaving no directly comparable finite-strain three-dimensional framework. Comparisons with existing work are therefore necessarily limited to qualitative trends or infinitesimal deformation responses, rather than constitutive or kinematic equivalence.
Nevertheless, to position the proposed formulation relative to existing models, we qualitatively compare it with the linear infinitesimal strain-gradient flexomagnetic model of \cite{Sidhardh2018}. Its energy functional for centrosymmetric purely flexomagnetic materials is given in the form
\begin{align}
    \Energy(\disp,\hmag) = \dfrac{1}{2}\int_\Vol \norm{\sym\Grad \disp}_{\Cm}^2 + \mu_\mathrm{e} \Lc^2\norm{\Grad (\sym \Grad \disp)}_{\widetilde{\Lm}}^2 - 2 \con{\hmag}{\widetilde{\Fm} \Grad (\sym \Grad \disp)} - \mu\norm{\hmag}^2 \, \dd \Vol \, , 
\end{align}
where $\Grad (\sym \Grad \disp): \Vol \to \R^{3 \times 3 \times 3}$ is the strain gradient with the corresponding material tensor $\widetilde{\Lm} \in \R^{3 \times 3 \times 3 \times 3 \times 3 \times 3}$ and coupling tensor $\widetilde{\Fm}\in\R^{3 \times 3 \times 3 \times 3}$, such that $\widetilde{\Lm}\Grad (\sym \Grad \disp) : \Vol \to \R^{3 \times 3 \times 3}$ and $\widetilde{\Fm}\Grad (\sym \Grad \disp) : \Vol \to \R^3$.
The tensor $\widetilde{\Lm}$ is simplified here to the identity map. For some arbitrary third order tensor $\mathbbm{T} \in \R^{3 \times 3\times 3}$, the flexomagnetic material tensor acts in the following manner
\begin{align}
    \widetilde{\Fm} \mathbbm{T} = \widetilde{f}_1 T_{ijj} \vb{e}_i + \widetilde{f}_2 T_{jij} \vb{e}_i + \widetilde{f}_3 T_{jji} \vb{e}_i \, ,
\end{align}
mapping to a vector in $\R^3$.
For cantilever deflection in the $z$-direction the dominant material coefficient is $\widetilde{f}_1$, which we vary in this example between $[10^{-4},10^{-1}]$ following the range from \cite{Sidhardh2018}. The two remaining coefficients are set to one $\widetilde{f}_2 = \widetilde{f}_3 = 1$. The infinitesimal flexomagnetic model is computed by defining the independent variable $\bm{\varepsilon} \in \CG^4(\Vol) \otimes \Sym(3)$ and incorporating the penalty term $p\norm{\sym \Grad \disp - \bm{\varepsilon}}^2$ with $p \to +\infty$ for the displacement field $\disp \in \CG^3(\Vol) \otimes \R^3$. Following the same domain and material coefficients as in the previous examples but explicitly with $\chi_\mathrm{m} = 10^{-2}$ using \cref{eq:flexoh}, we study the converse flexomagnetic effect under an imposed constant-valued magnetic field $\hmag = 10^{-2}\vb{e}_3$. The most directly comparable material parameter is $\eta_\mathrm{iso}$, which we vary in the same range as $\widetilde{f}_3$.
\begin{figure}
    \centering
    \begin{subfigure}{0.32\linewidth}
    \centering
        \input{figs/cvss}
        \caption{}
    \end{subfigure}
    \begin{subfigure}{0.32\linewidth}
    \centering
        \includegraphics[width=0.85\linewidth]{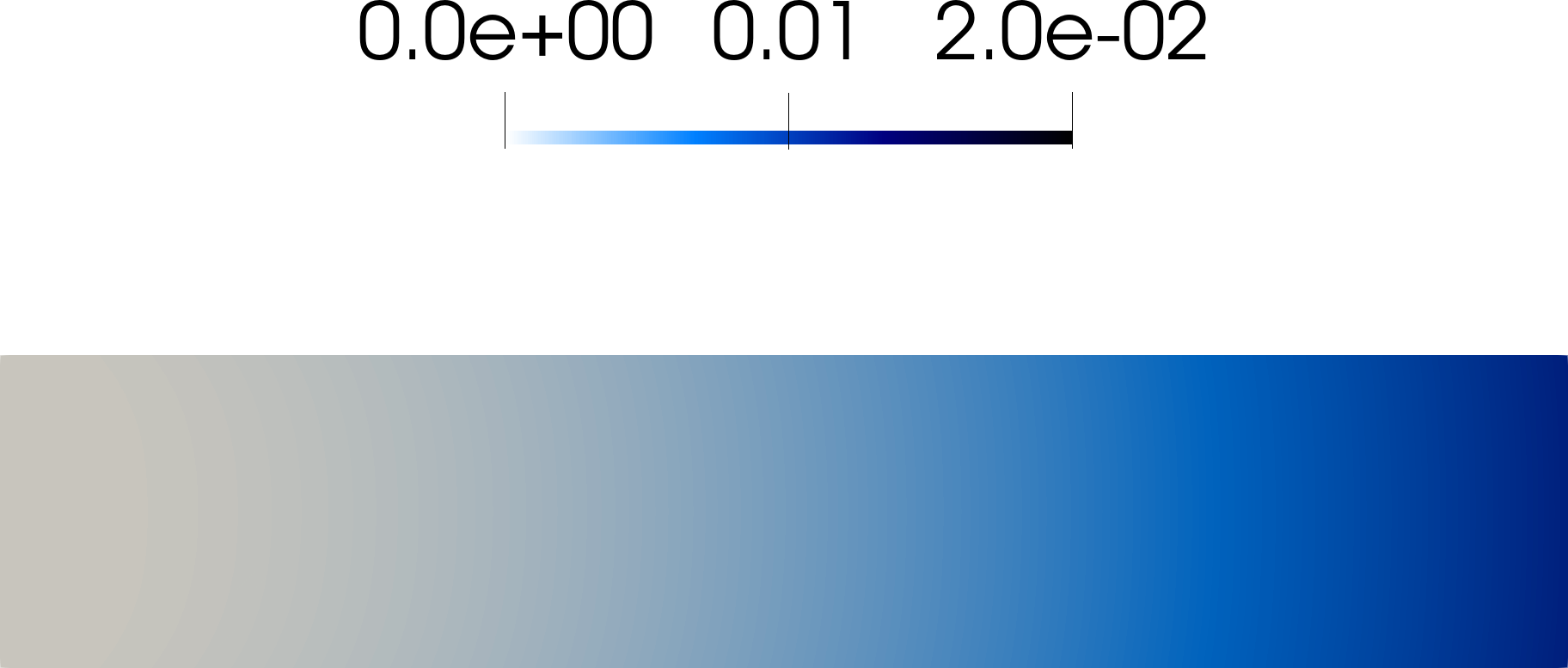}

        \vspace{3\baselineskip}
        
        \caption{}
    \end{subfigure}
    \begin{subfigure}{0.32\linewidth}
    \centering
        \includegraphics[width=0.85\linewidth]{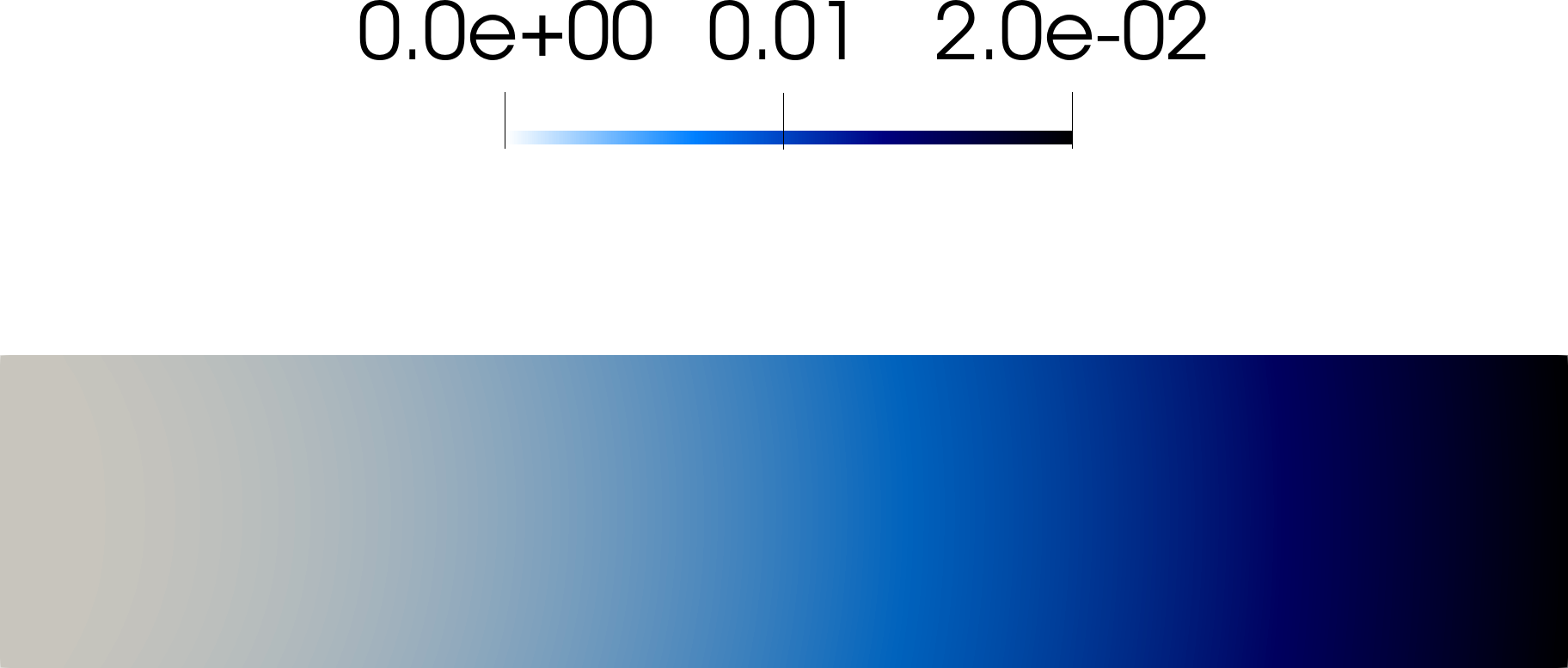}
        \vspace{3\baselineskip}
        
        \caption{}
    \end{subfigure}
    \caption{Comparison of the maximum displacement $\max\norm{\disp}$ as a function of the flexomagnetic parameters governing transverse deflection (a). Top view of the displacement field for the infinitesimal kinematics formulation (b) and the proposed finite-strain model (c).}
    \label{fig:cvss}
\end{figure}
The results in \cref{fig:cvss} show that, within the considered parameter range and small-strain regime, both models exhibit practically identical linear behaviour with a factor of two in-between. Accordingly, at the largest value, $\eta_\mathrm{iso} = \widetilde{f}_1 = 10^{-1}$
, the infinitesimal model predicts $\max\norm{\disp} = 0.011$, whereas the proposed model gives $\max\norm{\disp} = 0.02$. Comparisons far beyond this regime are not meaningful, since the proposed formulation is finite-strain, whereas the reference model is based on infinitesimal kinematics.

}

\section{Conclusions and outlook}
In this work we introduced a geometrically nonlinear (finite) phenomenological model for flexomagnetism based on the Cosserat micropolar theory, formulated in dislocation form. The descended couple-stress model is naturally derived from it by taking the limit of $\muc \to + \infty$, implying that no dislocations can occur.
The system can be formulated in terms of the magnetic field $\hmag$ as a min-max problem, or the total magnetic induction $\tbmag$ as a min-min problem. Specifically the latter is convenient for cases where the total permeability of the material approaches that of free space $\mu \approx \mu_0$, as only positive quadratic terms remain in the energy functional. In terms of simplicity of calibration, the couple-stress model entails the lowest amount of new material coefficients, such that under strict isotropy only the characteristic length-scale parameter $\Lc$ and the flexomagnetic coefficient $\gamma_\mathrm{iso}$ must be characterised. However, its computation requires $\C^1(\Vol)$-continuity and was therefore not considered here. As tested, for $\muc/\mu_\mathrm{e} \geq 1$ we do not observe much change when increasing the Cosserat couple modulus $\muc$ further, such that it can serve as a substitute. However, it is noteworthy that dislocations may be non-negligible at such small scales, such that a calibration of $\muc$ may prove necessary as well.
The benchmarks we pursued revealed some differences in the observed kinematics and resulting magnetic induction, in dependence of assumed isotropy or non-centrosymmetric cubic crystal symmetry. Nonetheless, these differences were subtle, such that a fitting or an optimisation algorithm may prove necessary to order to calibrate a single material. The numerical benchmarks verify that the model behaves consistently for the phenomena of flexomagnetism itself, being two-way and direction-dependent. \AS{The benchmarks further demonstrate the modelling assumption that non-uniform extension alone does not induce magnetic induction in the proposed model, whose response is instead governed purely by micro-dislocation or curvature effects}. The latter is in contrast to the prevalent models of flexoelectricity.

The aim of this work was to propose a phenomenologically consistent flexomagnetic continuum model for which material parameters can be meaningfully calibrated, thereby representing a step toward making flexomagnetism accessible for predictive simulations and future design applications. The model is motivated by the contrast to flexoelectricity, and is proposed on the foundation of the mathematical principles of symmetry and invariance. Crucially, both the proposed continuum formulation and the underlying physical phenomena remain relatively recent, such that corresponding material parameters are not yet available in the literature. Consequently, future work must focus on establishing calibration strategies based on \AS{inverse identification} via dedicated \AS{nanoscale} experiments or on estimates derived from first-principle approaches such as density functional theory in order to populate the model with physically meaningful parameters.
\AS{In particular, DFT-based computations of curvature-induced magnetisation in bent nanostructures could provide effective estimates for the isotropic and cubic-symmetric coupling coefficients introduced here.}
In this regard, the benchmarks presented in this work provide first indications of how the material parameters may be observed and isolated.



\section*{Acknowledgements}
Adam Sky acknowledges support by the European Commission within the \textit{Marie Skłodowska-Curie Actions} postdoctoral fellowship programme, project AGE2M, grant 101152236.
David Codony acknowledges support in the framework of the project PID2023-152533OB-I00, funded by MICIU/AEI/10.13039/501100011033 and FEDER, UE.
Patrizio Neff acknowledges support in the framework of the DFG-Priority Programme
2256 “Variational Methods for Predicting Complex Phenomena in Engineering Structures and Materials”, Neff 902/10-1, Project-No. 440935806. 

\bibliographystyle{spmpsci}   

\footnotesize
\bibliography{ref}   

\normalsize
\appendix

\section{Derivations and formulae}
In the following appendices we list helpful formulae and derivations.  

\subsection{Material tensors} \label{ap:fourth}
One can define three isotropic fourth order tensors 
\begin{align}
    &\mathbb{J} = \vb{e}_i \otimes \vb{e}_j \otimes \vb{e}_i \otimes \vb{e}_j \, , &&
    \mathbb{T} = \vb{e}_i \otimes \vb{e}_j \otimes \vb{e}_j \otimes \vb{e}_i  \, , && \mathbb{V} = \dfrac{1}{3} \one \otimes \one = \vb{e}_i \otimes \vb{e}_i \otimes \vb{e}_j \otimes \vb{e}_j  \, .
\end{align}
The first tensor is the fourth-order identity, the second yields a transposition, and the third returns the spherical part of a second-order tensor
\begin{align}
    \forall\, \bm{T} \in \R^{3 \times 3}: \quad \J \bm{T} = \id\bm{T} = \bm{T} \, , \quad \mathbb{T}\bm{T} = \bm{T}^T \, , \quad \mathbb{V} \bm{T} = \sph \bm{T} = \dfrac{1}{3}(\tr \bm{T}) \one \, .
\end{align}
From combinations of those tensors it is possible to construct three alternative isotropic fourth-order tensors
\begin{align}
    &\mathbb{D} = \J - \mathbb{V} \, , && \mathbb{S} = \dfrac{1}{2}(\J + \mathbb{T}) \, , && \mathbb{A} = \dfrac{1}{2}(\J - \mathbb{T}) \, ,
\end{align}
such that these tensors imply
\begin{align}
    \forall\, \bm{T} \in \R^{3 \times 3}: \quad \mathbb{D} \bm{T} = \dev\bm{T} = \bm{T} - \sph \bm{T} \, , \quad \mathbb{S}\bm{T} = \sym\bm{T} = \dfrac{1}{2}(\bm{T} + \bm{T}^T) \, , \quad \mathbb{A} \bm{T} = \skw \bm{T} = \dfrac{1}{2}(\bm{T} - \bm{T}^T) \, .
\end{align}
The only centrosymmetric third-order (pseudo-) tensor is the permutation tensor
\begin{align}
    \perm = \varepsilon_{ijk} \elec_i \otimes \elec_j \otimes \elec_k \in \R^{3\times3\times3} \, ,
\end{align}
where $\varepsilon_{ijk}$ is the Levi-Civita symbol. The double contraction of the tensor with any arbitrary second-order tensor $\bm{T} \in \R^{3\times3}$ extracts the axial vector of its skew-symmetric part
\begin{align}
    \perm \bm{T} = \varepsilon_{ijk} T_{jk} \elec_i = \perm(\sym\bm{T} + \skw\bm{T}) = -2 \axl (\skw \bm{T}) = -2 \axl\bm{T} \in \R^3    \, .  
\end{align}
Since for any fixed index $i$, the remaining part of the permutation tensor is skew-symmetric, its double contraction with symmetric tensors vanishes.
The tensor related to non-centrosymmetric cubic symmetric materials reads
\begin{align}
    \bm{\mathbbm{C}} = |\varepsilon_{ijk}| \elec_i \otimes \elec_j \otimes \elec_k \in \R^{3\times3\times3} \, .
\end{align}
Its double contraction with a skew-symmetric second-order tensor $\bm{A} \in \so(3)$ is given by
\begin{align}
    \bm{\mathbbm{C}} \bm{A} = |\varepsilon_{ijk}| A_{jk} \elec_i = (A_{23} + A_{32}) \elec_1 + (A_{31} + A_{13}) \elec_2 + (A_{12} + A_{21}) \elec_3 = 0 \, ,  
\end{align}
since $A_{ij} = -A_{ji}$. Thus, $\bm{\mathbbm{C}}$ extracts the symmetric off-diagonal components of an arbitrary tensor $\bm{T} \in \R^{3 \times 3}$
\begin{align}
    \bm{\mathbbm{C}} \bm{T} = \bm{\mathbbm{C}} \sym \bm{T} \in \R^3 \, .
\end{align}
Notably, no cubic-symmetric third-order tensor acts on the diagonal part of $\bm{T}$.

\subsection{Curvature measures} \label{ap:curl}
Let $\Coss:\Vol \to \SO(3)$ be a rotation field, $\vb{v} \in \R^3$ a constant unit vector in the reference configuration, and $\vb{v}_\varphi$ the rotated vector in the current configuration, there holds
\begin{align}
    &\vb{v}_\varphi = \Coss \vb{v} \, , && \vb{v} = \Coss^T \vb{v}_\varphi \, , && \norm{\vb{v}_\varphi} = \norm{\Coss\vb{v}} = \norm{\vb{v}} = 1 \, .
\end{align}
Any change in the orientation of $\vb{v}_\varphi$ is governed by $\Coss$, such that
\begin{align}
    \Grad \vb{v}_\varphi = \vb{v}_{\varphi,i} \otimes \elec_i = \Coss_{,i} \vb{v} \otimes \elec_i \, .
\end{align}
We observe that $\Coss_{,i}$ can be rewritten as
\begin{align}
    \Coss_{,i} = \Coss_{,i} \underbrace{\Coss^T \Coss}_{\one} = (\Coss_{,i} \Coss^T) \Coss = \axl(\Coss_{,i} \Coss^T) \times \Coss \, ,  
\end{align}
since the product $\Coss_{,i} \Coss^T$ is skew-symmetric 
\begin{align}
    &\Coss \, \Coss^T = \one \, , && (\Coss \, \Coss^T)_{,i} = \Coss_{,i} \Coss^T + \Coss \, \Coss^T_{,i} = 0 \quad \Rightarrow \quad \Coss_{,i} \Coss^T = - \Coss \, \Coss^T_{,i} \, ,
\end{align}
and can therefore be fully characterised by an axial vector. Hence, we can reformulate $\Grad \vb{v}_\varphi$ as
\begin{align}
    \Coss_{,i} \vb{v} \otimes \elec_i = \axl(\Coss_{,i} \Coss^T) \times \Coss \vb{v} \otimes \elec_i = \axl(\Coss_{,i} \Coss^T) \times \vb{v}_\varphi \otimes \elec_i =  - \vb{v}_\varphi \times \axl(\Coss_{,i} \Coss^T) \otimes \elec_i \, .
\end{align}
By the fact that $\axl(\Coss_{,i} \Coss^T)$ measures the rate of change in the orientation of $\vb{v}_\varphi$ while acting on $\vb{v}_\varphi$ it is clear that it represents a measure of curvature in the current configuration. To define a curvature measure in the reference configuration we rotate back to find
\begin{align}
    \Coss^T (\Coss_{,i} \Coss^T) \Coss = \Coss^T \Coss_{,i} \, , && \axl(\Coss^T \Coss_{,i}) &= \axl [\Coss^T (\Coss_{,i} \Coss^T) \Coss] = \axl (\Coss^T [\axl(\Coss_{,i} \Coss^T) \times \Coss]) 
    \notag \\
    &&&= \axl ([\Coss^T \axl(\Coss_{,i}\Coss^T)] \times [\Coss^T \Coss] ) = \axl ([\Coss^T \axl(\Coss_{,i}\Coss^T)] \times  \one ) \notag \\
    &&&= \axl (\Anti[\Coss^T \axl(\Coss_{,i}\Coss^T)]) = \Coss^T \axl(\Coss_{,i}\Coss^T) \, ,
\end{align}
such that 
\begin{align}
    \Coss^T[\axl(\Coss_{,i} \Coss^T) \times \vb{v}_\varphi] \otimes \elec_i &= [\Coss^T\axl(\Coss_{,i} \Coss^T)] \times [\Coss^T\vb{v}_\varphi] \otimes \elec_i 
    \notag \\
    &= \axl(\Coss^T\Coss_{,i}) \times \vb{v}\otimes \elec_i = - \vb{v}\times \axl(\Coss^T\Coss_{,i}) \otimes \elec_i \, .
\end{align}
Notably, $\axl (\bm{R}^T \Coss_{,i})$ acts on $\vb{v}$ and thus measures curvature in the reference configuration. The related tensor on the right is called the wryness tensor
\begin{align}
    \Wry = \axl(\Coss^T \Coss_{,i})\otimes \vb{e}_i = -\axl(\Coss_{,i}^T \Coss)\otimes \vb{e}_i \, , && \Wry:\Vol \to \R^{3 \times 3} \, .
\end{align}
It is connected to the gradient of the rotation via
\begin{align}
    \Grad \Coss = \Coss_{,i} \otimes \vb{e}_i = \underbrace{\Coss \Coss^T}_{\one} \Coss_{,i}  \otimes \vb{e}_i = \Coss \Anti \axl(\Coss^T\Coss_{,i}) \otimes \vb{e}_i \quad \Rightarrow \quad \Coss^T \Grad \Coss &= \Anti \axl(\Coss^T\Coss_{,i}) \otimes \vb{e}_i \notag \\
    &= \one \times \Wry \, ,
\end{align}
which is in fact simply the application of the algebraic $\Anti$-operator to the wryness tensor. Consequently, both tensors contain the exact same information.
The relation of the wryness tensor to the micro-dislocation tensor follows from
\begin{align}
\Curl(\underbrace{\Coss^T\Coss}_{\one}) &= \Coss^T\Curl \Coss - \Coss^T_{,i}\Coss (\Anti\vb{e}_i) = \Curva - [\axl(\Coss^T_{,i}\Coss) \times \vb{e}_i \times \vb{e}_j] \otimes \vb{e}_j
    \notag \\
    & = \Curva - \vb{e}_i \otimes \axl(\Coss^T_{,i}\Coss) + \tr[\vb{e}_i \otimes \axl(\Coss^T_{,i}\Coss)]\one = \Curva + \Wry^T - \tr(\Wry^T) \one = 0 \, .
\end{align}
By taking the trace we extract
\begin{align}
    \tr \Curva = 2 \tr (\Wry^T) \quad \Rightarrow \quad \tr (\Wry^T) =  \dfrac{1}{2}\tr\Curva = \dfrac{1}{2}\tr (\Curva^T) \, ,
\end{align}
such that the wryness tensor can be written as
\begin{align}
    \Wry = \dfrac{1}{2}\tr (\Curva^T)\one - \Curva^T \quad \Rightarrow \quad \Curva = \tr(\Wry^T)\one - \Wry^T \, .
\end{align}
Using that the relation between the wryness tensor and the gradient of the rotation can also be written in terms of the Levi-Civita permutation tensor
\begin{align}
    \axl(\Coss^T \Coss_{,i}) = -  \dfrac{1}{2} \perm (\Coss^T \Coss_{,i}) \quad \Rightarrow \quad \Wry = -  \dfrac{1}{2} \perm (\Coss^T \Grad \Coss) \, , 
\end{align}
where we insinuate a double contraction between $\perm$ and $\Coss^T \Grad \Coss$, the final relation between the micro-dislocation tensor and the gradient of the rotation reads
\begin{align}
    \boxed{
    \Curva = \dfrac{1}{2} \perm (\Coss^T \Grad \Coss) -\dfrac{1}{2}\tr[\perm (\Coss^T \Grad \Coss)]\one \, .
    }  
\end{align}

\subsection{Magnetostatics of magnetisable matter}\label{ap:electro}

The strong form of magnetostatics for magnetisable matter with no impressed or surface currents is given by
\begin{align}
    \begin{aligned}
        \curl \vb{h} &= 0 && \text{in} && \Vol \, , 
        \\
        \di \vb{b} &= 0 && \text{in} && \Vol \, , 
        \\
        \vb{b} &= \mu \vb{h} && \text{in} && \Vol \, .
    \end{aligned}
\end{align}
The corresponding energy can be derived using a vector potential $\vb{b} = \curl \vb{a}$ by multiplying the resulting first equation
\begin{align}
    \curl (\mu^{-1}\curl \vb{a}) = 0 \, ,
\end{align} 
with the virtual potential $\delta \vb{a}$, integrating, and applying the variational principle
\begin{align}
    \dfrac{1}{2}\int_\Vol \con{\curl \vb{a}}{\mu^{-1}\curl \vb{a}} \, \dd \Vol = \dfrac{1}{2}\int_\Vol \con{\vb{b}}{\vb{h}} \, \dd \Vol \, . 
\end{align}
From the latter, the energy density as a function of the magnetic induction field reads
\begin{align}
    \Psi_\mathrm{b}(\vb{b}) = \dfrac{1}{2\mu} \norm{\vb{b}}^2 = \dfrac{1}{2}\con{\vb{b}}{\vb{h}} \, . 
\end{align}
Applying the Legendre transform
\begin{align}
    \widetilde{\Psi}_\mathrm{b} = \inf_\vb{b}[\Psi_\mathrm{b}(\vb{b}) - \con{\vb{b}}{\vb{h}}] \, ,
\end{align}
yields the exact same term with an opposite sign, now as the negative enthalpy of the magnetic field 
\begin{align}
    \widetilde{\Psi}_\mathrm{b}(\vb{h}) = -\dfrac{1}{2}\con{\vb{b}}{\vb{h}} = -\dfrac{\mu}{2}\norm{\vb{h}}^2 = -\dfrac{\mu_0}{2}\con{\vb{h}}{\vb{h} + \chi_\mathrm{m}\vb{h}}   \, .
\end{align}
Clearly, the latter can be written as the sum of two enthalpies
\begin{align}
    \widetilde{\Psi}_\mathrm{b}(\vb{h}) = -\Psi_\mathrm{mag}(\vb{h}) 
    -\dfrac{\mu_0}{2}\con{\vb{h}}{\chi_\mathrm{m}\vb{h}} =
    -\Psi_\mathrm{mag}(\vb{h}) 
    -\dfrac{\mu_0}{2}\con{\vb{h}}{\vb{m}} \, , && \Psi_\mathrm{mag}(\vb{h}) = \dfrac{\mu_0}{2}\norm{\vb{h}}^2 \, .
\end{align}
The magnetisation $\vb{m} = \chi_\mathrm{m}\vb{h}$ governs the reaction of the material to magnetic fields and can be thought of as part of a more general constitutive law 
\begin{align}
    \vb{b} = \mu_0(\vb{h} + \vb{m}) \, .
\end{align}
Now, by defining its corresponding demagnetisation energy
\begin{align}
    \Psi_\mathrm{demag}(\vb{m}) = \dfrac{\mu_0}{2\chi_\mathrm{m}} \norm{\vb{m}}^2 \, , 
\end{align}
we immediately observe that the total enthalpy $\widetilde{\Psi}_\mathrm{b}(\vb{h})$ is given by the partial Legendre transform
\begin{align}
    \widetilde{\Psi}_\mathrm{b}(\vb{h}) = \inf_\mathrm{m} [\Psi_\mathrm{demag}(\vb{m}) - \Psi_\mathrm{mag}(\vb{h}) - \mu_0\con{\vb{m}}{\vb{h}}] \, . 
\end{align}
Thus, it is possible to start with the demagnetisation energy to retrieve the true magnetic enthalpy density function.

\subsection{Variation of Cosserat strain measures} \label{ap:variation}

Let $\delta \defmap$ and $\delta \rot$ be admissible virtual displacements and rotations
\begin{align}
    &\delta \defmap = \delta (\vb{x} + \disp) = \delta \disp  \, , && \delta \rot  = -\axl(\Coss \delta \Coss^T) = \axl(\delta \Coss \, \Coss^T) \, ,
\end{align}
the variation of the Biot-type stretch tensor reads
\begin{align}
    \delta \Stretch = \delta (\Coss^T \defgrad) = \delta \Coss^T \defgrad + \Coss^T \delta \defgrad  = \underbrace{\Coss^T \Coss}_{\one} \delta \Coss^T \defgrad + \Coss^T \Grad \delta \defmap= \Coss^T (\underbrace{\Coss \delta \Coss^T}_{-\Anti \delta \rot} \defgrad + \Grad \delta \defmap) = \Coss^T ( \Grad \delta \defmap - \delta \rot \times \defgrad ) \, ,
\end{align}
and the variation of the curvature yields
\begin{align}
    \delta \Curva &= \delta (\Coss^T \Curl \Coss) =  \delta \Coss^T \Curl \Coss + \Coss^T \Curl \delta \Coss = \underbrace{\Coss^T \Coss}_{\one} \delta \Coss^T \Curl \Coss + \Coss^T \Curl (\delta\Coss \underbrace{\Coss^T \Coss}_{\one})
    \notag \\
    &= \Coss^T [ \underbrace{\Coss \delta \Coss^T}_{-\Anti \delta \rot} \Curl \Coss + \Curl( \underbrace{\delta \Coss \, \Coss^T}_{\Anti \delta \rot} \Coss ) ] = \Coss^T[ \Curl(\delta \rot \times \Coss) - \delta \rot \times \Curl \Coss ] 
    \notag \\
    &= \Coss^T[ \underbrace{-\delta \rot_{,i} \times \Coss \times  \vb{e}_i + \delta \rot \times \Curl \Coss}_{\Curl(\delta \rot \times \Coss)} - \delta \rot \times \Curl \Coss] = -\Coss^T[\delta \rot_{,i} \times \Coss \times \vb{e}_i] 
    \notag \\
    &= - [ (\cof \Coss^{T}) \delta \rot_{,i} ] \times [(\cof \Coss^{T} ) \Coss ] \times \vb{e}_i = - (\Coss^T \delta \rot_{,i}) \times \one \times \vb{e}_i = - (\Coss^T \delta \rot_{,i}) \times (\Anti \vb{e}_i)
    \notag \\[1em]
    &= -(\Coss^T \delta \rot_{,i}) \times \vb{e}_i \times \one = -[(\Coss^T \delta \rot_{,i}) \times \vb{e}_i \times \vb{e}_j] \otimes \vb{e}_j = - [ \con{\Coss^T \delta \rot_{,i}}{\vb{e}_j} \vb{e}_i - \con{\Coss^T \delta \rot_{,i}}{\vb{e}_i} \vb{e}_j ] \otimes \vb{e}_j 
    \notag \\[1em]
    &= -  \vb{e}_i \otimes (\Coss^T \delta \rot_{,i}) \underbrace{\vb{e}_j \otimes \vb{e}_j}_{\one}  + \con{\Coss^T \delta \rot_{,i}}{\vb{e}_i} \vb{e}_j \otimes \vb{e}_j 
    =\con{\Coss^T \delta \rot_{,i}}{\vb{e}_i} \one  -  \underbrace{\vb{e}_i \otimes  \delta \rot_{,i}}_{(\Grad \delta \rot)^T} (\Coss \underbrace{\vb{e}_j \otimes \vb{e}_j}_{\one})  
    \notag \\
    & =\con{\Coss^T}{\underbrace{\vb{e}_i \otimes \delta \rot_{,i}}_{(\Grad \delta \rot)^T}} \one - (\Grad \delta \rot)^T \Coss 
     =\underbrace{\con{\Coss}{\Grad \delta \rot}}_{ \tr[(\Grad \delta \rot)^T \Coss] } \one - (\Grad \delta \rot)^T \Coss 
    \, ,
\end{align}
where we used the triple vector product expansion and that $\Anti \vb{e}_i = \vb{e}_i \times \one = \one \times \vb{e}_i$. 

\subsection{Gradients of tensor functions} \label{ap:tensorgrads}

Let $\Stretch:\Vol \to \in \GL(3)$, then there holds
\begin{align}
    \Stretch \, \Stretch^{-1} = \one \qquad \Rightarrow \qquad (\Stretch \, \Stretch^{-1})_{,i} = \Stretch_{,i}\Stretch^{-1} + \Stretch \, \Stretch^{-1}_{,i} = 0 \qquad \Rightarrow \qquad \Stretch_{,i}^{-1} = -\Stretch^{-1} \Stretch_{,i} \Stretch^{-1} \, ,
\end{align}
for some derivative of $\Stretch$ with respect to one its components.
Consequently, we find 
\begin{align}
    \Grad_{\Stretch} \con{\Stretch^{-T} \elec}{\Stretch^{-T} \elec} &= 2 \con{\Stretch^{-T} \elec}{\elec  \, \pder{\Stretch^{-1}}{\bar{U}_{ij}}} \vb{e}_i \otimes \vb{e}_j = - 2 \con{\Stretch^{-T} \elec}{\elec  \, \Stretch^{-1}\pder{\Stretch}{\bar{U}_{ij}}\Stretch^{-1}} \vb{e}_i \otimes \vb{e}_j 
    \notag \\
    &= - 2 \con{(\Stretch^{-T} \elec) \Stretch^{-T}}{(\Stretch^{-T}\elec)   \pder{\Stretch}{\bar{U}_{ij}}} \vb{e}_i \otimes \vb{e}_j = - 2 \langle {(\Stretch^{-T}\elec) \otimes (\Stretch^{-T} \elec) \Stretch^{-T}},\,\underbrace{{\pder{\Stretch}{\bar{U}_{ij}}} \rangle \vb{e}_i \otimes \vb{e}_j}_{\J}  \notag \\
    &= -2 [(\Stretch^{-T}\elec) \otimes (\Stretch^{-T} \elec) \Stretch^{-T}] \J = -2 (\Stretch^{-T}\elec) \otimes (\Stretch^{-T} \elec) \Stretch^{-T} \, .
\end{align}
Further, for $\det \Stretch$ we first observe that
\begin{align}
    \det (\one + \gamma \, \delta \Stretch) = 1 + \gamma (\tr \delta \Stretch) + \mathcal{O}(\gamma^2)  \, ,
\end{align}
by a Taylor-series expansion. Thus, the directional derivative of the identity tensor reads
\begin{align}
    \partial_{\delta\Stretch} \det \one = \pder{}{\gamma} \det(\one + \gamma\, \delta \Stretch) \at_{\gamma = 0} = \pder{}{\gamma}  [1 + \gamma (\tr \delta\Stretch) + \mathcal{O}(\gamma^2)] \at_{\gamma = 0} = \tr \delta \Stretch \, .
\end{align}
Consequently, there holds
\begin{align}
    \partial_{\delta\Stretch} \det \Stretch & = \pder{}{\gamma} \det(\Stretch + \gamma \, \delta \Stretch) \at_{\gamma = 0} = 
    \pder{}{\gamma} \det[\Stretch(\one + \gamma \, \Stretch^{-1} \delta \Stretch)] \at_{\gamma = 0} = (\det\Stretch) \pder{}{\gamma} \det(\one + \gamma \, \Stretch^{-1} \delta \Stretch) \at_{\gamma = 0} \notag \\
    &= (\det\Stretch) \pder{}{\gamma}  [1 + \gamma \, \tr (\Stretch^{-1} \delta\Stretch) + \mathcal{O}(\gamma^2)] \at_{\gamma = 0} = (\det \Stretch) \underbrace{\tr(\Stretch^{-1} \delta \Stretch)}_{\con{\Stretch^{-1} \delta \Stretch}{\one}} = \underbrace{(\det \Stretch) \langle{\Stretch^{-T}}}_{\cof \Stretch},\,{\delta \Stretch} \rangle \, .
\end{align}
Since the directional derivative can also be written as $\partial_{\delta\Stretch}\det \Stretch = \con{\Grad_{\Stretch} \det \Stretch}{\delta \Stretch}$, we recognise 
\begin{align}
    \Grad_{\Stretch} \det \Stretch = \cof \Stretch = (\det \Stretch) \Stretch^{-T} \, .
\end{align}

\end{document}

%% file: figs/mp.tex
\definecolor{qqqqff}{rgb}{0.,0.,1.}
\definecolor{uuuuuu}{rgb}{0.26666666666666666,0.26666666666666666,0.26666666666666666}
\definecolor{xfqqff}{rgb}{0.4980392156862745,0.,1.}
\definecolor{qqwwzz}{rgb}{0.,0.4,0.6}
\begin{tikzpicture}[scale = 0.35, line cap=round,line join=round,>=triangle 45,x=1.0cm,y=1.0cm]
\draw[line width=0.7pt,color=qqwwzz,fill=qqwwzz,fill opacity=0.05000000074505806] (3.1889826566981583,16.92469983125352)(4.405453779951308,17.376378797359347) -- (4.405454103776551,17.376378908308162) -- (4.405454751427061,17.37637913020579) -- (4.405456046728173,17.37637957400105) -- (4.405458637330771,17.376380461591545) -- (4.4054638185374495,17.37638223677246) -- (4.4054741809567455,17.37638578713399) -- (4.4054949058190935,17.376392887855854) -- (4.405536355638812,17.37640708929478) -- (4.405619255658328,17.376435492153433) -- (4.405785057217618,17.376492297793927) -- (4.406116666416761,17.376605908767353) -- (4.406779909133509,17.37683312948167) -- (4.408106491810578,17.377287565961957) -- (4.410760045896768,17.378196418983194) -- (4.416068707059454,17.380014044101) -- (4.42669222584283,17.3836489612993) -- (4.447963925215868,17.390917388837487) -- (4.490604978981223,17.405448018806858) -- (4.576269768942132,17.434479596930814) -- (4.749066577409425,17.492385774825415) -- (4.878000294859641,17.535018272928113) -- (5.0078784562677185,17.577461231508327) -- (5.1386424161144575,17.61967932436515) -- (5.270233528880657,17.661637225297675) -- (5.402593149047117,17.703299608104988) -- (5.535662631094636,17.74463114658619) -- (5.669383329504014,17.785596514540362) -- (5.8036965987560505,17.826160385766606) -- (5.938543793331544,17.86628743406401) -- (6.073866267711295,17.90594233323166) -- (6.209605376376101,17.945089757068658) -- (6.345702473806763,17.983694379374086) -- (6.482098914484079,18.021720873947043) -- (6.618736052888849,18.059133914586617) -- (6.755555243501874,18.0958981750919) -- (6.89249784080395,18.131978329261987) -- (7.02950519927588,18.167339050895965) -- (7.166518673398459,18.201945013792926) -- (7.30347961765249,18.23576089175197) -- (7.44032938651877,18.268751358572175) -- (7.5770093344781,18.300881088052645) -- (7.713460816011278,18.332114753992467) -- (7.8496251855991055,18.36241703019073) -- (7.985443797722379,18.391752590446533) -- (8.120858006861898,18.42008610855896) -- (8.255809167498464,18.447382258327107) -- (8.390238634112876,18.473605713550064) -- (8.524087761185932,18.498721148026924) -- (8.65729790319843,18.522693235556776) -- (8.789810414631173,18.545486649938717) -- (8.921566649964959,18.567066064971836) -- (9.052507963680586,18.587396154455224) -- (9.182575710258853,18.606441592187974) -- (9.311711244180561,18.624167051969174) -- (9.439855919926512,18.640537207597923) -- (9.566951091977497,18.655516732873306) -- (9.692938114814323,18.66907030159442) -- (9.817758342917788,18.68116258756035) -- (9.941353130768686,18.691758264570193) -- (10.063663832847825,18.700822006423042) -- (10.184631803635996,18.708318486917985) -- (10.304198397614005,18.714212379854118) -- (10.422304969262646,18.718468359030524) -- (10.538892873062721,18.721051098246306) -- (10.65390346349503,18.72192527130055) -- (10.767278095040371,18.721055551992343) -- (10.878958122179544,18.718406614120788) -- (10.988884899393348,18.713943131484967) -- (11.09699978116258,18.707629777883977) -- (11.203244121968044,18.699431227116907) -- (11.307559276290537,18.68931215298285) -- (11.409886598610857,18.677237229280898) -- (11.510167443409806,18.663171129810145) -- (11.608343165168181,18.64707852836968) -- (11.70435511836678,18.628924098758592) -- (11.798144657486407,18.608672514775975) -- (11.978821911411934,18.56173657889253) -- (12.149905762791155,18.505988111112067) -- (12.310927047470464,18.44114450182733) -- (12.461416601296257,18.36692314143105) -- (12.600905260114928,18.28304142031596) -- (12.72892385977287,18.189216728874793) -- (12.84500323611648,18.085166457500286) -- (12.94867422499215,17.97060799658517) -- (13.039467662246278,17.84525873652218) -- (13.116914383725252,17.708836067704052) -- (13.180552289216372,17.561067025657266) -- (13.23024707875549,17.40212622570804) -- (13.266323889850181,17.23281560098485) -- (13.28914219563299,17.053983966618908) -- (13.299061469236456,16.866480137741434) -- (13.299296304324866,16.76974140508937) -- (13.296441183793139,16.671152929483625) -- (13.290541041782841,16.57082081281561) -- (13.281640812435555,16.46885115697671) -- (13.269785429892831,16.365350063858337) -- (13.255019828296259,16.260423635351906) -- (13.237388941787383,16.154177973348773) -- (13.216937704507782,16.046719179740393) -- (13.193711050599031,15.938153356418141) -- (13.167753914202677,15.828586605273415) -- (13.139111229460308,15.718125028197633) -- (13.107827930513487,15.606874727082172) -- (13.073948951503777,15.494941803818456) -- (13.037519226572744,15.382432360297898) -- (12.99858368986196,15.269452498411859) -- (12.957187275512997,15.156108320051779) -- (12.913374917667408,15.04250592710902) -- (12.86719155046677,14.928751421475027) -- (12.818682108052666,14.81495090504118) -- (12.767891524566632,14.701210479698887) -- (12.714864734150257,14.587636247339518) -- (12.659646670945111,14.474334309854527) -- (12.60228226909275,14.361410769135283) -- (12.542816462734745,14.248971727073197) -- (12.481294186012658,14.13712328555966) -- (12.417760373068074,14.02597154648609) -- (12.352259958042549,13.915622611743885) -- (12.284837875077645,13.806182583224427) -- (12.215539058314942,13.697757562819149) -- (12.144408441896008,13.590453652419434) -- (12.071490959962393,13.484376953916673) -- (11.996831546655681,13.37963356920229) -- (11.920475136117428,13.27632960016767) -- (11.842466662489226,13.174571148704239) -- (11.762851059912624,13.07446431670337) -- (11.681673262529177,12.976115206056484) -- (11.59897820448047,12.879629918654956) -- (11.514810819908075,12.78511455639022) -- (11.34223880775846,12.602418014836694) -- (11.164316699212893,12.428874396527085) -- (10.981403967403875,12.265332516592636) -- (10.793857811550627,12.112624148901688) -- (10.698444434452536,12.040484619977697) -- (10.601977816467848,11.971149290913594) -- (10.504487452059482,11.904608554879992) -- (10.406002835690426,11.840852805047469) -- (10.306553461823565,11.779872434586643) -- (10.2061688249219,11.721657836668108) -- (10.104878419448344,11.666199404462468) -- (10.002711739865841,11.613487531140336) -- (9.899698280637349,11.563512609872319) -- (9.795867536225817,11.516265033828986) -- (9.691249001094153,11.471735196180973) -- (9.585872169705343,11.429913490098869) -- (9.479766536522316,11.390790308753267) -- (9.372961596008007,11.354356045314788) -- (9.265486842625382,11.320601092954043) -- (9.157371770837376,11.289515844841603) -- (9.04864587510691,11.26109069414808) -- (8.939338649896971,11.235316034044097) -- (8.829479589670473,11.212182257700228) -- (8.719098188890374,11.191679758287087) -- (8.608223942019603,11.173798928975302) -- (8.49688634352113,11.158530162935435) -- (8.38511488785787,11.1458638533381) -- (8.272939069492796,11.135790393353915) -- (8.160388382888836,11.128300176153473) -- (8.047492322508948,11.12338359490738) -- (7.934280382816044,11.121031042786221) -- (7.820782058273113,11.121232912960615) -- (7.707026843343054,11.123979598601153) -- (7.593044232488846,11.129261492878456) -- (7.478863720173439,11.137068988963115) -- (7.36451480085974,11.147392480025744) -- (7.250026969010714,11.160222359236926) -- (7.135429719089326,11.175549019767274) -- (7.02075254555849,11.193362854787374) -- (6.906024942881163,11.213654257467851) -- (6.791276405520288,11.236413620979299) -- (6.676536427938788,11.261631338492343) -- (6.561834504599662,11.289297803177533) -- (6.44720012996581,11.319403408205517) -- (6.33266279850019,11.35193854674688) -- (6.218252004665746,11.386893611972226) -- (6.103997242925405,11.424258997052156) -- (5.989928007742139,11.464025095157282) -- (5.876073793578892,11.506182299458203) -- (5.7624640948985855,11.550721003125503) -- (5.649128406164183,11.597631599329794) -- (5.536096221838636,11.646904481241691) -- (5.423397036384844,11.698530042031791) -- (5.3110603442658,11.752498674870708) -- (5.1991156399444405,11.808800772929025) -- (5.087592417883687,11.867426729377328) -- (4.976520172546496,11.928366937386265) -- (4.865928398395819,11.991611790126413) -- (4.755846589894612,12.05715168076837) -- (4.646304241505777,12.124977002482751) -- (4.537330847692313,12.195078148440132) -- (4.428955902917113,12.267445511811161) -- (4.321208901643163,12.342069485766402) -- (4.214119338333357,12.418940463476474) -- (4.1077167074507015,12.498048838111984) -- (4.002030503458104,12.579385002843509) -- (3.897090220818522,12.662939350841683) -- (3.792929441673863,12.7486988253616) -- (3.689660644948816,12.836583782428306) -- (3.587467621728166,12.926454392098435) -- (3.4865367289928315,13.018168658866443) -- (3.3870543237230493,13.111584587226957) -- (3.2892067628991697,13.206560181674604) -- (3.193180403502282,13.302953446703953) -- (3.099161602512794,13.400622386809346) -- (3.007336716910885,13.499425006485637) -- (2.917892103677474,13.599219310227227) -- (2.831014119792968,13.699863302528684) -- (2.7468891222379455,13.801214987884691) -- (2.6657034679930973,13.903132370789763) -- (2.5876435140387173,14.005473455738525) -- (2.512895617355497,14.108096247225603) -- (2.4416461349237295,14.210858749745398) -- (2.374081423724732,14.313618967792479) -- (2.3103878407382297,14.416234905861756) -- (2.2507517429449706,14.518564568447232) -- (2.195359487325703,14.62046596004393) -- (2.144397430861062,14.721797085146136) -- (2.0980519305311134,14.822415948248931) -- (2.0565093433169466,14.922180553846374) -- (2.0199560261987415,15.020948906433034) -- (1.9885783361572464,15.118579010503822) -- (1.9625626301728687,15.214928870552967) -- (1.9420952652262429,15.30985649107538) -- (1.9185509863686718,15.494877031517774) -- (1.9194363554284735,15.67250466778728) -- (1.9462422282504974,15.841603435840739) -- (2.0004594604876047,16.00103737173572) -- (2.08293711786348,16.1500087277808) -- (2.192008376022727,16.289045602756232) -- (2.3253900005339574,16.41900093678919) -- (2.480798756965555,16.54072767000696) -- (2.65595141088545,16.655078742537057) -- (2.7502181886435437,16.70975494983395) -- (2.8485647278620263,16.762907094506545) -- (2.9507056242370027,16.814641794071008) -- (3.0563554734645777,16.865065666043165) -- (3.1652288712408563,16.914285327938728) -- (3.277040413261261,16.962407397273864) -- (3.391504695222352,17.00953849156417) -- (3.5083363128195515,17.055785228326158) -- (3.6272498617501014,17.101254225074968) -- (3.74795993770897,17.14605209932722) -- (3.8701811363927163,17.190285468598404) -- (3.993628053497673,17.23406095040457) -- (4.118015284719263,17.277485162261655) -- (4.243057425753591,17.32066472168549) -- (4.368469072296762,17.363706246192237) -- (4.399846273351272,17.37445755360318) -- (4.403768570967372,17.375801411934503) -- (4.404749146296695,17.376137376171755) -- (4.4052394340056935,17.376305358273385) -- (4.405362005935444,17.376347353798224) -- (4.405423291899865,17.3763683515607) -- (4.405438613390629,17.37637360100109) -- (4.40544627413658,17.376376225721515) -- (4.405450104509669,17.37637753808167) -- (4.405452019695645,17.376378194261633) -- (4.405452977288974,17.376378522351843) -- (4.405453456085866,17.376378686396833)(3.188978813471408,16.924698154983275);
\draw[line width=0.7pt,color=qqwwzz,fill=qqwwzz,fill opacity=0.05000000074505806] (40.124959468497565,11.634150484937203)(36.95538678623991,18.796927831523796) -- (36.95538709171959,18.796927823721305) -- (36.95538770267893,18.796927808116248) -- (36.95538892459759,18.796927776905836) -- (36.95539136843484,18.79692771448382) -- (36.95539625610902,18.796927589635047) -- (36.95540603145615,18.79692733991848) -- (36.955425582145445,18.7969268404093) -- (36.95546468350422,18.79692584108674) -- (36.955542886142425,18.79692384122483) -- (36.95569929110149,18.79691983663394) -- (36.95601209974966,18.79691180798434) -- (36.95663771196224,18.79689567281785) -- (36.957888916020785,18.796863091047484) -- (36.960391242418595,18.796796682011507) -- (36.965395566314406,18.796658883992936) -- (36.97540288232599,18.79636338444425) -- (36.995412057781884,18.795692901499862) -- (37.035407546848184,18.79403504181664) -- (37.115298792986984,18.78946008002114) -- (37.27461608039502,18.775339848703716) -- (37.39192507145878,18.760713513286447) -- (37.50876485442521,18.742596186711832) -- (37.62507422030315,18.721049433619953) -- (37.74079196010144,18.696134818650886) -- (37.855856864828944,18.667913906444713) -- (37.97020772549449,18.636448261641505) -- (38.08378333310693,18.60179944888134) -- (38.19652247867511,18.564029032804303) -- (38.308363953207866,18.52319857805046) -- (38.41924654771406,18.479369649259905) -- (38.529109053202525,18.4326038110727) -- (38.63789026068211,18.382962628128933) -- (38.74552896116166,18.330507665068676) -- (38.85196394565002,18.275300486532007) -- (38.957134005156036,18.21740265715901) -- (39.06097793068856,18.156875741589754) -- (39.16343451325642,18.093781304464326) -- (39.264442543868476,18.028180910422794) -- (39.363940813533574,17.960136124105244) -- (39.46186811326055,17.889708510151745) -- (39.558163234058256,17.816959633202387) -- (39.652764966935536,17.741951057897236) -- (39.836643432964195,17.585401070779884) -- (40.01301383941729,17.42055106592031) -- (40.181386514365585,17.24789356043914) -- (40.34127178587985,17.067921071456997) -- (40.49217998203084,16.8811261160945) -- (40.564114654367216,16.78532412512081) -- (40.63362143088931,16.68800121147227) -- (40.700639102605976,16.589218939788946) -- (40.76510646052605,16.48903887471093) -- (40.82696229565837,16.387522580878283) -- (40.886145399011795,16.284731622931098) -- (40.94259456159516,16.180727565509446) -- (40.996248574417315,16.075571973253403) -- (41.047046228487105,15.969326410803053) -- (41.09492631481338,15.862052442798467) -- (41.13982762440498,15.753811633879726) -- (41.18168894827075,15.644665548686909) -- (41.22044907741954,15.53467575186009) -- (41.25604680286018,15.42390380803935) -- (41.28842091560154,15.312411281864765) -- (41.31751020665245,15.200259737976413) -- (41.34325346702176,15.087510741014373) -- (41.36558948771831,14.974225855618721) -- (41.38445705975095,14.860466646429536) -- (41.39979655980645,14.746296185673494) -- (41.4115867237688,14.63181401566403) -- (41.41984572623936,14.51715717515738) -- (41.42459354345354,14.402464415817121) -- (41.425850151646785,14.287874489306844) -- (41.42363552705451,14.173526147290135) -- (41.41796964591214,14.059558141430575) -- (41.4088724844551,13.946109223391757) -- (41.39636401891883,13.83331814483726) -- (41.38046422553876,13.721323657430673) -- (41.361193080550294,13.610264512835585) -- (41.33857056018887,13.500279462715575) -- (41.312616640689924,13.391507258734233) -- (41.28335129828886,13.28408665255515) -- (41.250794509221144,13.178156395841901) -- (41.21496624972217,13.073855240258077) -- (41.17588649602737,12.971321937467273) -- (41.133575224372166,12.870695239133056) -- (41.088052410992006,12.772113896919027) -- (41.0393380321223,12.675716662488766) -- (40.98745206399849,12.581642287505858) -- (40.87429690009143,12.400847190406985) -- (40.74903537161997,12.229888692026663) -- (40.61218389049102,12.068703222793216) -- (40.46425887432485,11.917227194332243) -- (40.305776740741734,11.77539701826935) -- (40.137253907361945,11.643149106230132) -- (39.95920679180575,11.520419869840195) -- (39.77215181169342,11.407145720725138) -- (39.67540775293504,11.35403443252869) -- (39.57660538464523,11.303263070510564) -- (39.47580925902652,11.254823686123958) -- (39.373083928281446,11.208708330822075) -- (39.26849394461253,11.164909056058114) -- (39.16210386022233,11.123417913285273) -- (39.05397822731336,11.084226953956756) -- (38.944181598088164,11.047328229525759) -- (38.83277852474927,11.012713791445487) -- (38.71983355949922,10.980375691169135) -- (38.60541125454054,10.950305980149906) -- (38.48957616207576,10.922496709840999) -- (38.372392834307426,10.896939931695616) -- (38.25392582343807,10.87362769716696) -- (38.13423968167021,10.852552057708223) -- (38.0133989612064,10.833705064772612) -- (37.89146821424916,10.817078769813325) -- (37.76851199300101,10.802665224283562) -- (37.64459484966454,10.790456479636518) -- (37.519781336442236,10.780444587325409) -- (37.39413600553662,10.772621598803422) -- (37.26772340915027,10.766979565523751) -- (37.14060809948568,10.763510538939617) -- (37.01285462874542,10.762206570504198) -- (36.884527549131995,10.763059711670714) -- (36.75569141284795,10.766062013892345) -- (36.62641077209583,10.77120552862231) -- (36.496750179078134,10.778482307313801) -- (36.36677429631263,10.787884210236456) -- (36.23655087580711,10.79939774336971) -- (36.106151101089736,10.813003465641529) -- (35.9756463346464,10.828681625834363) -- (35.84510793896297,10.846412472730638) -- (35.71460727652533,10.866176255112808) -- (35.58421570981936,10.887953221763329) -- (35.45400460133095,10.911723621464631) -- (35.32404531354598,10.937467702999172) -- (35.19440920895032,10.965165715149386) -- (35.06516765002988,10.994797906697727) -- (34.93639199927051,11.026344526426634) -- (34.808153619158105,11.059785823118567) -- (34.680523872178526,11.095102045555954) -- (34.553574120817686,11.13227344252126) -- (34.42737572756147,11.171280262796905) -- (34.302000054895714,11.212102755165361) -- (34.177518465306335,11.254721168409063) -- (34.05400232127923,11.299115751310467) -- (33.93152298530025,11.345266752651987) -- (33.81015181985529,11.393154421216089) -- (33.6899601874302,11.442759005785234) -- (33.5710194505109,11.494060755141845) -- (33.45340097158328,11.547039918068378) -- (33.33717611313317,11.601676743347276) -- (33.22241623764651,11.657951479760982) -- (33.109192707609154,11.715844376091958) -- (32.99757688550699,11.77533568112262) -- (32.88764013382587,11.836405643635445) -- (32.779453815051696,11.899034512412861) -- (32.673089291670365,11.963202536237311) -- (32.56861792616775,12.028889963891253) -- (32.46611108102975,12.09607704415712) -- (32.36564011874219,12.164744025817363) -- (32.267276401790994,12.234871157654439) -- (32.17109129266204,12.306438688450783) -- (31.98554234781439,12.453815942051051) -- (31.809564184086245,12.606717776877751) -- (31.643727701364693,12.764986183190452) -- (31.488603799536772,12.9284631512487) -- (31.344763378489542,13.09699067131207) -- (31.212777338110055,13.270410733640134) -- (31.09321657828539,13.44856532849245) -- (30.98665199890254,13.631296446128587) -- (30.893654499848605,13.818446076808115) -- (30.852421811659678,13.913628456495275) -- (30.814794981010603,14.009856210790602) -- (30.780845370387294,14.107109588476554) -- (30.750644342275617,14.205368838335609) -- (30.724263259161443,14.30461420915016) -- (30.701773483530683,14.404825949702698) -- (30.683246377869192,14.505984308775624) -- (30.66875330466284,14.60806953515143) -- (30.65836562639755,14.711061877612543) -- (30.652154705559195,14.81494158494138) -- (30.650191904633612,14.919688905920452) -- (30.65252963194618,15.025271000524157) -- (30.65911075402616,15.131579384566038) -- (30.66983887383853,15.238478460379639) -- (30.684617544027873,15.345832595549808) -- (30.703350317238943,15.45350615766111) -- (30.725940746116436,15.561363514298336) -- (30.752292383304763,15.669269033045993) -- (30.782308781448677,15.777087081488986) -- (30.81589349319293,15.884682027211767) -- (30.85295007118188,15.991918237799183) -- (30.89338206806022,16.0986600808358) -- (30.937093036472817,16.20477192390635) -- (30.983986529064197,16.31011813459557) -- (31.033966098478828,16.41456308048808) -- (31.086935297361578,16.517971129168558) -- (31.142797678356857,16.62020664822157) -- (31.201456794109475,16.721134005232017) -- (31.26281619726396,16.820617567784524) -- (31.326779440465003,16.91852170346354) -- (31.393250076357134,17.01471077985397) -- (31.462131657585104,17.10904916454055) -- (31.60674186662706,17.291631329140415) -- (31.759838488747647,17.46518313794067) -- (31.920649945104202,17.62861953161874) -- (32.08840465685367,17.780855450851874) -- (32.26233300201312,17.920812217930546) -- (32.44183914204888,18.04797788894561) -- (32.62663527306006,18.162845066961182) -- (32.72093702051339,18.215852860307706) -- (32.81646518810888,18.26600939746052) -- (32.91318774997944,18.313389134104995) -- (33.01107268025771,18.358066525926716) -- (33.110087953076544,18.400116028610825) -- (33.21020154256868,18.439612097842797) -- (33.31138142286693,18.476629189308056) -- (33.41359556810406,18.511241758692023) -- (33.51681195241284,18.54352426168012) -- (33.62099854992607,18.573551153957595) -- (33.72612333477653,18.601396891209987) -- (33.83215428109699,18.627135929122602) -- (33.93905936302025,18.650842723380975) -- (34.046806554678994,18.67259172967036) -- (34.15536383020617,18.692457403675945) -- (34.264699163734406,18.71051420108367) -- (34.37478052939656,18.726836577578723) -- (34.48557590132543,18.74149898884619) -- (34.59705325365371,18.754575890571772) -- (34.709180560514255,18.766141738440666) -- (34.821925796039835,18.776270988138464) -- (34.93525693436317,18.785038095350416) -- (35.04914194961705,18.792517515761944) -- (35.16354881593435,18.798783705058582) -- (35.2784455074478,18.803911118925527) -- (35.39379999829018,18.80797421304831) -- (35.5095802625942,18.811047443112358) -- (35.625754274492664,18.813205264802917) -- (35.742290008118516,18.814522133805298) -- (35.85915543760433,18.81507250580526) -- (35.97631853708302,18.814930836488003) -- (36.09374728068721,18.814171581538716) -- (36.211409642549796,18.812869196643106) -- (36.329273596803574,18.811098137486425) -- (36.44730711758126,18.808932859754094) -- (36.56547817901571,18.80644781913142) -- (36.68375475523962,18.80371747130394) -- (36.80210482038578,18.800816271957075) -- (36.920496348586994,18.797818676776075) -- (36.9500969562084,18.797062941407404) -- (36.95379704871061,18.796968436010985) -- (36.954722071939926,18.796944809420467) -- (36.95518458355966,18.796932996113497) -- (36.95530021146476,18.796930042786272) -- (36.955358025417326,18.796928566122716) -- (36.9553724789055,18.79692819695677) -- (36.95537970564956,18.796928012373797) -- (36.95538331902159,18.796927920082368) -- (36.955385125707636,18.796927873936482) -- (36.95538602905063,18.79692785086374) -- (36.955386480722154,18.796927839327367)(40.124981944560105,11.634166860628271);
\draw[line width=0.7pt,color=qqwwzz,fill=qqwwzz,fill opacity=0.05000000074505806] (20.88572790797199,8.386762508361862)(19.75860999842045,8.08873778854117) -- (19.758610316543518,8.088737922759291) -- (19.75861095278962,8.088738191195437) -- (19.758612225281702,8.088738728067348) -- (19.758614770265375,8.08873980180964) -- (19.758619860230773,8.088741949288107) -- (19.75863004015377,8.088746244220571) -- (19.758650399968545,8.088754833987634) -- (19.758691119473248,8.088772013130287) -- (19.75877255798326,8.088806369849722) -- (19.758935433005778,8.088875077025227) -- (19.75926117506132,8.089012466323815) -- (19.759912627218654,8.089287144719485) -- (19.761215403752345,8.089836100770379) -- (19.76382044596773,8.090932410434885) -- (19.769028489165756,8.093118624210907) -- (19.779436428032785,8.09746546312004) -- (19.800219854866707,8.10605705491899) -- (19.84165801866131,8.122834042858347) -- (19.92402849657612,8.154780423452639) -- (20.08681733225674,8.21238030230148) -- (20.205569055548978,8.249719353442815) -- (20.323038264767103,8.28274279268263) -- (20.439290788101975,8.311577604889374) -- (20.55439245374444,8.336350774931491) -- (20.66840908988536,8.357189287677432) -- (20.781406524715575,8.374220127995642) -- (20.893450586425946,8.387570280754568) -- (21.004607103207327,8.39736673082266) -- (21.114941903250568,8.403736463068359) -- (21.224520814746526,8.40680646236012) -- (21.33340966588605,8.406703713566385) -- (21.441674284859996,8.403555201555603) -- (21.549380499859215,8.39748791119622) -- (21.656594139074564,8.388628827356685) -- (21.76338103069689,8.377104934905445) -- (21.86980700291705,8.363043218710946) -- (21.975937883925894,8.346570663641637) -- (22.08183950191428,8.327814254565963) -- (22.187577685073062,8.306900976352372) -- (22.293218261593086,8.283957813869314) -- (22.39882705966521,8.259111751985232) -- (22.504469907480285,8.232489775568576) -- (22.610212633229168,8.204218869487791) -- (22.716121065102705,8.174426018611326) -- (22.822261031291756,8.143238207807627) -- (22.92869835998717,8.110782421945144) -- (23.035498879379805,8.077185645892321) -- (23.142728417660507,8.042574864517606) -- (23.250452803020135,8.007077062689449) -- (23.358737863649537,7.970819225276292) -- (23.467649427739573,7.933928337146588) -- (23.577253323481095,7.896531383168779) -- (23.687615379064948,7.858755348211316) -- (23.798801422681994,7.820727217142644) -- (23.910877282523078,7.7825739748312115) -- (24.02390878677906,7.7444226061454655) -- (24.137961763640796,7.706400095953853) -- (24.253091452186013,7.668624454227771) -- (24.369202503162686,7.631086058459612) -- (24.486087535191324,7.5936803323151665) -- (24.60353650109693,7.556300440041525) -- (24.721339353704508,7.51883954588577) -- (24.83928604583907,7.481190814094995) -- (24.95716653032562,7.4432474089162834) -- (25.07477075998916,7.4049024945967235) -- (25.1918886876547,7.366049235383403) -- (25.30831026614725,7.326580795523412) -- (25.423825448291804,7.286390339263834) -- (25.538224186913382,7.245371030851761) -- (25.65129643483698,7.203416034534276) -- (25.76283214488761,7.16041851455847) -- (25.872621269890278,7.116271635171429) -- (25.98045376266998,7.070868560620241) -- (26.08611957605174,7.024102455151993) -- (26.18940866286055,6.975866483013773) -- (26.290110975921422,6.92605380845267) -- (26.388016468059362,6.874557595715771) -- (26.482915092099375,6.821271009050161) -- (26.66285154718564,6.7088993709211655) -- (26.828239963780273,6.588084208042385) -- (26.977399964483315,6.4579708343905216) -- (27.108651171894824,6.317704563942275) -- (27.220313208614844,6.166430710674345) -- (27.31096784030302,6.003503381188361) -- (27.38074971509895,5.829513527378454) -- (27.43036315341675,5.645505835713019) -- (27.447827079730025,5.550071980817334) -- (27.460513341924198,5.452525682617223) -- (27.468510088332714,5.352997613165796) -- (27.471905467289066,5.2516184445162235) -- (27.47078762712673,5.148518848721666) -- (27.46524471617915,5.043829497835228) -- (27.45536488277983,4.937681063910086) -- (27.441236275262238,4.8302042189993895) -- (27.422947041959826,4.7215296351562674) -- (27.4005853312061,4.611787984433876) -- (27.3742392913345,4.5011099388853655) -- (27.34399707067853,4.389626170563874) -- (27.309946817571632,4.277467351522569) -- (27.27217668034731,4.164764153814568) -- (27.230774807339014,4.051647249493032) -- (27.185829346880222,3.9382473106110987) -- (27.137428447304398,3.824695009221936) -- (27.085660256945044,3.7111210173786753) -- (27.0306129241356,3.5976560071344346) -- (26.972374597209555,3.4844306505424236) -- (26.911033424500392,3.37157561965576) -- (26.846677554341564,3.2592215865275613) -- (26.77939513506656,3.1474992232110233) -- (26.709274315008837,3.036539201759249) -- (26.63640324250188,2.9264721942254135) -- (26.560870065879154,2.817428872662662) -- (26.48276293347416,2.709539909124127) -- (26.402169993620316,2.6029359756629753) -- (26.31917939465114,2.4977477443323175) -- (26.23387928490011,2.3941058871853422) -- (26.146357812700654,2.292141076275167) -- (26.056703126386278,2.191983983654952) -- (25.965003374290436,2.0937652813778436) -- (25.87134670474664,1.9976156414969872) -- (25.775821266088325,1.903665736065522) -- (25.678515206648964,1.812046237136601) -- (25.579516674762036,1.7228878167633632) -- (25.478913818761054,1.6363211469989687) -- (25.376794786979428,1.5524768998965635) -- (25.273247055535393,1.47148333658037) -- (25.168339603315275,1.3934023770597435) -- (25.062121049185784,1.3182229192608332) -- (24.95463896616524,1.2459301101298408) -- (24.845940927271997,1.1765090966129605) -- (24.736074505524385,1.1099450256563728) -- (24.625087273940743,1.0462230442062719) -- (24.51302680553941,0.9853282992088808) -- (24.399940673338733,0.927245937610337) -- (24.285876450357037,0.8719611063568777) -- (24.170881709612658,0.8194589523946831) -- (24.055004024123946,0.7697246226699406) -- (23.93829096690922,0.7227432641288587) -- (23.82079011098685,0.6785000237176106) -- (23.702549029375163,0.636980048382398) -- (23.583615295092464,0.5981684850694222) -- (23.46403648115713,0.5620504807248636) -- (23.343860160587475,0.5286111822949167) -- (23.223133906401863,0.49783573672578996) -- (23.1019052916186,0.46970929096365666) -- (22.980221889256043,0.4442169919547183) -- (22.858131272332535,0.42134398664516937) -- (22.735681013866404,0.40107542198119006) -- (22.612918686875993,0.38339644490899616) -- (22.489891864379622,0.3682922023747608) -- (22.36664811939567,0.3557478413246855) -- (22.24323502494243,0.34574850870495766) -- (22.119700154038263,0.3382793514617859) -- (21.99609107970152,0.33332551654132203) -- (21.872455374950498,0.33087215088981736) -- (21.74884061280357,0.3309044014534379) -- (21.625294366279057,0.3334074151783639) -- (21.50186420839531,0.33836633901080404) -- (21.378597712170674,0.3457663198969385) -- (21.255542450623444,0.3555925047829902) -- (21.13274599677201,0.36783004061511093) -- (21.010255923634666,0.38246407433952356) -- (20.888119804229795,0.39947975290240123) -- (20.76638521157571,0.4188622232499526) -- (20.645099718690716,0.4405966323283721) -- (20.52431089859322,0.46466812708382577) -- (20.40406632430151,0.4910618544625436) -- (20.284413568833948,0.5197629614106916) -- (20.16540020520886,0.5507565948744855) -- (20.04707380644458,0.5840279018000842) -- (19.929481945559466,0.6195620291337178) -- (19.812672195571835,0.6573441238215523) -- (19.696692129500036,0.6973593328097891) -- (19.581589320362397,0.7395928030446157) -- (19.467411341177275,0.7840296814722478) -- (19.354205764962984,0.8306551150388657) -- (19.24202016473788,0.8794542506906495) -- (19.130902113520293,0.9304122353738222) -- (19.02089918432857,0.9835142160345285) -- (18.91205895018105,1.0387453396190125) -- (18.80442898409604,1.0960907530734403) -- (18.698056859091906,1.1555356033440134) -- (18.592990148187006,1.217065037376905) -- (18.48927642439962,1.2806642021183308) -- (18.386963260748132,1.3463182445144852) -- (18.28609823025088,1.4140123115115415) -- (18.18672890592618,1.4837315500557224) -- (18.08890286079238,1.5554611070931728) -- (17.992667667867813,1.6291861295701437) -- (17.898070900170815,1.7048917644327801) -- (17.71398293253292,1.8621854590998836) -- (17.53701954202537,2.0272233666640602) -- (17.36756131279487,2.199886662694823) -- (17.205988828988147,2.380056522761734) -- (17.0526826747519,2.5676141224343496) -- (16.979248403768544,2.6641261919882453) -- (16.90802343423281,2.762440637282168) -- (16.839055339163046,2.862542605262277) -- (16.772396172244697,2.9644077888301013) -- (16.708142356971848,3.0679182622026815) -- (16.646415658535034,3.1729026295391236) -- (16.587338138749942,3.279188869129314) -- (16.531031859432517,3.386604959263366) -- (16.477618882399042,3.49497887823037) -- (16.42722126946512,3.604138604320667) -- (16.37996108244704,3.713912115823689) -- (16.335960383160568,3.8241273910294353) -- (16.295341233421595,3.9346124082276788) -- (16.258225695046292,4.045195145707851) -- (16.224735829850545,4.155703581760292) -- (16.194993699650297,4.2659656946742075) -- (16.169121366261493,4.375809462739824) -- (16.147240891500132,4.485062864246629) -- (16.129474337182216,4.593553877484567) -- (16.11594376512369,4.701110480743353) -- (16.10677123714038,4.8075606523125884) -- (16.1020788150484,4.912732370482615) -- (16.101988560663756,5.01645361354241) -- (16.106622535802273,5.1185523597823135) -- (16.11610280228001,5.218856587492098) -- (16.130551421912855,5.317194274961196) -- (16.15007255319432,5.41340973316116) -- (16.174618551882702,5.507485758293711) -- (16.23828803311045,5.689430342989368) -- (16.3205627094232,5.863452359628383) -- (16.420444840761718,6.029976671453909) -- (16.536936687066657,6.189428141709044) -- (16.66904050827867,6.342231633636885) -- (16.815758564338637,6.48881201048053) -- (16.976093115186757,6.629594135483302) -- (17.149046420764364,6.765002871888129) -- (17.333620741011202,6.895463082938221) -- (17.42995398786735,6.958970261218866) -- (17.528818335868607,7.02139963187679) -- (17.630089067507754,7.082804302817777) -- (17.73364146527723,7.143237381946875) -- (17.839350811669476,7.202751977169811) -- (17.947092389177158,7.261401196391745) -- (18.056741480292942,7.319238147518121) -- (18.16817336750961,7.3763159384544394) -- (18.281263333319373,7.432687677105861) -- (18.395886660215012,7.488406471377999) -- (18.51191863068908,7.543525429176128) -- (18.62923452723402,7.598097658405692) -- (18.747709632342833,7.6521762669719635) -- (18.867219228507395,7.705814362780615) -- (18.98763859822094,7.759065053736691) -- (19.108843023975794,7.811981447745808) -- (19.230707788264283,7.864616652713295) -- (19.353108173579642,7.917023776544596) -- (19.475919462413515,7.969255927144985) -- (19.599016937259705,8.02136621242002) -- (19.722275880609743,8.073407740274973) -- (19.75310123056812,8.086413602474437) -- (19.75695446376136,8.088039307805587) -- (19.7579177724636,8.088445733966267) -- (19.75839942683473,8.088648947038251) -- (19.758519840428164,8.088699750305835) -- (19.758580047224825,8.088725151939713) -- (19.758595098924218,8.088731502348168) -- (19.758602624773914,8.088734677552367) -- (19.75860638769865,8.088736265154495) -- (19.75860826916113,8.088737058955559) -- (19.758609209892313,8.08873745585609) -- (19.758609680257848,8.088737654306357)(20.885762128537593,8.386766124336134);
\draw [line width=0.7pt] (8.,15.)-- (8.,17.);
\draw [line width=0.7pt] (7.,16.)-- (9.,16.);
\draw [line width=0.7pt] (21.015716964635775,3.913406294724865)-- (21.015716964635775,5.913406294724865);
\draw [line width=0.7pt] (20.015716964635775,4.913406294724865)-- (22.015716964635775,4.913406294724865);
\draw [line width=0.7pt] (36.445929715504946,14.735235435611052)-- (38.04592971550494,13.135235435611051);
\draw [line width=0.7pt] (36.445929715504946,13.135235435611051)-- (38.04592971550494,14.735235435611052);
\draw [-to,line width=0.7pt] (10.,10.) -- (14.,6.);
\draw [-to,line width=0.7pt] (30.,6.) -- (34.,10.);
\draw [-to,line width=0.7pt] (18.,16.) -- (26.,16.);
\draw (22,16) node[anchor=south] {$\boldsymbol{F} = \Coss \, \Stretch$};
\draw (12,8) node[anchor=south west] {$\Stretch$};
\draw (32,8) node[anchor=south east] {$\Coss$};
\draw[line width=0.7pt,dashed,color=xfqqff, smooth,samples=100,domain=0.0:0.4951168640240539] plot[parametric] function{1.851438201145571*t**(3.0)+7.93012995336249*t+3.8489437636258463,-3.4914674216625166*t**(3.0)+7.516498923724005*t+12.70222554413047};
\draw[line width=0.7pt,dashed,color=xfqqff, smooth,samples=100,domain=0.4951168640240539:1.0] plot[parametric] function{-1.815624667901769*t**(3.0)+5.4468740037053065*t**(2.0)+5.233290777913776*t+4.29402728206764,3.423929772013722*t**(3.0)-10.271789316041167*t**(2.0)+12.602235037798089*t+11.862880972112391};
\draw[line width=0.7pt,dashed,color=xfqqff, smooth,samples=100,domain=0.0:0.4458981362513473] plot[parametric] function{-3.2023250620808863*t**(3.0)+8.698065320239955*t+4.40545345612604,-4.260132241701372*t**(3.0)-2.2397344399977457*t+17.37637868641052};
\draw[line width=0.7pt,dashed,color=xfqqff, smooth,samples=100,domain=0.4458981362513473:1.0] plot[parametric] function{2.57698244722123*t**(3.0)-7.73094734166369*t**(2.0)+12.1452803313451*t+3.8930845398895557,3.4282234928929807*t**(3.0)-10.284670478678942*t**(2.0)+2.346180958404446*t+16.69476164335922};
\draw[line width=0.7pt,dashed,color=xfqqff, smooth,samples=100,domain=0.0:0.38353560977730844] plot[parametric] function{4.8323757130460505*t**(3.0)+2.566840387682047*t+19.75860968029738,3.5929530608204003*t**(3.0)-8.807626241987853*t+8.088737654323012};
\draw[line width=0.7pt,dashed,color=xfqqff, smooth,samples=100,domain=0.38353560977730844:1.0] plot[parametric] function{-3.0064804961510503*t**(3.0)+9.019441488453152*t**(2.0)-0.8924366034425869*t+20.200861650357247,-2.235369090151665*t**(3.0)+6.706107270454996*t**(2.0)-11.379657183193853*t+8.417559472790195};
\draw[line width=0.7pt,dashed,color=xfqqff, smooth,samples=100,domain=0.0:0.4235739909332629] plot[parametric] function{1.0387367638750005*t**(3.0)+9.720051871082111*t+16.81961648578115,-3.110056111098751*t**(3.0)+5.330953129391901*t+2.891703459453802};
\draw[line width=0.7pt,dashed,color=xfqqff, smooth,samples=100,domain=0.4235739909332629:1.0] plot[parametric] function{-0.763292893941391*t**(3.0)+2.2898786818241734*t**(2.0)+8.750118819068847*t+16.956562623707597,2.285356417447782*t**(3.0)-6.856069252343347*t**(2.0)+8.235005744721803*t+2.4816764074019804};
\draw[line width=0.7pt,dashed,color=xfqqff, smooth,samples=100,domain=0.0:0.6264827270929108] plot[parametric] function{-0.41842082030037997*t**(3.0)+10.692338873904566*t+30.650246381322347,-1.639977496732701*t**(3.0)-0.94816696178928*t+14.932487133219139};
\draw[line width=0.7pt,dashed,color=xfqqff, smooth,samples=100,domain=0.6264827270929108:1.0] plot[parametric] function{0.7017973078836418*t**(3.0)-2.1053919236509255*t**(2.0)+12.011330547832786*t+30.374804547690548,2.750656125023898*t**(3.0)-8.251968375071694*t**(2.0)+4.22154868971009*t+13.852907946670397};
\draw[line width=0.7pt,dashed,color=xfqqff, smooth,samples=100,domain=0.0:0.4007223298080273] plot[parametric] function{-3.1562174201630437*t**(3.0)+2.55153149991141*t+36.4265682788719,0.30812865546894397*t**(3.0)+7.8161604114236285*t+10.783298164556761};
\draw[line width=0.7pt,dashed,color=xfqqff, smooth,samples=100,domain=0.4007223298080273:1.0] plot[parametric] function{2.1104854408862934*t**(3.0)-6.331456322658881*t**(2.0)+5.088687428605041*t+36.087669933927785,-0.20603810026923994*t**(3.0)+0.6181143008077198*t**(2.0)+7.5684682087162996*t+10.81638343007148};
\draw [-to,line width=0.7pt] (6.,4.) -- (8.,4.);
\draw [-to,line width=0.7pt] (6.,4.) -- (6.,6.);
\draw (8,4) node[anchor=west] {$x$};
\draw (6,6) node[anchor=south] {$y$};
\draw [color=qqwwzz](10.652216152257713,16.18842638209827) node[anchor=north west] {$V$};
\draw [color=qqwwzz](33.5,17.4) node[anchor=north west] {$V_\varphi$};
\draw [shift={(8.,16.)},line width=0.7pt,dotted,color=qqqqff]  plot[domain=0.45403061854450577:1.5707963267948966,variable=\t]({1.*0.6987700258055471*cos(\t r)+0.*0.6987700258055471*sin(\t r)},{0.*0.6987700258055471*cos(\t r)+1.*0.6987700258055471*sin(\t r)});
\draw [color=qqqqff](8,17.9) node[anchor=north west] {$\mu_c$};
\begin{scriptsize}
\draw [fill=uuuuuu] (6.,4.) circle (2.5pt);
\end{scriptsize}
\end{tikzpicture}

%% file: figs/mag.tex
\definecolor{qqwwzz}{rgb}{0.,0.4,0.6}
\definecolor{qqqqff}{rgb}{0.,0.,1.}
\definecolor{xfqqff}{rgb}{0.4980392156862745,0.,1.}
\begin{tikzpicture}[line cap=round,line join=round,>=triangle 45,x=1.0cm,y=1.0cm]
\fill[line width=0.7pt,dashed,color=qqwwzz,fill=qqwwzz,fill opacity=0.05000000074505806] (3.8,8.6) -- (3.8,11.8) -- (8.6,11.8) -- (8.6,8.6) -- cycle;
\fill[line width=0.7pt,dashed,color=qqwwzz,fill=qqwwzz,fill opacity=0.05000000074505806] (12.2,8.6) -- (17.,8.6) -- (17.,11.8) -- (12.2,11.8) -- cycle;
\draw [-to,line width=0.7pt,color=xfqqff] (4.4,9.4) -- (4.,9.4);
\draw [-to,line width=0.7pt,color=xfqqff] (4.6,9.6) -- (4.6,10.);
\draw [-to,line width=0.7pt,color=xfqqff] (4.8,9.4) -- (5.2,9.4);
\draw [-to,line width=0.7pt,color=xfqqff] (4.6,9.2) -- (4.6,8.8);
\draw [-to,line width=0.7pt,color=xfqqff] (6.2,9.6) -- (6.2,10.);
\draw [-to,line width=0.7pt,color=xfqqff] (6.4,9.4) -- (6.8,9.4);
\draw [-to,line width=0.7pt,color=xfqqff] (6.2,9.2) -- (6.2,8.8);
\draw [-to,line width=0.7pt,color=xfqqff] (6.,9.4) -- (5.6,9.4);
\draw [-to,line width=0.7pt,color=xfqqff] (7.6,9.4) -- (7.2,9.4);
\draw [-to,line width=0.7pt,color=xfqqff] (7.8,9.6) -- (7.8,10.);
\draw [-to,line width=0.7pt,color=xfqqff] (8.,9.4) -- (8.4,9.4);
\draw [-to,line width=0.7pt,color=xfqqff] (7.8,9.2) -- (7.8,8.8);
\draw [-to,line width=0.7pt,color=xfqqff] (7.8,10.8) -- (7.8,10.4);
\draw [-to,line width=0.7pt,color=xfqqff] (8.,11.) -- (8.4,11.);
\draw [-to,line width=0.7pt,color=xfqqff] (7.8,11.2) -- (7.8,11.6);
\draw [-to,line width=0.7pt,color=xfqqff] (7.6,11.) -- (7.2,11.);
\draw [-to,line width=0.7pt,color=xfqqff] (6.2,10.8) -- (6.2,10.4);
\draw [-to,line width=0.7pt,color=xfqqff] (6.4,11.) -- (6.8,11.);
\draw [-to,line width=0.7pt,color=xfqqff] (6.2,11.2) -- (6.2,11.6);
\draw [-to,line width=0.7pt,color=xfqqff] (6.,11.) -- (5.6,11.);
\draw [-to,line width=0.7pt,color=xfqqff] (4.6,10.8) -- (4.6,10.4);
\draw [-to,line width=0.7pt,color=xfqqff] (4.8,11.) -- (5.2,11.);
\draw [-to,line width=0.7pt,color=xfqqff] (4.6,11.2) -- (4.6,11.6);
\draw [-to,line width=0.7pt,color=xfqqff] (4.4,11.) -- (4.,11.);
\draw [line width=0.7pt,dashed,color=qqwwzz] (3.8,8.6)-- (3.8,11.8);
\draw [line width=0.7pt,dashed,color=qqwwzz] (3.8,11.8)-- (8.6,11.8);
\draw [line width=0.7pt,dashed,color=qqwwzz] (8.6,11.8)-- (8.6,8.6);
\draw [line width=0.7pt,dashed,color=qqwwzz] (8.6,8.6)-- (3.8,8.6);
\draw [-to,line width=0.7pt,color=xfqqff] (13.2,11.) -- (13.5,11.3);
\draw [-to,line width=0.7pt,color=xfqqff] (13.,11.3) -- (13.2,11.6);
\draw [-to,line width=0.7pt,color=xfqqff] (13.,10.8) -- (12.8,10.5);
\draw [-to,line width=0.7pt,color=xfqqff] (12.8,11.) -- (12.5,11.3);
\draw [-to,line width=0.7pt,color=xfqqff] (13.,9.6) -- (12.8,9.9);
\draw [-to,line width=0.7pt,color=xfqqff] (13.3,9.4) -- (13.6,9.6);
\draw [-to,line width=0.7pt,color=xfqqff] (13.,9.2) -- (12.8,8.9);
\draw [-to,line width=0.7pt,color=xfqqff] (12.8,9.4) -- (12.5,9.6);
\draw [-to,line width=0.7pt,color=xfqqff] (14.6,10.9) -- (14.6,10.5);
\draw [-to,line width=0.7pt,color=xfqqff] (14.4,11.) -- (14.1,11.3);
\draw [-to,line width=0.7pt,color=xfqqff] (14.8,11.) -- (15.1,11.3);
\draw [-to,line width=0.7pt,color=xfqqff] (14.6,11.3) -- (14.6,11.7);
\draw [-to,line width=0.7pt,color=xfqqff] (14.6,9.7) -- (14.6,10.1);
\draw [-to,line width=0.7pt,color=xfqqff] (14.8,9.4) -- (15.1,9.6);
\draw [-to,line width=0.7pt,color=xfqqff] (14.6,9.3) -- (14.6,8.9);
\draw [-to,line width=0.7pt,color=xfqqff] (14.4,9.4) -- (14.1,9.6);
\draw [-to,line width=0.7pt,color=xfqqff] (16.2,10.8) -- (16.4,10.5);
\draw [-to,line width=0.7pt,color=xfqqff] (16.,11.) -- (15.7,11.3);
\draw [-to,line width=0.7pt,color=xfqqff] (16.2,11.3) -- (16.1,11.7);
\draw [-to,line width=0.7pt,color=xfqqff] (16.4,11.) -- (16.7,11.3);
\draw [-to,line width=0.7pt,color=xfqqff] (16.2,9.6) -- (16.4,9.9);
\draw [-to,line width=0.7pt,color=xfqqff] (15.9,9.4) -- (15.6,9.6);
\draw [-to,line width=0.7pt,color=xfqqff] (16.2,9.2) -- (16.4,8.9);
\draw [-to,line width=0.7pt,color=xfqqff] (16.4,9.4) -- (16.7,9.6);
\draw [line width=0.7pt,dashed,color=qqwwzz] (12.2,8.6)-- (17.,8.6);
\draw [line width=0.7pt,dashed,color=qqwwzz] (17.,8.6)-- (17.,11.8);
\draw [line width=0.7pt,dashed,color=qqwwzz] (17.,11.8)-- (12.2,11.8);
\draw [line width=0.7pt,dashed,color=qqwwzz] (12.2,11.8)-- (12.2,8.6);
\draw [-to,line width=0.7pt] (9,11.) -- (11.8,11.);
\draw (10.4,11) node[anchor=south] {$\Coss: \quad \Psi_\mathrm{curv} \gg 0$};
\draw (5.4,10.2) node[color=xfqqff] {$\norm{\vb{m}} = 0$};
\draw (13.5,10.2) node[color=xfqqff] {$|\con{\vb{m}}{\vb{e}_2}| > 0$};
\draw [-to,line width=0.7pt] (10,9) -- (10.8,9);
\draw (10.8,9) node[anchor=west] {$x$};
\draw [-to,line width=0.7pt] (10,9) -- (10,9.8);
\draw (10,9.8) node[anchor=south] {$y$};
\end{tikzpicture}

%% file: figs/nanobeam.tex
\definecolor{xfqqff}{rgb}{0.4980392156862745,0.,1.}
\definecolor{qqwwzz}{rgb}{0.,0.4,0.6}
\begin{tikzpicture}[scale=0.25,line cap=round,line join=round,>=triangle 45,x=1.0cm,y=1.0cm]
\fill[line width=0.7pt,fill=black,fill opacity=0.25] (2.,2.) -- (10.,3.) -- (10.,2.) -- (2.,1.) -- cycle;
\fill[line width=0.7pt,color=qqwwzz,fill=qqwwzz,fill opacity=0.10000000149011612] (2.,1.) -- (32.,-6.) -- (32.,-5.) -- (2.,2.) -- cycle;
\fill[line width=0.7pt,color=xfqqff,fill=xfqqff,fill opacity=0.10000000149011612] (32.,-6.) -- (40.,-5.) -- (40.,-4.) -- (32.,-5.) -- cycle;
\fill[line width=0.7pt,color=qqwwzz,fill=qqwwzz,fill opacity=0.10000000149011612] (2.,2.) -- (10.,3.) -- (40.,-4.) -- (32.,-5.) -- cycle;
\draw [line width=0.7pt] (2.,2.)-- (10.,3.);
\draw [line width=0.7pt] (10.,3.)-- (10.,2.);
\draw [line width=0.7pt] (10.,2.)-- (2.,1.);
\draw [line width=0.7pt] (2.,1.)-- (2.,2.);
\draw [line width=0.7pt,color=qqwwzz] (2.,1.)-- (32.,-6.);
\draw [line width=0.7pt,color=qqwwzz] (32.,-6.)-- (32.,-5.);
\draw [line width=0.7pt,color=qqwwzz] (32.,-5.)-- (2.,2.);
\draw [line width=0.7pt,color=black] (2.,2.)-- (2.,1.);
\draw [line width=0.7pt,color=xfqqff] (32.,-6.)-- (40.,-5.);
\draw [line width=0.7pt,color=xfqqff] (40.,-5.)-- (40.,-4.);
\draw [line width=0.7pt,color=xfqqff] (40.,-4.)-- (32.,-5.);
\draw [line width=0.7pt,color=xfqqff] (32.,-5.)-- (32.,-6.);
\draw [line width=0.7pt,color=qqwwzz] (10.,3.)-- (40.,-4.);
\draw [line width=0.7pt,color=qqwwzz] (32.,-5.)-- (2.,2.);
\draw [line width=0.7pt,color=qqwwzz] (10.,2.)-- (40.,-5.);
\draw [color=qqwwzz](16.8,3.4) node[anchor=north west] {$\bar{V} = [0,5000] \times [-500,500]\times[0,100]$};
\draw (4.6,5) node[anchor=north west] {$A_{D}^{\boldsymbol{\varphi}}$};
\draw [color=xfqqff](35.25,-5.5) node[anchor=north west] {$A_{N_{f}}^{\boldsymbol{\varphi}}$};
\end{tikzpicture}

%% file: figs/lcbend.tex
\definecolor{asb}{rgb}{0.,0.4,0.6}
\definecolor{asl}{rgb}{0.4980392156862745,0.,1.}
\begin{tikzpicture}[scale = 0.6, spy using outlines={magnification = 3, circle, size=1.65cm, black, dotted, connect spies}]
			\begin{semilogxaxis}[
				/pgf/number format/1000 sep={},
				axis lines = left,
				xlabel={$\Lc$},
				ylabel={$\max \norm{\disp}$},
				xmin=0.85, xmax=125,
				ymin=650, ymax=4250,
				xtick={1, 10, 100},
				ytick={700, 1400, 2100, 2800,3500, 4200},
				legend pos=south west,
				ymajorgrids=true,
				grid style=dotted,
				]
				\addplot[
				color=blue,
				mark=diamond,
				]
				coordinates {
                        (3.16227766, 4208.130121765515)
                        (10 , 4086.1765721524334)
                        (17.7827941004, 3843.732252911684)
                        (31.6227766, 3211.222332992371)
                        (56.234132519, 2069.7902527036)
                        (1e+2 , 1100.4494842420156)
				};
				\addlegendentry{$\muc/\mu_\mathrm{e} = 10^{-2}$}
				\addplot[
				color=asb,
				mark=triangle,
				]
				coordinates {
                        (1, 4205.564314414478)
                        (3.16227766, 4193.958927019422)
                        (10 , 4080.4885810897995)
                        (17.7827941004, 3823.953495310776)
                        (31.6227766, 3136.639244800965)
                        (56.234132519, 1868.304303810721)
                        (1e+2 , 811.5094568874146)
				};
				\addlegendentry{$\muc/\mu_\mathrm{e} = 10^{-1}$}
				\addplot[
				color=violet,
				mark=pentagon,
				]
				coordinates {
                        (1, 4205.514815411869)
                        (3.16227766, 4193.89772153927)
                        (10 , 4079.577579822768)
                        (17.7827941004, 3818.361390008312)
                        (31.6227766, 3111.7630625242296)
                        (56.234132519, 1803.9953185916102)
                        (1e+2 , 724.7471998320682)
				};
				\addlegendentry{$\muc/\mu_\mathrm{e} = 1$}
				\addplot[
				color=asl,
				mark=square,
				]
				coordinates {
                        (1, 4205.362136482542)
                        (3.16227766, 4193.745034856679)
                        (10 , 4079.324408006833)
                        (17.7827941004, 3817.3501879486694)
                        (31.6227766, 3106.2974341089102)
                        (56.234132519, 1787.1335653080628)
                        (1e+2 , 701.7001936173909)
				};
				\addlegendentry{$\muc/\mu_\mathrm{e} = 10$}
				\addplot[
                dashed,
				color=black,
                mark=none
				]
				coordinates {
                        ( 1e-2 , 4186.7961082624)
                        ( 1e+2 , 4186.7961082624)
				};

                \coordinate (c1) at (axis cs: 9.5, 560);
                \coordinate (c2) at (axis cs: 1.25, 1850);
                \spy on (c1) in node at (c2);

				\addplot[
				color=black,
                mark=none
				]
				coordinates {
                        ( 1e+2 , 1067)
                        ( 1e+2 , 709)
				};
			\end{semilogxaxis}
            \draw (6.5,6.25) node[anchor=north east]{$_{ \muc/ \mu_\mathrm{e} \geq 1, \, \Lc = 0}$};

            \draw (1.85,3.75) node[anchor=west]{$_{\Delta \norm{\disp} = 399}$};
		\end{tikzpicture}

%% file: figs/lctwist.tex
\definecolor{asb}{rgb}{0.,0.4,0.6}
\definecolor{asl}{rgb}{0.4980392156862745,0.,1.}
\begin{tikzpicture}[scale = 0.6, spy using outlines={magnification = 1.75, circle, size=1.65cm, black, dotted, connect spies}]
			\begin{semilogxaxis}[
				/pgf/number format/1000 sep={},
				axis lines = left,
				xlabel={$\Lc$},
				ylabel={$\max \norm{\disp}$},
				xmin=0.85, xmax=125,
				ymin=190, ymax=750,
				xtick={1, 10, 100},
				ytick={200, 300, 400, 500, 600, 700},
				legend pos=south west,
				ymajorgrids=true,
				grid style=dotted,
				]
				\addplot[
				color=blue,
				mark=diamond,
				]
				coordinates {
                        (1, 745.3279289239471)
                        (3.16227766, 743.7271806541945)
                        ( 10 , 735.9584826078268)
                        (17.7827941004, 719.5027622696118)
                        (31.6227766, 672.8088882846106)
                        (56.234132519, 553.3952801955758)
                        ( 1e+2 , 339.46068309160677)
				};
				\addlegendentry{$\muc/\mu_\mathrm{e} = 10^{-2}$}
				\addplot[
				color=asb,
				mark=triangle,
				]
				coordinates {
                        (1, 743.9908137002813)
                        (3.16227766, 743.185532011826)
                        ( 10 , 735.2725219199123)
                        (17.7827941004, 716.6734156038467)
                        (31.6227766, 660.0333460741346)
                        (56.234132519, 502.92732380567185)
                        ( 1e+2 , 238.63602753801248)
				};
				\addlegendentry{$\muc/\mu_\mathrm{e} = 10^{-1}$}
				\addplot[
				color=violet,
				mark=pentagon,
				]
				coordinates {
                        (1, 743.9761236584789)
                        (3.16227766, 743.1667501628237)
                        ( 10 , 735.1326475983036)
                        (17.7827941004, 716.0698658458682)
                        (31.6227766, 657.277796938927)
                        (56.234132519, 490.4279454991494)
                        ( 1e+2 , 213.6508441491632)
				};
				\addlegendentry{$\muc/\mu_\mathrm{e} = 1$}
				\addplot[
				color=asl,
				mark=square,
				]
				coordinates {
                        (1, 743.9133136990191)
                        (3.16227766, 743.1035001851121)
                        ( 10 , 735.0386725971979)
                        (17.7827941004, 715.8488663952116)
                        (31.6227766, 656.474704789179)
                        (56.234132519, 487.1803635846731)
                        ( 1e+2 , 208.33239403386892)
				};
				\addlegendentry{$\muc/\mu_\mathrm{e} = 10$}
				\addplot[
                dashed,
				color=black,
                mark=none
				]
				coordinates {
                        ( 1e-2 , 741)
                        ( 1e+2 , 741)
				};

                \coordinate (c1) at (axis cs: 9.5, 200);
                \coordinate (c2) at (axis cs: 1.25, 375);
                \spy on (c1) in node at (c2);

				\addplot[
				color=black,
                mark=none
				]
				coordinates {
                        ( 1e+2 , 206)
                        ( 1e+2 , 335)
				};
			\end{semilogxaxis}
            \draw (6.5,6.25) node[anchor=north east]{$_{ \muc/ \mu_\mathrm{e} \geq 1, \, \Lc = 0}$};

            \draw (1.85,3.75) node[anchor=west]{$_{\Delta \norm{\disp} = 131}$};
		\end{tikzpicture}

%% file: figs/gphi.tex
\definecolor{asb}{rgb}{0.,0.4,0.6}
\definecolor{asl}{rgb}{0.4980392156862745,0.,1.}
\begin{tikzpicture}[scale = 0.6]
			\begin{axis}[
				/pgf/number format/1000 sep={},
				axis lines = left,
				xlabel={$\gamma$},
				ylabel={$\max \norm{\disp}$},
				xmin=-1.1e+3, xmax=1.1e+3,
				ymin=3450, ymax=4150,
				xtick={-1000,-500,0,500,1000,2000},
				ytick={3500,3650,3800,3950,4100},
				legend pos=north west,
				ymajorgrids=true,
				grid style=dotted,
				]
				\addplot[
				color=blue,
				mark=diamond,
				]
				coordinates {
                        (-1000, 3535.590485995321)
                        (-750, 3711.781423386873)
                        (-500, 3883.9808530082382)
                        (-250, 4022.9729962095544)
                        (-100, 4070.054823965246)
                        (-10, 4079.481393353536)
                        (0, 4079.577587247499)
                        (10, 4079.481393353535)
                        (100, 4070.0548239652535)
                        (250, 4022.972996209557)
                        (500, 3883.9808530082387)
                        (750, 3711.7814233868858)
                        (1000, 3535.590485995316)
				};
            \addlegendentry{$\gamma_\mathrm{iso}$}

				\addplot[
				color=asb,
				mark=triangle,
				]
				coordinates {
                        (-1000, 3662.7912088573603)
                        (-750, 3825.146335715924)
                        (-500, 3955.432736556902)
                        (-250, 4041.476280839316)
                        (-100, 4070.968995212766)
                        (-10, 4079.4622402069563)
                        (0, 4079.5775872474956)
                        (10, 4079.4622402069595)
                        (100, 4070.9689952127637)
                        (250, 4041.4762808393302)
                        (500, 3955.4327365568965)
                        (750, 3825.146335715941)
                        (1000, 3662.7912088573603)
				};
            \addlegendentry{$\gamma_\mathrm{cub}$}

            
			\end{axis}
		\end{tikzpicture}

%% file: figs/gb.tex
\definecolor{asb}{rgb}{0.,0.4,0.6}
\definecolor{asl}{rgb}{0.4980392156862745,0.,1.}
\begin{tikzpicture}[scale = 0.6]
			\begin{axis}[
				/pgf/number format/1000 sep={},
				axis lines = left,
				xlabel={$\gamma$},
            ylabel={$\ext \con{\tbmag}{\vb{e}_3}$},
				xmin=-1.1e+3, xmax=1.1e+3,
				ymin=-0.3, ymax=0.3,
				xtick={-1000,-500,0,500,1000,2000},
				ytick={-0.25,-0.125,0,0.125,0.25},
				legend style={at={(0.03,0.5)},anchor=west},
				ymajorgrids=true,
				grid style=dotted,
				]
				\addplot[
				color=blue,
				mark=diamond,
				]
				coordinates {
                        (-1000, 0.23485436007498384)
                        (-750, 0.17990081668920138)
                        (-500, 0.12300541501685075)
                        (-250, 0.0633114670743007)
                        (-100, 0.025787849847364538)
                        (-10, 0.002592630513684062)
                        (0, 0)
                        (10, -0.002592630513684059)
                        (100, -0.02578784984736473)
                        (250, -0.06331146707430112)
                        (500, -0.12300541501685068)
                        (750, -0.17990081668920133)
                        (1000, -0.23485436007498303)
				};
            \addlegendentry{$\gamma_\mathrm{iso}$}

				\addplot[
				color=asb,
				mark=triangle,
				]
				coordinates {
                        (-1000, -0.2802550553829114)
                        (-750, -0.2143078409051356)
                        (-500, -0.1477351077074482)
                        (-250, -0.07612825607926257)
                        (-100, -0.03074200291672062)
                        (-10, -0.003085649257736697)
                        (0, 0)
                        (10, 0.0030856492577367033)
                        (100, 0.03074200291672068)
                        (250, 0.07612825607926306)
                        (500, 0.1477351077074476)
                        (750, 0.21430784090513733)
                        (1000, 0.28025505538291207)
				};
            \addlegendentry{$\gamma_\mathrm{cub}$}

			\end{axis}
		\end{tikzpicture}

%% file: figs/twigphi.tex
\definecolor{asb}{rgb}{0.,0.4,0.6}
\definecolor{asl}{rgb}{0.4980392156862745,0.,1.}
\begin{tikzpicture}[scale = 0.6]
			\begin{axis}[
				/pgf/number format/1000 sep={},
				axis lines = left,
				xlabel={$\gamma$},
				ylabel={$\max \norm{\disp}$},
				xmin=-1.1e+3, xmax=1.1e+3,
				ymin=718, ymax=737,
				xtick={-1000,-500,0,500,100},
				ytick={720,725,730,735},
				legend style={at={(0.5,0.03)},anchor=south},
				ymajorgrids=true,
				grid style=dotted,
				]
				\addplot[
				color=blue,
				mark=diamond,
				]
				coordinates {
                        (-2000, 696.3246950880534)
                        (-1000, 722.281044900731)
                        (-750, 727.6585848259098)
                        (-500, 731.720238487981)
                        (-250, 734.2639828544704)
                        (-100, 734.9930123847168)
                        (-10, 735.1313893426953)
                        (0, 735.1327885925028)
                        (10, 735.1313893426947)
                        (100, 734.9930123847123)
                        (250, 734.2639828544725)
                        (500, 731.7202384879758)
                        (750, 727.6585848259147)
                        (1000, 722.2810449007238)
                        (2000, 696.3246950880495)
				};
            \addlegendentry{$\gamma_\mathrm{iso}$}

				\addplot[
				color=asb,
				mark=triangle,
				]
				coordinates {
                        (-2000, 734.4176726340238)
                        (-1000, 734.5763036268904)
                        (-750, 734.6103340017451)
                        (-500, 734.6538242923016)
                        (-250, 734.7631214758943)
                        (-100, 734.9735098467326)
                        (-10, 735.1293914799752)
                        (0, 735.1327885924873)
                        (10, 735.1293914799735)
                        (100, 734.9735098467323)
                        (250, 734.7631214759037)
                        (500, 734.6538242923117)
                        (750, 734.6103340017507)
                        (1000, 734.5763036268964)
                        (2000, 734.4176726340244)
				};
            \addlegendentry{$\gamma_\mathrm{cub}$}

			\end{axis}
		\end{tikzpicture}

%% file: figs/twigb.tex
\definecolor{asb}{rgb}{0.,0.4,0.6}
\definecolor{asl}{rgb}{0.4980392156862745,0.,1.}
\begin{tikzpicture}[scale = 0.6]
			\begin{axis}[
				/pgf/number format/1000 sep={},
				axis lines = left,
				xlabel={$\gamma$},
            ylabel={$\max \norm{\tbmag}$},
				xmin=-1.1e+3, xmax=1.1e+3,
				ymin=-0.05, ymax=0.35,
				xtick={-1000,-500,0,500,100},
				ytick={0,0.1,0.2,0.3},
				legend pos=south west,
				ymajorgrids=true,
				grid style=dotted,
				]
				\addplot[
				color=blue,
				mark=diamond,
				]
				coordinates {
                        (-2000, 1.156997400867491)
                        (-1000, 0.33222074797321527)
                        (-750, 0.24444517810288036)
                        (-500, 0.16014079495547592)
                        (-250, 0.0803422536295229)
                        (-100, 0.03245038527903728)
                        (-10, 0.003254923758894494)
                        (0, 0)
                        (10, 0.0032549237588945096)
                        (100, 0.03245038527903762)
                        (250, 0.08034225362952188)
                        (500, 0.1601407949554726)
                        (750, 0.24444517810290156)
                        (1000, 0.3322207479732014)
                        (2000, 1.1569974008674502)
				};
            \addlegendentry{$\gamma_\mathrm{iso}$}

				\addplot[
				color=asb,
				mark=triangle,
				]
				coordinates {
                        (-2000, 0.6037096994217915)
                        (-1000, 0.30619861699962686)
                        (-750, 0.23044981707236212)
                        (-500, 0.15407878369750255)
                        (-250, 0.07721589241084162)
                        (-100, 0.0309140552746497)
                        (-10, 0.003095886621876087)
                        (0, 0)
                        (10, 0.0030958866218760736)
                        (100, 0.030914055274649795)
                        (250, 0.07721589241084208)
                        (500, 0.15407878369750347)
                        (750, 0.23044981707236314)
                        (1000, 0.30619861699962464)
                        (2000, 0.6037096994217991)
				};
            \addlegendentry{$\gamma_\mathrm{cub}$}

			\end{axis}
		\end{tikzpicture}

%% file: figs/nanobeamtop.tex
\definecolor{xfqqff}{rgb}{0.4980392156862745,0.,1.}
\definecolor{qqwwzz}{rgb}{0.,0.4,0.6}
\begin{tikzpicture}[scale=0.25,line cap=round,line join=round,>=triangle 45,x=1.0cm,y=1.0cm]
\fill[line width=0.7pt,fill=black,fill opacity=0.25] (2.,2.) -- (10.,3.) -- (10.,2.) -- (2.,1.) -- cycle;
\fill[line width=0.7pt,color=qqwwzz,fill=qqwwzz,fill opacity=0.10000000149011612] (2.,1.) -- (32.,-6.) -- (32.,-5.) -- (2.,2.) -- cycle;
\fill[line width=0.7pt,color=xfqqff,fill=qqwwzz,fill opacity=0.10000000149011612] (32.,-6.) -- (40.,-5.) -- (40.,-4.) -- (32.,-5.) -- cycle;
\fill[line width=0.7pt,color=qqwwzz,fill=xfqqff,fill opacity=0.10000000149011612] (2.,2.) -- (10.,3.) -- (40.,-4.) -- (32.,-5.) -- cycle;
\draw [line width=0.7pt] (2.,2.)-- (10.,3.);
\draw [line width=0.7pt] (10.,3.)-- (10.,2.);
\draw [line width=0.7pt] (10.,2.)-- (2.,1.);
\draw [line width=0.7pt] (2.,1.)-- (2.,2.);
\draw [line width=0.7pt,color=qqwwzz] (2.,1.)-- (32.,-6.);
\draw [line width=0.7pt,color=qqwwzz] (32.,-6.)-- (32.,-5.);
\draw [line width=0.7pt,color=xfqqff] (32.,-5.)-- (2.,2.);
\draw [line width=0.7pt,color=black] (2.,2.)-- (2.,1.);
\draw [line width=0.7pt,color=qqwwzz] (32.,-6.)-- (40.,-5.);
\draw [line width=0.7pt,color=qqwwzz] (40.,-5.)-- (40.,-4.);
\draw [line width=0.7pt,color=xfqqff] (40.,-4.)-- (32.,-5.);
\draw [line width=0.7pt,color=qqwwzz] (32.,-5.)-- (32.,-6.);
\draw [line width=0.7pt,color=xfqqff] (10.,3.)-- (40.,-4.);
\draw [line width=0.7pt,color=xfqqff] (32.,-5.)-- (2.,2.);
\draw [line width=0.7pt,color=qqwwzz] (10.,2.)-- (40.,-5.);
\draw [color=xfqqff](18.5,3.4) node[anchor=north west] {$A_{N_{k}}^{\tbmag}$};
\draw (4.6,5) node[anchor=north west] {$A_{D}^{\boldsymbol{\varphi}}$};
\end{tikzpicture}

%% file: figs/cvss.tex
\definecolor{asb}{rgb}{0.,0.4,0.6}
\definecolor{asl}{rgb}{0.4980392156862745,0.,1.}
\begin{tikzpicture}[scale = 0.6]
			\begin{loglogaxis}[
				/pgf/number format/1000 sep={},
				axis lines = left,
				ylabel={$\max \norm{\disp}$},
				xmin=0.5e-4, xmax=2e-1,
				ymin=0.05e-4, ymax=0.5,
				xtick={1e-4,1e-3,1e-2,1e-1},
				ytick={1e-5,1e-4,1e-3,1e-2,1e-1},
				legend pos=north west,
				ymajorgrids=true,
				grid style=dotted,
				]
				\addplot[
				color=blue,
				mark=diamond,
				]
				coordinates {
                        (1e-4, 0.000020287019999166176)
                        (1e-3, 0.0002)
                        (1e-2, 0.002)
                        (1e-1, 0.02)
				};
            \addlegendentry{$\chi_\mathrm{m}\mu_0\eta_\mathrm{iso}$}

				\addplot[
				color=asb,
				mark=triangle,
				]
				coordinates {
                        (1e-4, 0.000011317186371833368)
                        (1e-3, 0.00011)
                        (1e-2, 0.0011)
                        (1e-1, 0.011)
				};
            \addlegendentry{$\widetilde{f}_1$}

            
			\end{loglogaxis}
		\end{tikzpicture}